\documentclass[journal]{IEEEtran}

\ifx\Red\relax\def\Red#1{#1}\fi
\def\Red#1{#1}

\usepackage{epsfig}
\usepackage{amssymb}
\usepackage{verbatim}
\usepackage{delarray}
\usepackage{graphics}
\usepackage{amsmath}

\newcommand{\inner}[2]{\langle{#1},{#2}\rangle}

\newcommand{\A}{{\mathcal{A}}}

\newcommand{\CC}{{\mathcal{C}}}

\newcommand{\FF}{{\mathcal{F}}}
\newcommand{\G}{{\mathcal{G}}}

\newcommand{\I}{{\mathcal{I}}}
\newcommand{\J}{{\mathcal{J}}}

\newcommand{\SSS}{{\mathcal{S}}}
\newcommand{\T}{{\mathcal{T}}}

\newcommand{\W}{{\mathcal{W}}}

\newcommand{\C}{{\mathbb{C}}}
\newcommand{\F}{{\mathbb{F}}}
\newcommand{\R}{{\mathbb{R}}}
\newcommand{\Z}{{\mathbb{Z}}}
\newcommand{\zerob}{{\mathbf 0}}
\newcommand{\oneb}{{\mathbf 1}}
\newcommand{\ab}{{\mathbf a}}
\newcommand{\bb}{{\mathbf b}}
\newcommand{\cb}{{\mathbf c}}
\newcommand{\eb}{{\mathbf e}}
\newcommand{\fb}{{\mathbf f}}

\newcommand{\pb}{{\mathbf p}}

\renewcommand{\sb}{{\mathbf s}}

\newcommand{\xb}{{\mathbf x}}
\newcommand{\yb}{{\mathbf y}}
\newcommand{\zb}{{\mathbf z}}

\newcommand{\Bf}{{\mathfrak{B}}}

\newcommand{\ie}{{\em i.e., }}
\newcommand{\eg}{{\em e.g., }}

\newcommand{\im}{\mathrm{im~}}

\newcommand{\openbox}{\leavevmode
     \hbox to.77778em{%
     \hfil\vrule
     \vbox to.675em{\hrule width.6em\vfil\hrule}%
     \vrule\hfil}}
\newcommand{\qed}{\hspace*{1cm}\hspace*{\fill}\openbox}

\usepackage{tikz}\usetikzlibrary{positioning}
\usetikzlibrary{decorations.markings}  
\newlength{\davesep}
\newlength{\franksep}
\tikzset{
   >=latex,
   dot/.style={circle,draw,fill=black,inner sep=0pt,minimum size=3.5pt},
   box/.style={rectangle,draw,inner sep=1pt,minimum size=3ex},
   var/.style={inner sep=1pt},
   invert/.style={circle,draw,inner sep=0pt,minimum size=3.5pt},
   mid arrowa/.style={postaction={decorate,decoration={markings,mark=at position .6 with {\arrow[rotate=\aangle,cm={\ca,-\sa,0,1,(0,0)}]{>}}}}},
   mid revarrowa/.style={postaction={decorate,decoration={markings,mark=at position .4 with {\arrowreversed[rotate=\aangle,cm={\ca,-\sa,0,1,(0,0)}]{>}}}}},
   mid arrowb/.style={postaction={decorate,decoration={markings,mark=at position .6 with {\arrow[rotate=-\bangle,cm={\cb,\sb,0,1,(0,0)}]{>}}}}},
   mid revarrowb/.style={postaction={decorate,decoration={markings,mark=at position .4 with {\arrowreversed[rotate=-\bangle,cm={\cb,\sb,0,1,(0,0)}]{>}}}}},
   mid arrowc/.style={postaction={decorate,decoration={markings,mark=at position .6 with {\arrow{>}}}}},
   mid revarrowc/.style={postaction={decorate,decoration={markings,mark=at position .4 with {\arrowreversed{>}}}}},
   daveline/.style={black},
}
\newcommand{\updown}{+(0,1ex) -- +(0,-1ex) +(0,0)}
\newcommand{\leftright}{+(-1ex,0) -- +(1ex,0) +(0,0)}
\begin{document}

\title{Codes on Graphs:  Models for Elementary \\ Algebraic Topology and Statistical Physics}

\author{G. David Forney, Jr., \textit{Life Fellow, IEEE}\thanks{
The author is with the
Laboratory for Information and Decision Systems,
Massachusetts Institute of Technology,
Cambridge, MA 02139
 (email: forneyd@comcast.net).  
 
  \emph{IEEE Trans.\ Inf.\ Theory}, vol. 64, pp.\ 7465--7487, Dec.\ 2018.
 Copyright (c) 2018 IEEE. Personal use of this material is permitted.  However, permission to use this material for any other purposes must be obtained from the IEEE by sending a request to pubs-permissions@ieee.org.  DOI 10.1109/TIT.2018.2866577.
}}
 \date{}

\maketitle

\begin{abstract}  This paper is mainly a semi-tutorial introduction to elementary algebraic topology and  its applications to Ising-type models of statistical physics, using graphical models of linear and group codes.  It contains new material  on systematic $(n,k)$  group codes and their information sets; normal realizations of homology and cohomology spaces; dual and hybrid models;  and connections with system-theoretic concepts such as observability, controllability, and  input/output realizations. \end{abstract}


\section{Introduction}  Algebraic topology goes back to Kirchhoff's circuit laws \cite{Bamberg};  however, it  is not very familiar to most engineers and scientists.  A major purpose of this paper is to provide an introduction to elementary algebraic topology using  graphical models that have arisen in coding theory--- namely, normal realizations (NRs) \cite{F01} and normal factor graphs (NFGs) \cite{L04}--- which turn out to be very well suited to this purpose.  

This work was directly stimulated by that of Al-Bashabsheh and Vontobel \cite{AV16}, who as far as we know were the first  to use NFGs to model algebraic topology spaces. They apply these models to computing partition functions of Ising-type models of statistical physics, which had been shown by Al-Bashabsheh and Mao \cite{AM11} and Forney and Vontobel \cite{FV11} to be nicely modeled by NFGs.  Some differences in our approach are:
\begin{itemize}
\item  We use NRs rather than NFGs to model the principal spaces of elementary algebraic topology.
\item We focus on the group case rather than the field case, although we treat both.
\item  In particular, algebraic topology spaces are regarded as ``systematic $(n,k)$ group codes."
\item We make connections to system-theoretic notions such as observability, controllability and  input/output (I/O) realizations.
\end{itemize}

Another stimulus was the work of Molkaraie \emph{et al}. \cite{M15, M16, ML13}, who have used dual NFGs for Monte Carlo evaluations of partition functions of Ising-type models.  Our work lays an algebraic foundation for such evaluations, and systematically presents  alternative approaches to carrying them out.

In Section \ref{Sec2}, we develop the main results of elementary (one-dimensional) algebraic topology.  
We describe the topology of a  graph $\G = (V,E)$ by its \emph{incidence matrix} $M$ \cite{Bollobas}, and show how this allows us to treat the group and vector space cases in a common framework.  

We introduce the concept of a \emph{systematic $(n,k)$ group code} over a group alphabet $\A$, generalizing a linear $(n,k)$ block code, and  show that the principal spaces of elementary algebraic topology are systematic $(n,k)$ group codes.  We model all of these spaces by normal realizations.  Using system-theoretic concepts such as observability and controllability, we then reduce these realizations to nonredundant (observable and controllable) input/output (I/O) realizations.

We begin  with cohomology (coboundary operators, etc.), which we regard as more basic than homology (boundary operators, etc.), and then obtain  dual  results using an  elementary Adjoint Homomorphism Lemma, as well as normal realization duality.  We  exhibit ``bases" of principal spaces  that are based on  cut sets and cycles of $\G$ in the primal and dual cases, respectively \cite{Bollobas}.
Finally, we give simple dual normal realizations that we believe capture the essences of the zeroth and first (co)homology spaces of $\G$.

In Section \ref{Sec3}, following \cite{AM11, AV16, FV11, ML13}, we show how to model  partition functions of an Ising-type (\eg Ising or Potts) model by ``edge-weighted normal factor graphs" based on normal realizations of algebraic-topology spaces as in Section \ref{Sec2}.  Some of the I/O realizations of Section \ref{Sec2} are  simpler than the straightforward normal realizations that have been used previously, and may be more suitable for simulations.  

Since the partition function of such a model is just a number, it is equal to its Fourier transform, which is represented by the dual NFG, up to scale.  As observed in \cite{ML13}, computations based on the dual NFG may be simpler, as in the case of a single-cycle graph, or  may behave better at low temperatures.  We generalize  the well-known high-temperature expansion for Ising models ($\A = \Z_2$) to Potts models ($\A = \Z_q$) and to generalized Ising-type models whose spin alphabet $\A$ may be any finite abelian group.

In the presence of an external field, we show that a realization of the partition function using the dual NFG is generally more complicated (higher-dimensional) than using the primal NFG.  To reduce this increased complexity, we suggest a novel hybrid model, with part in the primal domain, part in the dual domain, and  a Fourier transform between them.
 
In Section \ref{Sec4},  following \cite{AV16}, we give an introduction to two-dimensional algebraic topology, using planar graphs to illustrate two-dimensional complexes.  We introduce dual graphs, and show (as in \cite{AV16, ML13}) that there are in general four different ways to represent the partition function of an Ising-type model on a planar graph $\G$, involving either $\G$ or its dual graph $\hat{\G}$, and either the original interaction weights or their Fourier transforms (in which  temperature is dualized).  

In an Appendix, we give a very simple proof of the Normal Factor Graph Duality Theorem, including scale factors, and show how the scale factor must be modified when the NFG is based on a normal realization.  Interestingly, this leads to an alternative proof of the Controllability Test of \cite{FGL12}.

\section{Introduction to  algebraic topology}\label{Sec2}

  The concepts of  elementary (one- and two-dimensional) algebraic topology are often phrased in scary mathematical jargon, but they actually  involve only some elementary graph theory and  linear algebra, or, more fundamentally, the algebra of abelian groups.  
  
 This section is a tutorial introduction to these concepts, with the following unusual features:
  \begin{itemize}
  \item we  treat the field and group cases in a common setting;
  \item we give graphical models (normal realizations) of all important  spaces;
  \item we begin with cochains and coboundary operators rather than chains and boundary operators; consequently, our primal model is the dual of the usual primal model, and \emph{vice versa};
  \item we use system-theoretic properties such as observability, controllability and I/O realizations.
  \end{itemize}
  
  \subsection{Elementary graph theory}
  
  A finite undirected graph $\G = (V, E)$ is specified by a finite \emph{vertex set} $V$, a finite \emph{edge set} $E$, and a specification of which two vertices in $V$ are incident on each edge $e \in E$.
  
  Rather than specifying the  topology of $\G$  as usual by its adjacency matrix $A$, indexed by $V \times V$, we will use instead its \emph{incidence matrix} $M$, indexed by $E \times V$, which is defined as follows \cite{Bollobas}.

We first give each edge $e \in E$ an orientation, perhaps arbitrary.   This orientation is merely a technical device  to resolve  ambiguities;  we are still thinking of $\G$ as an undirected graph.  

We then define $M_{ev} =1$ if $v = h(e)$, the \emph{head} of $e$;  $M_{ev} =-1$ if $v = t(e)$, the \emph{tail} of $e$; and $M_{ev} =0$ otherwise.       (We assume that there are no self-loops;  \ie that $h(e) \neq t(e)$.) 

Each edge $e\in E$ is thus associated with a $\{0, \pm1\}$-valued vector  $M_e = (M_{ev}, v \in V)$, namely the $e$th row of $M$,  which has precisely two nonzero components, namely $M_{eh(e)} = +1$ and $M_{et(e)} = -1$.  (As we will see, it makes no difference whether we take $M_e$ or $-M_e$ as the $e$th row of $M$.)
  
  Thus each of the $|E|$ rows of the incidence matrix $M$ has two nonzero values, namely $\pm 1$.  The number of nonzero values in the $v$th column $M_v$ of $M$ is  the number of edges whose initial or final vertex is $v$, namely the \emph{degree} $d_v$ of $v$.

\begin{figure}[h]
\setlength{\unitlength}{5pt}
\centering
%%%%%%%%%%%%%%
% Figure 1
%%%%%%%%%%%%%%
\begin{tikzpicture}
\node[dot,label=left:$v_1$]                  (V1) {};
\node[dot,right=8ex of V1,label=above:$v_2$] (V2) {};
\node[dot,right=8ex of V2,label=right:$v_3$] (V3) {};
\node[dot,below=8ex of V2,label=right:$v_4$] (V4) {};
\node[dot,below=8ex of V1,label=left:$v_5$]  (V5) {};
\draw[->] (V1) -- node[above] {$e_1$} (V2);
\draw[->] (V2) -- node[above] {$e_2$} (V3);
\draw[->] (V1) -- node[left] {$e_3$} (V5);
\draw[->] (V2) -- node[left] {$e_4$} (V4);
\draw[->] (V3) -- node[right] {$e_5$} (V4);
\draw[->] (V4) -- node[above] {$e_6$} (V5);
\end{tikzpicture}
\caption{Graph $\G$ of Example 1.}
\label{E1}
\end{figure}
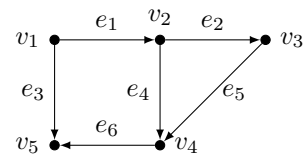

\noindent
\textbf{Example 1}. Consider the directed graph $\G$ of Figure \ref{E1}.  $\G$ has $|V| = 5$ vertices and $|E| = 6$ edges.  Its incidence matrix is 
$$
M = \left[
\begin{array}{rrrrr}
-1 & +1 & 0 & 0 & 0 \\
0 & -1 & +1 & 0 & 0 \\
-1 & 0 & 0 & 0 & +1 \\
0 & -1 & 0 & +1 & 0 \\
0 & 0 & -1 & +1 & 0 \\
0 & 0 & 0 & -1 & +1  
\end{array}
\right].  \qed
$$ 

The key graphical parameters of a finite  graph $\G = (V, E)$ are: 
\begin{itemize}
\item the number $|V|$ of its vertices; 
\item the number $|E|$ of its edges; 
\item the number $\beta_0(\G)$ of its connected components, also called its \emph{zeroth Betti number}; 
\item its \emph{cyclomatic number}\footnote{The term ``cyclomatic number" was apparently coined by James Clerk Maxwell \cite{Giblin}.} $\beta_1(\G) = |E| - |V| + \beta_0(\G)$, also called  its \emph{first Betti number}. 
\end{itemize}
For example, for our example graph $\G$, we have $|E| = 6$, $|V| = 5$, and $\beta_0(\G) = 1$, so $\beta_1(\G) = 6-5+1 = 2$.

Evidently $\beta_0(\G) \ge 0$, with equality if and only if $\G$ is the empty graph with $|V| = |E| = 0$.  We will assume that $\G$ is nonempty; \ie $\beta_0(\G) \ge 1$.

If $\G = (V, E)$ has $\beta_0(\G)$ connected components $\G_i = (V_i,E_i)$, then $V = \bigcup_i V_i$ and $E = \bigcup_i E_i$.  If the vertices and edges are ordered to reflect this partition, then the incidence matrix of $\G$ is $M = \textrm{diag}\{M_1, M_2, \ldots\}$, where $M_i$ is the incidence matrix of $\G_i$.  The component graphs $\G_i$ may then be analyzed independently.  Therefore, without essential loss of generality, we will assume from now on that $\G$ is connected;  \ie $\beta_0(\G) = 1$ and $\beta_1(\G) = |E| - |V| + 1$.

It is well known in elementary graph theory that if $\G = (V, E)$ is connected, then $\G$ contains a \emph{spanning tree} $\T = (V, E_{\T})$, namely a {cycle-free} connected subgraph of $\G$ that comprises all $|V|$ vertices of $\G$ and $|E_{\T}| = |V|-1$ of its edges.  The  number of edge deletions required to reduce $\G$ to a spanning tree  $\T$ of $\G$ is thus  $|E| - |E_\T| = |E| - |V| + 1 = \beta_1(\G)$, the cyclomatic number of $\G$.  It follows that $\beta_1(\G) \ge 0$, with equality if and only if $\G$ is cycle-free.  We will see that $\beta_1(\G)$ measures the number of independent cycles in $\G$.

For example, given the graph $\G$ of Example 1, we may obtain a  spanning tree $\T \subseteq \G$ by removing  $\beta_1(\G) = 2$ edges; \eg $e_3$ and $e_5$.
Here $\G$ contains three cycles, any two of which may be chosen as ``independent;" the third cycle is then the ``sum" of the other two.  (These terms will be  defined more satisfactorily later.)

\subsection{Elements of algebraic topology}

In algebraic topology \cite{Bamberg}, a finite graph $\G = (V,E)$ is called a \emph{1-dimensional complex}.  Its vertices $v \in V$ are called its \emph{0-dimensional objects}, and its edges $e \in E$ are called  its \emph{1-dimensional objects}. As we shall see, its incidence matrix $M$  is the matrix of a certain operator.

Let $\A$ be some abelian group alphabet.  The reader will not be misled if he or she  assumes that $\A$  is a field;   however, we assert that all of our results hold if $\A$ is  any abelian group with a well-defined dual group (\ie character group) $\hat{\A}$;  \eg any finite abelian group.  We will use notation that is appropriate when $\A$ is a finite abelian group.

We consider the  spaces $C^0 = \A^V$ and $C^1 = \A^E$ of all $\A$-valued functions defined on $V$ and $E$, respectively.  In algebraic topology, the elements of $\A^V$ and $\A^E$ are called \emph{0-cochains} and \emph{1-cochains}, respectively. We will regard the elements $\xb \in \A^V$ as sets $\xb = (x_{v_1}, x_{v_2}, \ldots, x_{v_{|V|}})$ of $|V|$ elements of $\A$ indexed by the elements of $V$, and similarly the elements $\yb \in \A^E$  as sets $\yb = (y_{e_1}, y_{e_2}, \ldots, y_{e_{|E|}})$ of $|E|$ elements of $\A$ indexed by the elements of $E$.  When $\xb$ and $\yb$ are used in conjunction with the incidence matrix $M$, we may regard them as column vectors, and we will call them  \emph{vertex vectors} $\xb$ and \emph{edge vectors} $\yb$, respectively.  

In elementary algebraic topology,  the \emph{coboundary operator} $d$ is defined as the homomorphism whose matrix is $M$;  \ie
$$d: \A^V \to \A^E, \quad \xb \mapsto M\xb.$$
This definition makes sense because the ``product" $za$ is well-defined for all $z \in \{0, \pm 1\}, a \in \A$, for any abelian group $\A$;  \ie we   regard $M$ as an integer matrix.\footnote{
More precisely, any group homomorphism $\varphi:  \A^V \to \A^E$ may be characterized by a matrix of group homomorphisms  $\{\varphi_{ev}:  \A \to \A\}$.  For the coboundary operator $d$,  the constituent homomorphisms $d_{ev}$ are all zero, identity, or negative identity homomorphisms, as  indicated by the entries $M_{ev} \in \{0, \pm 1\}$ of the incidence matrix $M$.} 
If $\A$ contains a unit element $1$--- \eg if $\A$ is a field or a ring with unity--- then $M$ may alternatively be thought of as an $\A$-matrix.  

The image   of the coboundary operator $d$ is called the \emph{coboundary space} $B^1 =  \im d$. Thus $B^1$ is  the subspace/subgroup of $C^1 = \A^E$ that is generated by the \emph{elementary coboundary vectors} $\{M_v a \mid v \in V, a \in \A\}$.  If $\A$ is a field, then $B^1$ is the column space of  $M$.

The kernel of $d$ is called the \emph{zero-coboundary space} $Z^0 = \ker d$.  We have immediately:

\vspace{1ex}
\noindent
\textbf{Theorem 1(a)} ($Z^0$).  If $\G = (V,E)$ is a connected graph, then its zero-coboundary space $Z^0 = \ker d$ is the repetition code $\CC_= = \{(a, a, \ldots, a) \mid a \in \A\} \subseteq \A^V$ over $\A$.

\vspace{1ex}
\noindent
\textit{Proof}:  The constraint $M\xb = \zerob$ implies that for every edge $e \in E$, $M_e \xb = 0$.  Since the row vector $M_e$ has precisely two nonzero values of opposite sign, this implies $x_{h(e)} = x_{t(e)}$ for all $e \in E$;  \ie the vertex values at the two ends of edge $e$ must be equal.  Since $\G$ is connected,  these edge constraints propagate throughout $\G$, implying that all vertex values $x_v$ must be equal.  \qed \vspace{1ex}

A graphical illustration of this proof will be given in Section \ref{Sec2.4}; see Figure \ref{NRZ0}.

Theorem 1(a) has the following corollary:

\vspace{1ex}
\noindent
\textbf{Theorem 1(b)} ($B^1$).    If $\G = (V,E)$ is a connected graph, then its coboundary space $B^1 = \im d$ is isomorphic to $\A^{|V|-1}$.

\vspace{1ex}
\noindent
\textit{Proof}:  By the fundamental theorem of homomorphisms, we have
$B^1 = \im d \cong {C^0}/{\ker d} = {C^0}/{Z^0}.$

For any $v \in V$, let $(C^0)_{:\bar{v}}$ denote the subset $\A^{V\setminus\{v\}} \times \{0\}^{\{v\}}$ of $C^0 = \A^V$ such that $x_v = 0$.  Evidently $(C^0)_{:\bar{v}}$  is isomorphic to $\A^{|V|-1}$.  Moreover, $(C^0)_{:\bar{v}}$ is a set of coset representatives for ${C^0}/{Z^0}$, and the corresponding one-to-one map  $(C^0)_{:\bar{v}} \leftrightarrow {C^0}/{Z^0}$ is an isomorphism.   Therefore $B^1 \cong {C^0}/{Z^0} \cong (C^0)_{:\bar{v}} \cong \A^{|V|-1}$. \qed \vspace{1ex}

A graphical illustration of this proof will be given in Section \ref{Sec2.6}; see Figure \ref{NRB1IO}.

\vspace{-1ex}
\subsection{Systematic $(n,k)$ group codes}\label{SCIS}

In coding theory, if $\A$ is a field, then $\A^n$ is a vector space, and any subspace $\CC \subseteq \A^n$ of dimension $k$ is called a \emph{linear $(n,k)$ block code} over $\A$.  Thus in the field case $Z^0 = \ker d$ is a linear $(|V|, 1)$ block code over $\A$, and $B^1 = \im d$ is a linear $(|E|, |V| - 1)$ block code over $\A$.  

Moreover, every  linear $(n,k)$  block code $\CC$ has a \emph{systematic} $k \times n$ generator matrix that contains a $k \times k$ identity matrix as a submatrix, whose column positions are said to be an \emph{information set} of $\CC$;  the remaining column positions are called a \emph{check set}.

  For example, if $\A$ is a field and $\I$ is an index set of size $|\I| = n$, then the repetition code $\CC_= \subseteq \A^\I$ is a linear $(n,1)$ block code over $\A$, the all-1 codeword is a systematic $1 \times n$ generator matrix, and for any $i \in \I$ the singleton subset $\{i\} \subseteq \I$ is an information set.

We now extend this nomenclature by defining \emph{systematic $(n,k)$ group codes}.  We will ultimately show that all of the important spaces of algebraic topology are systematic $(n,k)$ group codes.

\vspace{1ex}
\noindent
\textbf{Definition} (\emph{systematic $(n,k)$ group codes}). If $\A$ is an abelian group and $\I$ is an index set of size $|\I| = n$, then a subgroup  $\CC \subseteq \A^\I$ that is isomorphic to $\A^k$  will be called an \textbf{$(n,k)$ group code} over  $\A$.  Moreover, such a code will be called a \textbf{systematic $(n,k)$ group code} over  $\A$ if the projection of $\CC$ onto $\A^\J$ for some subset $\J \subseteq \I$ of size $|\J| = k$ is an isomorphism to $\A^\J \cong \A^k$.  Any such subset $\J$ will be called an \textbf{information set} of $\CC$, and its complement $\bar{\J}$ a \textbf{check set} of $\CC$. \qed \vspace{1ex}

  For example, if $\A$ is a group, then the \textbf{repetition code} $\CC_= \subseteq \A^\I$ is a systematic $(n,1)$ group code over $\A$, and for any $i \in \I$ the singleton subset $\{i\}$ is an information set, since the projection of $\CC_=$ onto any coordinate is an isomorphism.  

In the spirit of this definition, for any $(n,k)$ group code $\CC$ over an abelian group $\A$, we may call $n$ its ``length" and $k$ its ``dimension," even though, strictly speaking, the term ``dimension" is inappropriate when $\A$ is not a field.  We may also call $\A^\I$ a ``space" and $\CC \subseteq \A^\I$ a ``subspace."

For another example, a \textbf{zero-sum code} $\CC_+ \subseteq \A^\I$ of length $n = |\I|$ over $\A$ is defined as
$$\CC_+ = \left\{\ab \in \A^\I ~\middle|~ \sum_{i \in \I} a_i = 0\right\}.$$
$\CC_+$ is evidently a strict subgroup of $\A^\I$.  Moreover, since we may choose any $n - 1$ elements $a_i$ of a codeword $\ab \in \CC_+$ freely from $\A$, and then choose the remaining element so that $\sum_{\I} a_i = 0$, it is evident that the projection of $\CC_+$ onto any subset $\J \subseteq \I$ of size $|\J| = n - 1$ is $\A^\J \cong \A^{n-1}$.  Thus a zero-sum code of length $n$ over $\A$ is  a systematic $(n,n-1)$ group code over $\A$, and every subset of $\I$ of size $n-1$ is an information set.

We now show that every systematic $(n,k)$ group code has an \emph{I/O map}, which will allow us to represent all of our realizations as I/O behaviors.

\vspace{1ex}
\noindent
\textbf{I/O Map Lemma}.
If $\CC$ is a systematic $(n,k)$ group code over an abelian group $\A$ with information set $\J \subseteq \I$ and check set $\bar{\J} = \I \setminus \J$, then there exists a homomorphism $\varphi_\CC:  \A^\J \to \A^{\bar{\J}}$ such that 
$$\CC = \left\{(\ab_\J, \varphi_\CC(\ab_\J)) ~\middle|~ \ab_\J \in \A^\J\right\}.$$

\vspace{1ex}
\noindent
\textit{Proof}:  
Write the codewords $\cb \in \A^\I$ as $\cb = (\cb_{\J}, \cb_{\bar{\J}})$, where $\cb_{\J}$ and $\cb_{\bar{\J}}$ are the projections of $\cb$ onto $\A^\J$ and $\A^{\bar{\J}}$, respectively.  Then the required homomorphism   $\varphi_\CC:  \A^\J \to \A^{\bar{\J}}$ exists, since there is an isomorphism  $\CC \leftrightarrow \CC_{|\J} \cong \A^\J$, and the projection of $\CC$ onto $\A^{\bar{\J}}$ is a homomorphism.  \qed \vspace{1ex}

In coding theory, an encoder that maps an \emph{information sequence} $\ab_\J \in \A^\J$ to a \emph{check sequence} $\varphi_\CC(\ab_\J) \in \A^{\bar{\J}}$ and transmits both as the codeword $(\ab_\J, \varphi_\CC(\ab_\J))$ is called a \emph{systematic encoder}.  

In  system theory, a behavior of the form $\{(\ab_\J, \varphi_\CC(\ab_\J)) \mid \ab_\J \in \A^\J\}$ is called an \emph{I/O behavior}, where $\ab_\J \in \A^\J$ is regarded as the \emph{input}, and $\varphi_\CC(\ab_\J) \in \A^{\bar{\J}}$ as the \emph{output}.  In this context, we will call the homomorphism $\varphi_\CC:  \A^\J \to \A^{\bar{\J}}$ an \emph{I/O map}.

To  construct a ``generator matrix" for a systematic $(n,k)$ group code $\CC$ with information set $\J$, we observe that $\CC$ has a set of $k$ one-dimensional subcodes that may be regarded as a systematic ``basis" of $\CC$, as follows.  For each of the $k$ coordinates $i \in \J$, consider the subcode $\CC_i \subseteq \CC$ consisting of the codewords in $\CC$ that are all-zero in the remaining $k-1$ coordinates $\J \setminus \{i\}$.  By the isomorphism between $\A^\J$ and $\CC$, $\CC_i$ must be isomorphic to $\A$ via projection onto the $i$th coordinate.  Moreover, $\CC$ is evidently the direct sum of the subcodes $\CC_i$: 
$$\CC =\bigoplus_{i \in \J} \CC_i \cong \A^\J.$$

For  example, for the {zero-sum code} $\CC_+ \subseteq \A^\I$,  any subset $\{i\} \subseteq \I$ of size 1 is a check set, and the set of $|\I| - 1$ ``one-dimensional" subcodes $(\CC_+)_j, j \neq i,$ that consist of all codewords $\cb \in \CC_+$ such that $\cb_j = a, \cb_i = -a$, and all remaining coordinates are zero forms a ``basis" for $\CC_+$.

Finally, the \emph{support} $\SSS \subseteq \I$ of an $(n,k)$ group code $\CC \subseteq \A^\I$ is the set of indices $i \in \SSS$ such that the projection $\CC_{|i}$ of $\CC$ onto the \emph{i}th coordinate is nontrivial.  If $\CC$ has support $\SSS$ (or less), then we may say that $\CC$ is \emph{effectively} a $(|\SSS|, k)$ group code, and that its \emph{effective length} is $|\SSS|$.  For example, the subcodes $\CC_i$ defined in the previous paragraph are effectively $(n-k+1,1)$ group codes with supports $\{i\} \cup \bar{\J}$ (or less).

\subsection{Elementary normal realizations}\label{Sec2.4}

We will now construct normal realizations for the zero-coboundary space $Z^0 = \ker d$ and the coboundary space $B^1 = \im d$ that will help to visualize and prove their properties.

In general, a \emph{normal realization} \cite{F01} is a graphical model based on a graph $G = (V, E, H)$, in which the vertices (or ``nodes") $v \in V$ represent constraint codes $\CC_v$,  the edges $e \in E$ represent internal variables, and the \emph{half-edges} $h \in H$ represent external variables.  The constraint code $\CC_v$ is the set of all permissible (``valid") values of the variables corresponding to the edges and half-edges that are incident on vertex $v$. The set of all valid variable configurations--- \ie the configurations that are compatible with all constraints--- is called the \emph{behavior} $\Bf$ of the realization, and the projection of the behavior onto the half-edge variables is called its \emph{external behavior} $\CC$; alternatively, $\CC$ is called the \emph{code} that is realized.  If all constraint codes are linear, then $\Bf$ and $\CC$ are vector spaces;  if all constraint codes  are abelian groups, then $\Bf$ and $\CC$ are abelian groups. 

For this paper, we  need only \emph{elementary normal realizations}, namely normal realizations that satisfy the following restrictions:
\begin{itemize}
\item All internal and external variables have a common alphabet $\A$, which is either a field $\F$, or more generally an abelian group $\A$ with a well-defined dual (character) group $\hat{\A}$;
\item  All constraint codes are either \emph{repetition codes} $\CC_=$, which constrain all incident variables to be equal, or \emph{zero-sum codes} $\CC_+$, which constrain the sum of all incident variables to be zero.  Thus if the degree of  vertex $v$ is $n$, then $\CC_v$ is either the $(n, 1)$ {repetition code} $\CC_=$ over $\A$, or the $(n, n-1)$ {zero-sum code} $\CC_+$ over $\A$.
\end{itemize}

We will indicate repetition constraints by the symbol 
\begin{picture}(2,1)
\put(0.25,-0.25){\framebox(1.5,1.5){$=$}}
\end{picture},
and zero-sum constraints by the symbol
\begin{picture}(2,1)
\put(0.25,-0.25){\framebox(1.5,1.5){$+$}}
\end{picture}.
We may also employ the following simple manipulations and special symbols:
\begin{itemize}
\item A repetition constraint of degree 2 may be simply replaced by an edge, since 
\begin{picture}(11,1)
\put(0, 0.5){\line(1,0){1}}
\put(1,-0.25){\framebox(1.5,1.5){$=$}}
\put(2.5, 0.5){\line(1,0){1}}
\put(5, 0.2){$=$}
\put(7.5, 0.5){\line(1,0){3}}
\end{picture}.
\item A zero-sum constraint of degree 2 may be replaced by an edge with a small circle representing a \emph{sign inverter}: 
\begin{picture}(11,1)
\put(0, 0.5){\line(1,0){1}}
\put(1,-0.25){\framebox(1.5,1.5){$+$}}
\put(2.5, 0.5){\line(1,0){1}}
\put(5, 0.2){$\Rightarrow$}
\put(7.5, 0.5){\line(1,0){1.2}}
\put(8.7, 0.15){$\circ$}
\put(9.25, 0.5){\line(1,0){1.25}}
\end{picture}.
We will sometimes call this an \emph{inverting edge}.
\item A repetition constraint of degree 2 plus a sign inverter may be replaced by an inverting edge, since 
\begin{picture}(11,1)
\put(0, 0.5){\line(1,0){1}}
\put(1,-0.25){\framebox(1.5,1.5){$=$}}
\put(2.5, 0.15){$\circ$}
\put(3.3, 0.5){\line(1,0){1}}
\put(5, 0.2){$=$}
\put(7.5, 0.5){\line(1,0){1.2}}
\put(8.7, 0.15){$\circ$}
\put(9.25, 0.5){\line(1,0){1.25}}
\end{picture}.
\item A zero-sum constraint of degree 2 plus a sign inverter may be replaced simply by an  edge, since
\begin{picture}(11,1)
\put(0, 0.5){\line(1,0){1}}
\put(1,-0.25){\framebox(1.5,1.5){$+$}}
\put(2.5, 0.15){$\circ$}
\put(3.3, 0.5){\line(1,0){1}}
\put(5, 0.2){$=$}
\put(7.5, 0.5){\line(1,0){3}}
\end{picture}.
\item We may optionally put an arrow on an edge if we wish to indicate the direction of a cause-and-effect relationship, as we will  illustrate shortly below.
\end{itemize}

Using normal realizations, we now wish to study the image and kernel of the coboundary operator $d: \A^V \to \A^E, \xb \mapsto M\xb$ of a graph $\G = (V, E)$ with incidence matrix $M$.  The \emph{input/output (I/O) behavior} of $d$ will be defined as  $W^{01} = \{(\xb, \yb) \in \A^V \times \A^E \mid \yb = M\xb)\}$, which is evidently a systematic $(|V| + |E|, |V|)$ group code over $\A$ with information set $V$. 

\begin{figure}[h]
\setlength{\unitlength}{5pt}
\centering
%%%%%%%%%%%%%%
% Figure 2
%%%%%%%%%%%%%%
\begin{tikzpicture}
\node[box]                  (EQ1) {$=$};
\node[box,right=\davesep of EQ1] (PL1) {$+$};
\node[box,right=\davesep of PL1] (EQ2) {$=$};
\node[box,right=\davesep of EQ2] (PL2) {$+$};
\node[box,right=\davesep of PL2] (EQ3) {$=$};
\node[box,below=\davesep of EQ1] (PL3) {$+$};
\node[box,below=\davesep of EQ2] (PL4) {$+$};
\node[box,below=\davesep of PL2] (PL5) {$+$};
\node[box,below=\davesep of PL3] (EQ5) {$=$};
\node[box,right=\davesep of EQ5] (PL6) {$+$};
\node[box,right=\davesep of PL6] (EQ4) {$=$};
\node[invert,anchor=west]  (IPL1) at (PL1.east)  {};
\node[invert,anchor=west]  (IPL2) at (PL2.east)  {};
\node[invert,anchor=north] (IPL3) at (PL3.south) {};
\node[invert,anchor=north] (IPL4) at (PL4.south) {};
\node[invert,anchor=north east] (IPL5) at (PL5.south west) {};
\node[invert,anchor=east] (IPL6) at (PL6.west) {};
\node[var,left=\davesep of EQ1] (X1) {$x_1$};
\node[var,above=\davesep of EQ2] (X2) {$x_2$};
\node[var,right=\davesep of EQ3] (X3) {$x_3$};
\node[var,right=\davesep of EQ4] (X4) {$x_4$};
\node[var,left=\davesep of EQ5] (X5) {$x_5$};
\node[var,above=\davesep of PL1] (Y1) {$y_1$};
\node[var,above=\davesep of PL2] (Y2) {$y_2$};
\node[var,left=\davesep of PL3] (Y3) {$y_3$};
\node[var,left=\davesep of PL4] (Y4) {$y_4$};
\node[var,right=\davesep of PL5] (Y5) {$y_5$};
\node[var,below=\davesep of PL6] (Y6) {$y_6$};
\draw[->] (X1.east) \updown -- (EQ1);
\draw[->] (X2.south) \leftright -- (EQ2);
\draw[->] (X3.west) \updown -- (EQ3);
\draw[->] (X4.west) \updown -- (EQ4);
\draw[->] (X5.east) \updown -- (EQ5);
\draw[<-] (Y1.south) \leftright -- (PL1);
\draw[<-] (Y2.south) \leftright -- (PL2);
\draw[<-] (Y3.east) \updown -- (PL3);
\draw[<-] (Y4.east) \updown -- (PL4);
\draw[<-] (Y5.west) \updown -- (PL5);
\draw[<-] (Y6.north) \leftright -- (PL6);
\draw[->] (EQ1) -- (PL1);
\draw[->] (EQ1) -- (PL3);
\draw[->] (EQ2) -- (IPL1);
\draw[->] (EQ2) -- (PL2);
\draw[->] (EQ2) -- (PL4);
\draw[->] (EQ3) -- (IPL2);
\draw[->] (EQ3) -- (PL5);
\draw[->] (EQ4) -- (PL6);
\draw[->] (EQ4) -- (IPL4);
\draw[->] (EQ4) -- (IPL5);
\draw[->] (EQ5) -- (IPL3);
\draw[->] (EQ5) -- (IPL6);
\end{tikzpicture}
\caption{Normal realization of  I/O behavior $W^{01} = \{(\xb, M\xb) \mid \xb \in \A^V\}$ for graph $\G$ of Figure \ref{E1}.}
\label{NRIOD}
\end{figure}
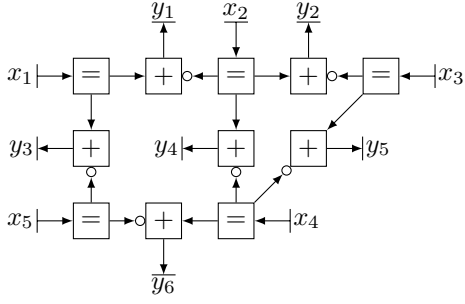

Figure \ref{NRIOD} shows an elementary normal realization of the I/O behavior  $W^{01}$ for our example graph $\G$. (We observe that the graph $G = (V_G, E_G, H_G)$ of this normal realization actually has $|V_G| = |V| + |E|$ vertices, $|E_G| = 2|E|$ edges, and $|H_G| = |V| + |E|$ half-edges.)

In Figure \ref{NRIOD}, a set $\xb = \{x_v, v \in V\}$ of $|V|$ external input variables  is associated with the vertices $v \in V$ of $\G$.  Each vertex variable $x_v$ is replicated $d_v$ times via a repetition constraint, and passed on (with a sign inversion if $v = h(e)$) to a zero-sum constraint associated with one of the $d_v$ adjacent edges $e \in E(v)$, thus making the input to the zero-sum constraint  $ -M_{ev}x_v$.  For each edge $e \in E$, a zero-sum constraint on all the incident signed vertex variables and the output variable $y_e$ at edge $e$ enforces the constraint $y_e + (- M_e\xb) = 0$;  thus $\yb = M\xb$, as desired.  Thus every constraint in the realization is realized as a little I/O behavior.  Arrows on all  edges indicate the directions of these cause-and-effect relationships.

Now, to obtain a realization of the zero-coboundary space $Z^0 = \ker d = \{\xb \mid M\xb = \zerob\}$, we  constrain the external edge variables $y_e$ to equal zero, which simply removes them from the realization.\footnote{In other words, $Z^0$ is the cross-section $(W^{01})_{:\A^V} = \{\xb \in \A^V \mid (\xb, \zerob) \in W^{01}\}$ of $W^{01}$ on $\A^V$.}  Also, since \begin{picture}(10,1)
\put(0, 0.5){\line(1,0){1}}
\put(1,-0.25){\framebox(1.5,1.5){$+$}}
\put(2.5, 0.15){$\circ$}
\put(3.25, 0.5){\line(1,0){1}}
\put(5.5, 0.2){$=$}
\put(7.5, 0.5){\line(1,0){2}}
\end{picture},
we may simply use the latter realization for each edge.  Thus we obtain the extremely simple realization of $Z^0$ that is shown in Figure \ref{NRZ0}.
We have removed the arrows because  the effects of the constraints now flow in all directions.

\begin{figure}[h]
\setlength{\unitlength}{5pt}
\centering
%%%%%%%%%%%%%%
% Figure 3
%%%%%%%%%%%%%%
\begin{tikzpicture}
\node[box]                   (EQ1) {$=$};
\node[box,right=\franksep of EQ1 ] (EQ2) {$=$};
\node[box,right=\franksep of EQ2 ] (EQ3) {$=$};
\node[box,below=\franksep of EQ2 ] (EQ4) {$=$};
\node[box,below=\franksep of EQ1 ] (EQ5) {$=$};
\node[var,left=\davesep of EQ1] (X1) {$x_1$};
\node[var,above=\davesep of EQ2] (X2) {$x_2$};
\node[var,right=\davesep of EQ3] (X3) {$x_3$};
\node[var,right=\davesep of EQ4] (X4) {$x_4$};
\node[var,left=\davesep of EQ5] (X5) {$x_5$};
\draw (X1.east) \updown -- (EQ1);
\draw (X2.south) \leftright -- (EQ2);
\draw (X3.west) \updown -- (EQ3);
\draw (X4.west) \updown -- (EQ4);
\draw (X5.east) \updown -- (EQ5);
\draw (EQ1) -- (EQ2);
\draw (EQ1) -- (EQ5);
\draw (EQ2) -- (EQ3);
\draw (EQ2) -- (EQ4);
\draw (EQ3) -- (EQ4);
\draw (EQ4) -- (EQ5);
\end{tikzpicture}
\caption{Normal realization of $Z^0 = \ker d$  for graph $\G$ of Figure \ref{E1}.}
\label{NRZ0}
\end{figure}

It is obvious from this realization that $Z^0$ is the $(|V|, 1)$ repetition code  $\CC_= \subseteq \A^V$;  \ie Figure \ref{NRZ0} gives a pictorial proof of Theorem 1(a).

Moreover, if $\G$ were a disconnected graph with $\beta_0(\G) > 1$ connected components, then the corresponding realization of $Z_0 = \ker d$ as in Figure \ref{NRZ0} would evidently consist of $\beta_0(\G)$ disconnected (and therefore independent) repetition codes.  This is the main reason why we have started in this paper with  the coboundary operator $d$ rather than the usual boundary operator $\partial$, since we regard connectedness as the most elementary  concept of topology.

\begin{figure}[h]
\setlength{\unitlength}{5pt}
\centering
%%%%%%%%%%%%%%
% Figure 4
%%%%%%%%%%%%%%
\begin{tikzpicture}
\node[box]                  (EQ1) {$=$};
\node[box,right=\davesep of EQ1] (PL1) {$+$};
\node[box,right=\davesep of PL1] (EQ2) {$=$};
\node[box,right=\davesep of EQ2] (PL2) {$+$};
\node[box,right=\davesep of PL2] (EQ3) {$=$};
\node[box,below=\davesep of EQ1] (PL3) {$+$};
\node[box,below=\davesep of EQ2] (PL4) {$+$};
\node[box,below=\davesep of PL2] (PL5) {$+$};
\node[box,below=\davesep of PL3] (EQ5) {$=$};
\node[box,right=\davesep of EQ5] (PL6) {$+$};
\node[box,right=\davesep of PL6] (EQ4) {$=$};
\node[invert,anchor=west]  (IPL1) at (PL1.east)  {};
\node[invert,anchor=west]  (IPL2) at (PL2.east)  {};
\node[invert,anchor=north] (IPL3) at (PL3.south) {};
\node[invert,anchor=north] (IPL4) at (PL4.south) {};
\node[invert,anchor=north east] (IPL5) at (PL5.south west) {};
\node[invert,anchor=east] (IPL6) at (PL6.west) {};
\node[var,above=\davesep of PL1] (Y1) {$y_1$};
\node[var,above=\davesep of PL2] (Y2) {$y_2$};
\node[var,left=\davesep of PL3] (Y3) {$y_3$};
\node[var,left=\davesep of PL4] (Y4) {$y_4$};
\node[var,right=\davesep of PL5] (Y5) {$y_5$};
\node[var,below=\davesep of PL6] (Y6) {$y_6$};
\draw (Y1.south) \leftright -- (PL1);
\draw (Y2.south) \leftright -- (PL2);
\draw (Y3.east) \updown -- (PL3);
\draw (Y4.east) \updown -- (PL4);
\draw (Y5.west) \updown -- (PL5);
\draw (Y6.north) \leftright -- (PL6);
\draw (EQ1) -- (PL1);
\draw (EQ1) -- (PL3);
\draw (EQ2) -- (IPL1);
\draw (EQ2) -- (PL2);
\draw (EQ2) -- (PL4);
\draw (EQ3) -- (IPL2);
\draw (EQ3) -- (PL5);
\draw (EQ4) -- (PL6);
\draw (EQ4) -- (IPL4);
\draw (EQ4) -- (IPL5);
\draw (EQ5) -- (IPL3);
\draw (EQ5) -- (IPL6);
\end{tikzpicture}
\caption{Normal realization of $B^1 = \im d$  for graph $\G$ of Figure \ref{E1}.}
\label{NRB1}
\end{figure}
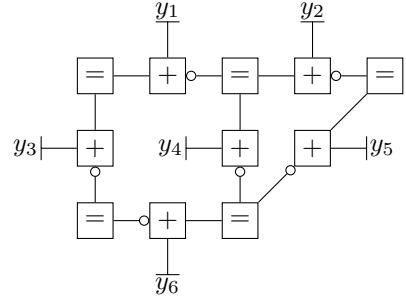

Similarly, to get a realization of the coboundary space $B^1 =  \im d$, we simply remove the external vertex variables ${x}_v$ from Figure \ref{NRIOD}, while leaving the $d_v$ internal replica variables representing the values $x_v$.\footnote{In other words, $B^1$ is the projection $(W^{01})_{|\A^E} = \{\yb \in \A^E \mid \exists \xb: (\xb, \yb) \in W^{01}\}$ of $W^{01}$ onto $\A^E$.}  Thus we obtain the realization of $B^1$ shown in Figure \ref{NRB1}.  

\subsection{Partitions, cut sets, and bases for $B^1$}\label{PCS}

We have seen in  Theorem 1(b) that the coboundary space $B^1 =  \im d$ of a graph $\G = (V,E)$ is a systematic $(|E|, |V| - 1)$ group code over $\A$.  We will now show that for any spanning tree $\T$ of $\G$, $B^1$ has an information set corresponding to the $(|V| - 1)$-element edge set $E_\T$ of $\T$, and a corresponding ``basis" consisting of certain  \emph{cut-set vectors}, to be defined below. 

Let $P = V_1 \sqcup V_2$ be a nontrivial disjoint partition  of the vertex set $V$ into two  subsets,  $V_1$ and $V_2 = \bar{V}_1$.  $P$ induces a partition of the edge set $E$ into three subsets:  a subset $E_1$ of edges whose ends are both in $V_1$, a subset $E_2$ of edges whose ends are both in $V_2$, and the remaining subset $E_P$ of edges that have one end in $V_1$ and one in $V_2$, which is called the \emph{cut set} of the partition $P$.  We assume that the two subgraphs $\G_1 = (V_1, E_1)$ and $\G_2 = (V_2, E_2)$ are both connected.

We  define a \emph{cut-set vector} \cite{Bollobas} of the partition $P = V_1 \sqcup V_2$ as the image $\yb = M\xb$ of any vertex vector $\xb = (\xb_1, \xb_2)$ such that $\xb_1$ is any element of the repetition code $\CC_= \subseteq \A^{V_1}$ defined on the vertex subset $V_1$, and $\xb_2$ is the all-zero vector on the complementary subset $V_2$.  Evidently $\yb \in B^1$.

\vspace{1ex}
\noindent
\textbf{Cut-Set Lemma}.  If $\yb \in \A^E$ is a cut-set vector of a partition $P = V_1 \sqcup V_2$, then $y_e = 0$ if $e \in E_1$ or $e \in E_2$.  Moreover, if $e \in E_P$, then  $y_e = \pm a$ for some $a \in \A$, where the sign depends on whether the edge $e$ goes from $V_1$ to $V_2$, or \emph{vice versa}. 

\vspace{1ex}
\noindent
\emph{Proof}.  The first statement follows from the fact that $\xb_1$ and $\xb_2$ are elements of the repetition codes over $V_1$ and $V_2$, respectively, and therefore are in the kernels of the coboundary operators $d_1:  \A^{V_1} \to \A^{E_1}$ and $d_2:  \A^{V_2} \to \A^{E_2}$ of the subgraphs $\G_1 = (V_1, E_1)$ and $\G_2 = (V_2, E_2)$, respectively.  The second statement follows from the fact that if $x_v = a$ for all $v \in V_1$ and $x_v = 0$ for all $v \in V_2$, and $e \in E_P$ connects a vertex in $V_1$ to a vertex in $V_2$, then $y_e = \pm a$, with the sign depending on whether the head or the tail of $e$ is in $V_1$.  \qed \vspace{1ex}

\begin{figure}[h]
\setlength{\unitlength}{5pt}
\centering
%%%%%%%%%%%%%%
% Figure 5
%%%%%%%%%%%%%%
\begin{tikzpicture}
\node[box]                  (EQ1) {$=$};
\node[box,right=\davesep of EQ1] (PL1) {$+$};
\node[box,right=\davesep of PL1] (EQ2) {$=$};
\node[box,right=\davesep of EQ2] (PL2) {$+$};
\node[box,right=\davesep of PL2] (EQ3) {$=$};
\node[box,below=\davesep of EQ1] (PL3) {$+$};
\node[box,below=\davesep of EQ2] (PL4) {$+$};
\node[box,below=\davesep of PL2] (PL5) {$+$};
\node[box,below=\davesep of PL3] (EQ5) {$=$};
\node[box,right=\davesep of EQ5] (PL6) {$+$};
\node[box,right=\davesep of PL6] (EQ4) {$=$};
\node[invert,anchor=west]  (IPL1) at (PL1.east)  {};
\node[invert,anchor=west]  (IPL2) at (PL2.east)  {};
\node[invert,anchor=north] (IPL3) at (PL3.south) {};
\node[invert,anchor=north] (IPL4) at (PL4.south) {};
\node[invert,anchor=north east] (IPL5) at (PL5.south west) {};
\node[invert,anchor=east] (IPL6) at (PL6.west) {};
\node[var,left=\davesep of EQ1] (X1) {$a$};
\node[var,above=\davesep of EQ2] (X2) {$a$};
\node[var,right=\davesep of EQ3] (X3) {$a$};
\node[var,right=\davesep of EQ4] (X4) {$0$};
\node[var,left=\davesep of EQ5] (X5) {$0$};
\node[var,above=\davesep of PL1] (Y1) {$0$};
\node[var,above=\davesep of PL2] (Y2) {$0$};
\node[var,left=\davesep of PL3] (Y3) {$a$};
\node[var,left=\davesep of PL4] (Y4) {$a$};
\node[var,right=\davesep of PL5] (Y5) {$a$};
\node[var,below=\davesep of PL6] (Y6) {$0$};
\draw[->] (X1.east) \updown -- (EQ1);
\draw[->] (X2.south) \leftright -- (EQ2);
\draw[->] (X3.west) \updown -- (EQ3);
\draw[->] (X4.west) \updown -- (EQ4);
\draw[->] (X5.east) \updown -- (EQ5);
\draw[<-] (Y1.south) \leftright -- (PL1);
\draw[<-] (Y2.south) \leftright -- (PL2);
\draw[<-] (Y3.east) \updown -- (PL3);
\draw[<-] (Y4.east) \updown -- (PL4);
\draw[<-] (Y5.west) \updown -- (PL5);
\draw[<-] (Y6.north) \leftright -- (PL6);
\draw[->] (EQ1) -- (PL1);
\draw[->] (EQ1) -- (PL3);
\draw[->] (EQ2) -- (IPL1);
\draw[->] (EQ2) -- (PL2);
\draw[->] (EQ2) -- (PL4);
\draw[->] (EQ3) -- (IPL2);
\draw[->] (EQ3) -- (PL5);
\draw[->] (EQ4) -- (PL6);
\draw[->] (EQ4) -- (IPL4);
\draw[->] (EQ4) -- (IPL5);
\draw[->] (EQ5) -- (IPL3);
\draw[->] (EQ5) -- (IPL6);
\coordinate (A) at (barycentric cs:EQ1=0.5,PL3=0.5);
\draw[dashed] (X1.west |- A) -- (X3 |- A) -- node[above,near end] (G1) {$\mathcal{G}_1$} ++(4ex,0);
\coordinate (B) at (barycentric cs:EQ5=0.6,PL3=0.4);
\draw[dashed] (X1.west |- B) -- (X3 |- B) -- node[below,near end] (G2) {$\mathcal{G}_2$} ++(4ex,0);
\node at (barycentric cs:G1=0.5,G2=0.5) {$E_P$};
\end{tikzpicture}
\caption{Typical  cut-set vector $\yb = M\xb$ for partition $P = \{v_1, v_2, v_3\} \sqcup \{v_4, v_5\}$.}
\label{NRIOC}
\end{figure}
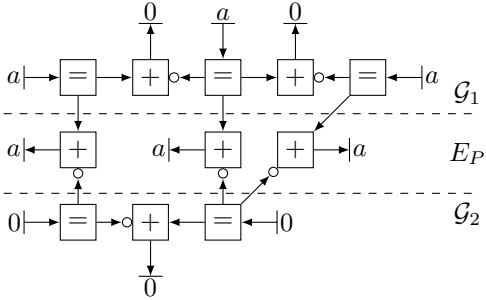

To illustrate, in our example graph $\G$, the edges $E_P = \{e_3, e_4, e_5\}$ form a cut set corresponding to the partition $P = \{v_1, v_2, v_3\} \sqcup \{v_4, v_5\}$ of $V$, as shown in Figure \ref{NRIOC}.  Given an input vector $\xb$ that has the constant value $a$ on $V_1$ and $0$ on $V_2$, the coboundary operator $d$ produces an output vector $d(\xb) = M\xb$ that is all-zero on $E_1$ and $E_2$, and equal to either $a$ or $-a$ (depending on the edge orientation) on the cut set $E_P$.  

For any nontrivial partition $P$ of $V$, the set of all cut-set vectors $\yb \in B^1$ as $\xb_1$ ranges through all elements of the repetition code $\CC_= \subseteq \A^{V_1}$ will be called  the \emph{cut-set subspace}   $(B^1)_P \subseteq B^1$.  Evidently $(B^1)_P$ is  a one-dimensional linear or group code with support $E_P \subseteq E$.
 
We  now  find a set of $|V| - 1$ \emph{fundamental cut sets} $E_P$ such that the corresponding one-dimensional cut-set subspaces $(B^1)_{P}$ generate the $(|V| - 1)$-dimensional space $B^1$;  \ie we  find a ``basis" for $B^1$ (where the quotes may be removed in the linear case).  
We start with any spanning tree $\T = (V, E_\T)$ of $\G$, where $|E_\T| = |V| - 1$. 
Since $\T$ is a tree, every edge $e \in E_\T$ is a cut set of $\T$ that partitions the vertex set $V$ of $\T$ into two connected subsets, say $V_1(e)$ and $V_2(e)$. The cut set $E_{P(e)}$ of $\G$ that corresponds to the same partition $P(e)$  of the vertex set $V$ of $\G$ will be defined as our $e$th fundamental cut set of $\G$.  
The cut set $E_{P(e)}$ must include $e$, but cannot include any other edges in $E_\T$, since their ends are  either both in $V_1(e)$ or both in $V_2(e)$.  Thus $E_{P(e)}  \subseteq  \{e\} \cup E_{\bar{\T}}$.  

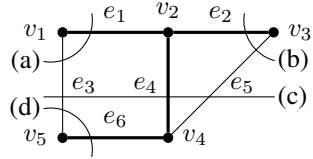
\begin{figure}[h]
\setlength{\unitlength}{5pt}
\centering
%%%%%%%%%%%%%%
% Figure 6
%%%%%%%%%%%%%%
\begin{tikzpicture}
\node[dot,label=left:$v_1$]                  (V1) {};
\node[dot,right=8ex of V1,label=above:$v_2$] (V2) {};
\node[dot,right=8ex of V2,label=right:$v_3$] (V3) {};
\node[dot,below=8ex of V2,label=right:$v_4$] (V4) {};
\node[dot,below=8ex of V1,label=left:$v_5$]  (V5) {};
\draw[very thick] (V1) -- node[above] {$e_1$} (V2);
\draw[very thick] (V2) -- node[above] {$e_2$} (V3);
\draw[very thick] (V2) -- node[left] {$e_4$} (V4);
\draw[very thick] (V4) -- node[above] {$e_6$} (V5);
\draw[] (V1) -- node[right] {$e_3$} (V5);
\draw[] (V3) -- node[right] {$e_5$} (V4);
\draw (V1) ++(-1ex,1ex) ++(10:3.5ex) arc (10:-100:3.5ex) node[anchor=east,inner sep=1pt] (A) {(a)};
\draw (V5) ++(-1ex,-1ex) ++(-10:3.5ex) arc (-10:100:3.5ex) node[anchor=east,inner sep=1pt] (D) {(d)};
\draw (V3) ++(1ex,1ex) ++(170:3.5ex) arc (170:255:3.5ex) node[anchor=west,inner sep=1pt] (B) {(b)};
\coordinate (AE) at (A.east);
\coordinate (DE) at (D.east);
\coordinate (X) at (barycentric cs:AE=0.25,DE=0.75);
\draw (X) -- (X-|B.west) node[anchor=west,inner sep=1pt] {(c)};
\end{tikzpicture}
\vspace{-1ex}
\caption{Spanning tree $\T$ of $\G$, and cut sets of $\G$ including (a) $e_1$; (b) $e_2$; (c) $e_4$; (d) $e_6$.}
\label{B1X}
\end{figure}

For example, for our example graph $\G$, deleting the $\beta_1(\G) = 2$ edges $e_3$ and $e_5$ yields a spanning tree $\T$ with $E_{\bar{\T}} = \{e_3, e_5\}$. The respective cut sets in $\{e_1\} \cup E_{\bar{\T}}$, $\{e_2\} \cup E_{\bar{\T}}$, $\{e_4\} \cup E_{\bar{\T}}$, and $\{e_6\} \cup E_{\bar{\T}}$ are $(e_1, e_3), (e_2, e_5), (e_4, e_3, e_5),$ and $(e_6, e_3)$, as shown in Figure \ref{B1X}.  

\noindent
\textbf{Theorem 2} (\textit{``basis" for} $B^1$).  For any graph $\G$ and any spanning tree $\T \subseteq \G$ with cut set $E_\T$, the $|V| - 1$  cut-set subspaces $\{(B^1)_{P(e)}, e \in E_\T\}$ are a set of one-dimensional subspaces of $B^1$ that form a systematic ``basis" for $B^1$.  Thus $E_\T$ is an information set for $B^1$.

\vspace{1ex}
\noindent
\textit{Proof}.
The $|V| - 1$ cut-set subspaces $\{(B^1)_{P(e)}, e \in E_\T\}$ are all independent, since their supports are completely disjoint on $E_\T$.  Since $B^1$ is $(|V|-1)$-dimensional, it follows that every element of $B^1$ is a unique  combination of elements of $\{(B^1)_{P(e)}, e \in E_\T\}$.  \qed \vspace{1ex}

Thus $B^1$ is  sometimes  called the \emph{cut space} of $\G$, and $\dim B^1 = |V| - 1$ the \emph{rank} of $\G$ \cite{Bollobas}.

 \subsection{Nonredundant I/O realizations}\label{Sec2.6}
 
We have seen that $Z^0$ and $B^1$ are systematic $(n,k)$ group codes, and we have identified their information sets.  We will now construct I/O realizations with these information sets.  Moreover, in order to obtain I/O realizations in which every edge may be labeled with a cause-and-effect arrow, we first analyze the unobservability (generator redundancy) and uncontrollability (constraint redundancy) properties of our previous realizations, and eliminate the corresponding redundancies.

As we have  seen,  $Z^0 = \ker d$ is a $(|V|, 1)$ repetition code over ${\A}$, so the set $\{v\}$ comprising any single vertex $v \in V$ may be taken as an information set, and then the remaining vertices comprise the corresponding check set.  

\begin{figure}[h]
\setlength{\unitlength}{5pt}
\centering
%%%%%%%%%%%%%%
% Figure 7
%%%%%%%%%%%%%%
\begin{tikzpicture}
\node[box]                   (EQ1) {$=$};
\node[box,right=\franksep of EQ1 ] (EQ2) {$=$};
\node[box,right=\franksep of EQ2 ] (EQ3) {$=$};
\node[box,below=\franksep of EQ2 ] (EQ4) {$=$};
\node[box,below=\franksep of EQ1 ] (EQ5) {$=$};
\node[var,left=\davesep of EQ1] (X1) {$x_1$};
\node[var,above=\davesep of EQ2] (X2) {$x_2$};
\node[var,right=\davesep of EQ3] (X3) {$x_3$};
\node[var,right=\davesep of EQ4] (X4) {$x_4$};
\node[var,left=\davesep of EQ5] (X5) {$x_5$};
\draw[->] (X1.east) \updown -- (EQ1);
\draw[<-] (X2.south) \leftright -- (EQ2);
\draw[<-] (X3.west) \updown -- (EQ3);
\draw[<-] (X4.west) \updown -- (EQ4);
\draw[<-] (X5.east) \updown -- (EQ5);
\draw[->] (EQ1) -- (EQ2);
\draw[->] (EQ2) -- (EQ3);
\draw[->] (EQ2) -- (EQ4);
\draw[->] (EQ4) -- (EQ5);
\end{tikzpicture}
\caption{I/O realization of $Z^0 = \ker d$ on $E_\T = \{e_1, e_2, e_4, e_6\}$, with $\{{x}_1\}$ as information set.}
\label{NRZ0IO}
\end{figure}
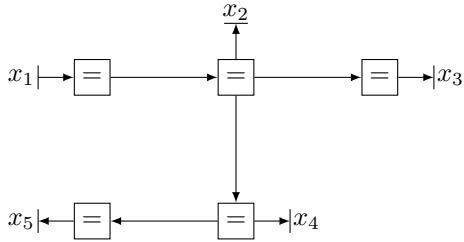

However, we can see that specifying an information set does not suffice to determine cause-and-effect relationships on the internal edges.  The fundamental reason is that the  realization of $Z^0$ in Figure \ref{NRZ0} contains $\beta_1(\G) = 2$ redundant edge constraints.  Indeed, it is easy to see that it would suffice to  propagate the value of any single input vertex variable through a spanning tree $\T$ of $\G$ in order to generate all other output vertex variables correctly;  moreover, in such a realization every edge in $E_\T$  would have a definite direction, namely the direction ``away" from the input vertex.  Such a nonredundant realization of $Z^0$ is illustrated in Figure  \ref{NRZ0IO}, where we choose the information set $\{v_1\}$ and $E_\T = \{e_1, e_2, e_4, e_6\}$.    

The realization of Figure \ref{NRZ0} thus has $\beta_1(\G) = 2$ ``degrees of redundancy."  This kind of constraint redundancy  is called \emph{uncontrollability} in  \cite{F14, FGL12}. More precisely, in \cite{FGL12} ``uncontrollable" is defined as ``having dependent constraints,"  which in this case are the $\beta_1(\G) = 2$ redundant edge constraints.  (``Local reductions" such as those leading to Figure \ref{NRZ0IO} are discussed more generally in \cite{FGL12}.)

Next, we will develop an I/O realization for the coboundary space $B^1 \subseteq \A^E$.  We have seen that  $B^1$ is a systematic $(|E|, |V| - 1)$ group code over $\A$, and that  for any spanning tree $\T$ of $\G$, $E_\T \subseteq E$ is an information set for $B^1$. 

However, we again observe  that  specifying an information set for the realization of $B^1$ in Figure \ref{NRB1} does not suffice to determine cause-and-effect relationships on the internal edges.  In this case, the basic reason is that this  realization has $\beta_0(\G) = 1$ internal degree of freedom corresponding to the \emph{unobservable behavior} $\Bf^u$ of this realization, namely the internal behavior when the external variables are fixed to zero.  This kind of redundancy  is called \emph{unobservability} in  \cite{F14, FGL12}.  (Alternatively, ``unobservable" means ``having  redundant generators.") 

\begin{figure}[h]
\setlength{\unitlength}{5pt}
\centering
%%%%%%%%%%%%%%
% Figure 8
%%%%%%%%%%%%%%
\begin{tikzpicture}
\node[box]                   (EQ1) {$=$};
\node[box,right=\franksep of EQ1 ] (EQ2) {$=$};
\node[box,right=\franksep of EQ2 ] (EQ3) {$=$};
\node[box,below=\franksep of EQ2 ] (EQ4) {$=$};
\node[box,below=\franksep of EQ1 ] (EQ5) {$=$};
\draw (EQ1) -- (EQ2);
\draw (EQ1) -- (EQ5);
\draw (EQ2) -- (EQ3);
\draw (EQ2) -- (EQ4);
\draw (EQ3) -- (EQ4);
\draw (EQ4) -- (EQ5);
\end{tikzpicture}
\caption{Realization of the unobservable behavior $\Bf^u$ of the Figure \ref{NRB1} realization of $B^1$.}
\label{UBF}
\end{figure}
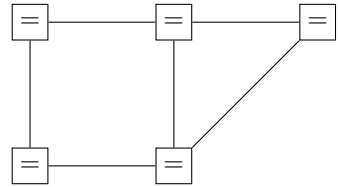

For the realization of $B^1$  in Figure \ref{NRB1}, the unobservable behavior $\Bf^u \subseteq \A^E$ is the behavior  of the realization shown in Figure \ref{UBF}.  This behavior is evidently the same as the internal behavior of the (observable) realization of $Z^0 = \ker d$ in Figure \ref{NRZ0}, since both may be obtained from the realization of $W^{01}$ in Figure \ref{NRIOD} by deleting the vertex variables $x_v$ and setting all edge variables $y_e$ to zero.  Again, since $\G$ is connected, $\Bf^u$ is obviously a repetition code with dimension $\beta_0(\G) =  1$.  

\pagebreak
Since the unobservable behavior $\Bf^u$ of Figure \ref{NRB1} is a repetition code, it follows that adding the same element of $\A$ to every vertex value does not change the output configuration in $B^1$.  (In physics, this kind of unobservability is sometimes called ``global symmetry.")
To obtain an observable realization of $B^1$, we may therefore fix any internal vertex variable in Figure \ref{NRB1} to zero,  breaking the global symmetry.  (Again, this is an example of a ``local reduction" as in \cite{FGL12}.)

\begin{figure}[h]
\setlength{\unitlength}{5pt}
\centering
%%%%%%%%%%%%%%
% Figure 9
%%%%%%%%%%%%%%
\begin{tikzpicture}
\node[box] (PL1) {$+$};
\node[box,right=\davesep of PL1] (EQ2) {$=$};
\node[box,right=\davesep of EQ2] (PL2) {$+$};
\node[box,right=\davesep of PL2] (EQ3) {$=$};
\node[box,below=\davesep of EQ2] (PL4) {$+$};
\node[box,below=\davesep of PL2] (PL5) {$+$};
\node[box,below=\davesep of PL4] (EQ4) {$=$};
\node[box,left=\davesep of EQ4] (PL6) {$+$};
\node[box,left=\davesep of PL6] (EQ5) {$=$};
\node[box,above=\davesep of EQ5] (PL3) {$+$};
\node[invert,anchor=west]  (IPL1) at (PL1.east)  {};
\node[invert,anchor=west]  (IPL2) at (PL2.east)  {};
\node[invert,anchor=north] (IPL3) at (PL3.south) {};
\node[invert,anchor=north] (IPL4) at (PL4.south) {};
\node[invert,anchor=north east] (IPL5) at (PL5.south west) {};
\node[invert,anchor=east] (IPL6) at (PL6.west) {};
\node[var,above=\davesep of PL1] (Y1) {$y_1$};
\node[var,above=\davesep of PL2] (Y2) {$y_2$};
\node[var,left=\davesep of PL3] (Y3) {$y_3$};
\node[var,left=\davesep of PL4] (Y4) {$y_4$};
\node[var,right=\davesep of PL5] (Y5) {$y_5$};
\node[var,below=\davesep of PL6] (Y6) {$y_6$};
\draw[->] (Y1.south) \leftright -- (PL1);
\draw[->] (Y2.south) \leftright -- (PL2);
\draw[<-] (Y3.east) \updown -- (PL3);
\draw[->] (Y4.east) \updown -- (PL4);
\draw[<-] (Y5.west) \updown -- (PL5);
\draw[->] (Y6.north) \leftright -- (PL6);
\draw[<-] (EQ2) -- (IPL1);
\draw[->] (EQ2) -- (PL2);
\draw[->] (EQ2) -- (PL4);
\draw[<-] (EQ3) -- (IPL2);
\draw[->] (EQ3) -- (PL5);
\draw[->] (EQ4) -- (PL6);
\draw[<-] (EQ4) -- (IPL4);
\draw[->] (EQ4) -- (IPL5);
\draw[->] (EQ5) -- (IPL3);
\draw[<-] (EQ5) -- (IPL6);
\end{tikzpicture}
\caption{I/O realization of    $B^1 = \im d$, fixing $x_1 = 0$, and using $\{{y}_1, {y}_2, {y}_4, {y}_6\}$ as information set.}
\label{NRB1IO}
\end{figure}
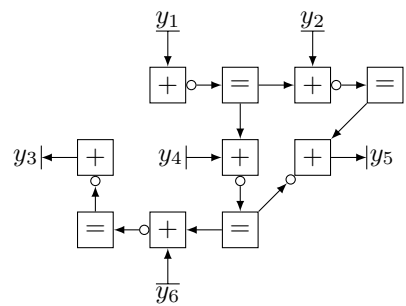

We illustrate such a realization in Figure \ref{NRB1IO}, where we have chosen $E_\T = \{e_1, e_2, e_4, e_6\}$ as the information set, and fixed $x_1 = 0$.  Such an I/O realization could be used in a simulation to generate all elements of $B^1$, by letting  $\{y_1, y_2, y_4, y_6\}$ run through all $|\A|^{4}$ possible configurations.

This argument is essentially the same as the algebraic argument used in our proof of Theorem 1(b), and Figure \ref{NRB1IO} may be regarded as an illustration of that proof.

\subsection{Elements of duality theory}

We  now introduce some elementary duality theory for both the group and field cases.  (See the Appendix for more  on  duality theory for abelian groups, focussing on Fourier transforms.)

\pagebreak
We  stipulate that $\A$ must have a well-defined \textbf{dual group}  $\hat{\A}$, for which the inner product $\inner{\hat{a}}{{a}}$ is well defined for all $\hat{a} \in \hat{\A}, {a} \in {\A}$.\footnote{As explained in the Appendix, we assume an additive inner product with $\inner{\hat{a}}{{a}} \in \R/\Z$, rather than a multiplicative inner product with $\inner{\hat{a}}{{a}}$ in the complex unit circle $\mathbb{T}$.}  If $\A$ is a field $\F$, then $\hat{\A}$ may also be taken as $\F$;  if $\A$ is  $\Z_q$, then $\hat{\A}$ may also be taken as $\Z_q$;  more generally, if $\A$ is a finite abelian group, then $\A$ has a well-defined dual group $\hat{\A}$ (called its \emph{character group}) that is isomorphic to $\A$.  Again, we will assume that $\A$ is a finite abelian group (or finite field) in this paper.
 
 Given a finite coordinate index set $\I$, the dual group to $\A^\I$ is then $\hat{\A}^\I$, with the inner product between $\hat{\ab} \in \hat{\A}^\I$ and ${\ab} \in {\A}^\I$ defined in standard coordinatewise fashion: $\inner{\hat{\ab}}{{\ab}} = \sum_{i \in \I} \inner{\hat{a}_i}{{a}_i}$.  Two vectors $\hat{\ab} \in \hat{\A}^\I$ and ${\ab} \in {\A}^\I$ are \textbf{orthogonal} if $\inner{\hat{\ab}}{{\ab}} = 0$.  The \textbf{dual code} (or \emph{orthogonal code}) $\CC^\perp$ to a group code $\CC \subseteq \A^\I$ is the set of all $\hat{\ab} \in \hat{\A}^\I$ that are orthogonal to all $\ab \in \CC$.  
 
 Given a group code $\CC \subseteq \A^\I$ and an index subset $\J \subseteq \I$ with complement $\bar{\J} = \I \setminus \J$, the \emph{projection} of $\CC$ onto $\J$ is defined as $\CC_{|\J} = \{\ab_\J \in \A^\J \mid \exists (\ab_\J, \ab_{\bar{\J}}) \in \CC\}$, and the \emph{cross-section} of $\CC$ on $\J$ is defined as $\CC_{:\J} = \{\ab_\J \in \A^\J \mid (\ab_\J, \zerob_{\bar{\J}}) \in \CC\}$.  If $\CC^\perp$ is the dual code to $\CC$, then by \emph{projection/cross-section duality} \cite{F01}, $(\CC_{|\J})^\perp = (\CC^\perp)_{:\J}$ and $(\CC_{:\J})^\perp = (\CC^\perp)_{|\J}$.
 
Given a homomorphism $\varphi: \A^\I \to \A^\J$, its  \textbf{adjoint homomorphism} is defined as the unique homomorphism $\hat{\varphi}: \hat{\A}^{{\J}} \to \hat{\A}^\I$ such that $\inner{\hat{\varphi}(\hat{\ab})}{\ab} = \inner{\hat{\ab}}{\varphi(\ab)}$ for all $\hat{\ab} \in \hat{\A}^{{\J}}, {\ab} \in \A^\I$.  

We define the \emph{I/O code} of $\varphi$ as $\W = \{(\ab, \varphi(\ab)) \in \A^\I \times \A^\J \mid \ab \in \A^\I\}$, and its \emph{dual I/O code} as $\hat{\W} = \{(-\hat{\varphi}(\hat{\ab}), \hat{\ab}) \in \hat{\A}^{\I} \times \hat{\A}^ \J \mid \hat{\ab} \in \hat{A}^\J\}$.  Evidently $\hat{\W}$ is essentially the I/O code of $\hat{\varphi}$, up to sign inversion and coordinate ordering.

Adjoint homomorphisms and their I/O codes have the following  properties:
 
 \vspace{1ex}
 \noindent
 \textbf{Adjoint Homomorphism Lemma}.  Given a homomorphism $\varphi: \A^\I \to \A^\J$ and its adjoint homomorphism $\hat{\varphi}: \hat{\A}^{{\J}} \to \hat{\A}^\I$:
 
 (a)  The I/O code $\W$ of $\varphi$ is a systematic $(|\I| + |\J|, |\I|)$ group code,  with information set $\I$.

 (b)  The dual I/O code $\hat{\W}$ is the orthogonal systematic $(|\I| + |\J|, |\J|)$ group code $\W^\perp$ with information set $\J$.
 
 (c)  The kernel of $\varphi$ and the image of $\hat{\varphi}$ are orthogonal codes:  $(\ker \varphi)^\perp = \im  \hat{\varphi}$.  
 
 (d) Similarly, $(\im \varphi)^\perp = \ker \hat{\varphi}$.
 
  \vspace{1ex}
 \noindent
 \textit{Proof}:  
 (a) The projection of the I/O code $\W = \{(\ab, \varphi(\ab)) \in \A^\I \times \A^\J \mid \ab \in \A^\I\}$ onto $\I$ is an isomorphism $\W \leftrightarrow \A^\I$ with image $\W_{|\I}  = \A^\I$.
 
 (b) The projection of the dual I/O code $\hat{\W} = \{(-\hat{\varphi}(\hat{\ab}), \hat{\ab}) \in \hat{\A}^{\I} \times \hat{\A}^ \J \mid \hat{\ab} \in \hat{\A}^\J\}$ onto $\J$ is an isomorphism $\hat{\W} \leftrightarrow \hat{\A}^\J$ with image $\hat{\W}_{|\J} = \hat{A}^\J$.
  Because the inner product is defined coordinatewise, the inner product between  $(-\hat{\varphi}(\hat{\ab}), \hat{\ab}) \in \hat{\W}$ and $(\ab, \varphi(\ab)) \in \W$ is  $-\inner{\hat{\varphi}(\hat{\ab})}{\ab} + \inner{\hat{\ab}}{\varphi(\ab)} = 0$ for all $\hat{\ab} \in \hat{\A}^{\J}, \ab \in \A^\I$, by the definition of  $\hat{\varphi}$.  Therefore $\W$ and $\hat{\W}$ are orthogonal.  Indeed, they are dual codes, since $\dim \hat{\W} + \dim \W = |\J| + |\I|$.
 
 (c)  The cross-section $\W_{:\I} = \{\ab \in \A^\I \mid (\ab, \zerob) \in \W\}$ of $\W$ on $\A^\I$ is precisely $\ker \varphi$, whereas the projection $(\W^\perp)_{|\I} = \{\hat{\bb} \in \hat{\A}^\I \mid \exists \hat{\ab} \in \hat{\A}^\J: (\hat{\bb}, \hat{\ab}) \in \W^\perp\}$ of $\W^\perp = \hat{\W}$ onto $\hat{\A}^\I$ is precisely $\im \hat{\varphi}$.  By projection/cross-section duality,  $\W_{:\I} = \ker \varphi$ and $(\W^\perp)_{|\I} = \im \hat{\varphi}$ are thus dual codes.
 
 (d) \emph{Mutatis mutandis}, $\W_{|\J} = \im \varphi$ and $(\W^\perp)_{:\J} = \ker \hat{\varphi}$ are dual codes.
 \qed \vspace{1ex}
 
In view of the I/O Map Lemma for systematic $(n,k)$ group codes, we  thus  have the following generalization of well-known results for the field case:
 
 \vspace{1ex}
 \noindent
 \textbf{Theorem 3} (\emph{dual systematic group codes}).  If $\CC \subseteq \A^\I$ is a systematic $(n,k)$ group code over $\A$, then its dual code $\CC^\perp \subseteq \hat{\A}^\I$ is a systematic $(n,n-k)$ group code over the dual group $\hat{\A}$.  Moreover, the information sets of $\CC^\perp$ are the check sets of $\CC$, and \emph{vice versa}. 
 
  \vspace{1ex}
 \noindent
 \textit{Proof}:  By the I/O Map Lemma, if $\CC$ is a systematic $(n,k)$ group code  with information set $\J \subseteq \I$, then there exists a homomorphism $\varphi_\CC:  \A^\J \to \A^{\bar{\J}}$ such that $\CC$ is the I/O code of $\varphi_\CC:  \A^\J \to \A^{\bar{\J}}$;  \ie
$\CC = \{(\ab_\J, \varphi_\CC(\ab_\J)) \mid \ab_\J \in \A^\J\}.$  By the Adjoint Homomorphism Lemma, if $\hat{\varphi}_\CC: \hat{\A}^{\bar{\J}} \to \hat{\A}^\J$ is the adjoint homomorphism to ${\varphi}_\CC$, then the dual I/O code $\{(-\hat{\varphi}_\CC(\hat{\ab}), \hat{\ab}) \in \hat{\A}^{\J} \times \hat{\A}^{\bar{\J}} \mid \hat{\ab} \in \hat{\A}^{\bar{\J}}\}$ is the orthogonal code $\CC^\perp$.  Evidently $\CC^\perp$ is  a systematic $(n,n-k)$ group code over $\hat{\A}$ with an information set $\bar{\J}$ that is equal to the complement of the information set $\J$ of $\CC$.  \qed \vspace{1ex}

 For example, if $\CC_= \subseteq \A^\I$ is an $(n, 1)$ repetition code over $\A$, then its dual code $(\CC_=)^\perp$ is  the $(n, n-1)$ 
 zero-sum code $\CC_+ \subseteq \hat{\A}^\I$.  The information sets of $\CC$ are the singleton sets $\{\{i\}, i \in \I\}$, whereas the information sets of $\CC^\perp$ are the complementary sets $\{\I \setminus \{i\}, i \in \I\}$.

\subsection{Duality in elementary algebraic topology}\label{DSO}

In algebraic topology, for historical reasons, the dual  space to the space $C^0 = \A^V$ of 0-cochains is called the \emph{space of 0-chains}, denoted by $C_0 = \hat{\A}^V$.  Similarly,  the dual $\hat{\A}^E$ to the space $C^1 = \A^E$ of 1-cochains is called the \emph{space of 1-chains},  denoted by $C_1 = \hat{\A}^E$.  

\begin{figure*}[ht]
\setlength{\unitlength}{5pt}
\centering
\begin{picture}(80,13)(-1, 0)
\put(0,0){$B_0 = \im \partial:  (|V|, |V|-1)$}
\put(32,0){$\hat{\A}^V \longleftarrow \hat{\A}^E$}
\put(37,1.5){$\partial$}
\put(50,0){$Z_1 = \ker \partial:  (|E|, |E| - |V| + 1)$}
\put(4,5){$\Updownarrow$}
\put(6,5){$\perp$}
\put(37,5){$\Updownarrow$}
\put(39,5){adjoint}
\put(54,5){$\Updownarrow$}
\put(56,5){$\perp$}

\put(0,9){$Z^0 = \ker d:  (|V|,1)$}
\put(32,9){$\A^V \longrightarrow \A^E$}
\put(37,10.5){$d$}
\put(50,9){$B^1 = \im d:  (|E|, |V| -1)$}
\end{picture}
\caption{Duality of coboundary operator $d: \A^V \to \A^E$ and boundary operator $\partial: \hat{\A}^E \to \hat{\A}^V$.}
\label{DR}
\end{figure*}

We will regard the elements $\hat{\xb} \in \hat{\A}^V$ as sets $\hat{\xb} = (\hat x_{v_1}, \hat x_{v_2}, \ldots, \hat x_{v_{|V|}})$ of $|V|$ elements of $\hat \A$ indexed by the elements of $V,$ and similarly the elements $\hat \yb \in \hat{\A}^E$  as sets $\hat \yb = (\hat y_{e_1}, \hat y_{e_2}, \ldots, \hat y_{e_{|E|}})$ of $|E|$ elements of $\hat \A$ indexed by the elements of $E$.  When $\hat \xb$ and $\hat \yb$ are used in conjunction with the incidence matrix $M$, we may regard them as row vectors, and call them  \emph{dual vertex vectors} $\hat \xb$ and \emph{dual edge vectors} $\hat \yb$, respectively.  
We define inner products  in  standard componentwise fashion;  \eg for $\xb \in \A^V$ and $\hat{\xb} \in \hat{\A}^V$, $\inner{\hat{\xb}}{\xb} = \sum_V\inner{ \hat{x}_v }{{x}_v}$. 

The \emph{boundary operator} $\partial: \hat{\A}^E \to \hat{\A}^V$ is then defined as the homomorphism that maps $\hat{\yb} \in \hat{\A}^E$ to $\partial(\hat{\yb}) = \hat{\yb}M \in \hat{\A}^V$, where  $M$ is again the incidence matrix of $\G$.
  Evidently, for any $\hat{\yb} \in \hat{\A}^E, {\xb} \in \hat{\A}^V$ we have 
$$\inner{\partial(\hat{\yb})}{\xb} =  \hat{\yb} M{\xb} =  \inner{\hat{\yb}}{d(\xb)}.$$
It follows that:
 
 \vspace{1ex}
 \noindent
 \textbf{Boundary Operator Lemma}. The boundary operator $\partial:  \hat{\A}^E \to \hat{\A}^V$ is the {adjoint homomorphism} $\hat{d}$  to the coboundary operator $d: \A^V \to \A^E$. \qed \vspace{1ex}
 
By the Adjoint Homomorphism Lemma, this implies that the kernel of $\partial$, called the \emph{zero-boundary space} $Z_1 = \ker \partial$,  is the orthogonal code $(B^1)^\perp$ 
to the image $B^1 =  \im d$ of $d$; and conversely  the image   of $\partial$, called the \emph{boundary space} $B_0 =  \im \partial$, is the orthogonal code $(Z^0)^\perp$  to the kernel $Z^0 = \ker d$ of $d$.  Therefore we have:

\vspace{1ex}
\noindent
\textbf{Theorem 4} ($B_0, Z_1$).  For a  connected graph $\G = (V, E)$:

(a) The boundary space $B_0 =  \im \partial$  is a $(|V|, |V| - 1)$ zero-sum code  $\CC_+ \subseteq \hat{\A}^V$.

(b) The zero-boundary space $Z_1 = \ker \partial$ is a systematic $(|E|, \beta_1(\G))$ group code over $\hat{\A}$.

\vspace{1ex}
\noindent
\textit{Proof}. (a) By Theorem 1(a), $Z^0$ is the $(|V|, 1)$ repetition code over $\A$, so its dual code $B_0$ is the $(|V|, |V| - 1)$ zero-sum code over $\hat{\A}$.

(b) By Theorem 1(b), $B^1 \subseteq \A^E$ is a systematic $(|E|, |E| - \beta_1(\G))$  group code over $\A$, so its dual code $Z_1$ is a systematic $(|E|, \beta_1(\G))$ group code over $\hat{A}$. \qed \vspace{1ex}

Figure \ref{DR} summarizes the duality relationships between the adjoint operators $d: \A^V \to \A^E$ and $\partial: \hat{\A}^E \to \hat{\A}^V$, and between their kernels and images.

\subsection{Dual normal realizations}\label{DNR}

We will now construct normal realizations of the boundary space $B_0 = \im \partial$ and the zero-boundary space $Z_1 = \ker \partial$ of $\G$ as  dual realizations to our earlier realizations of the zero-coboundary space $Z_0 = (B_0)^\perp$ and the coboundary space $B^1 = (Z_1)^\perp$ of $\G$, respectively.

In general, the dual realization to an elementary normal realization is obtained as follows:
\begin{itemize}
\item The variable alphabet $\A$ is replaced by its dual alphabet $\hat{\A}$;
\item Repetition constraints 
(\begin{picture}(2,1)
\put(0.25,-0.25){\framebox(1.5,1.5){$=$}}
\end{picture})
are replaced by zero-sum constraints
(\begin{picture}(2,1)
\put(0.25,-0.25){\framebox(1.5,1.5){$+$}}
\end{picture}),
 and \emph{vice versa};
 \item Edges 
(\begin{picture}(2,1)
\put(0, 0.5){\line(1,0){2}}
\end{picture})
 are replaced by inverting edges
(\begin{picture}(3,1)
\put(0, 0.5){\line(1,0){1.2}}
\put(1.2, 0.15){$\circ$}
\put(1.75, 0.5){\line(1,0){1.25}}
\end{picture}),
and \emph{vice versa}.
\end{itemize}

By the \emph{normal realization duality theorem} \cite{F01}, if the  external behavior of the original normal realization is  $\CC$, then the external behavior of the dual normal realization is  $\CC^\perp$.  (If $\A$ is a finite abelian group, then this theorem is a corollary to the normal factor graph duality theorem;  see Appendix.)

 We  start with the dual I/O behavior $W_{10} = \{(-\hat{\yb}M, \hat{\yb}) \mid \hat{\yb} \in \hat{\A}^E\}$ of the boundary operator $\partial$, which by the Adjoint Homomorphism Lemma is the dual code to the I/O behavior $W^{01} = \{(\xb, M\xb) \mid \xb \in \A^V\}$ of the coboundary operator $d$;  \ie ${W}_{10} = (W^{01})^\perp$.  Thus ${W}_{10}$ is a systematic $(|V| + |E|, |E|)$ group code over $\hat{\A}$ with information set $E$.  
 
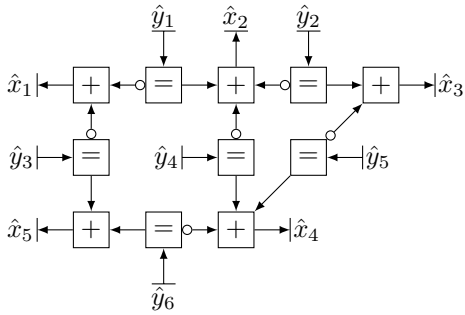
\begin{figure}[h]
\setlength{\unitlength}{5pt}
\centering
%%%%%%%%%%%%%%
% Figure 11
%%%%%%%%%%%%%%
\begin{tikzpicture}
\node[box]                  (PL1) {$+$};
\node[box,right=\davesep of PL1] (EQ1) {$=$};
\node[box,right=\davesep of EQ1] (PL2) {$+$};
\node[box,right=\davesep of PL2] (EQ2) {$=$};
\node[box,right=\davesep of EQ2] (PL3) {$+$};
\node[box,below=\davesep of PL1] (EQ3) {$=$};
\node[box,below=\davesep of PL2] (EQ4) {$=$};
\node[box,below=\davesep of EQ2] (EQ5) {$=$};
\node[box,below=\davesep of EQ3] (PL5) {$+$};
\node[box,right=\davesep of PL5] (EQ6) {$=$};
\node[box,right=\davesep of EQ6] (PL4) {$+$};
\node[invert,anchor=east]  (IEQ1) at (EQ1.west)  {};
\node[invert,anchor=east]  (IEQ2) at (EQ2.west)  {};
\node[invert,anchor=south] (IEQ3) at (EQ3.north) {};
\node[invert,anchor=south] (IEQ4) at (EQ4.north) {};
\node[invert,anchor=south west] (IEQ5) at (EQ5.north east) {};
\node[invert,anchor=west] (IEQ6) at (EQ6.east) {};
\node[var,left=\davesep of PL1] (X1) {$\hat{x}_1$};
\node[var,above=\davesep of PL2] (X2) {$\hat{x}_2$};
\node[var,right=\davesep of PL3] (X3) {$\hat{x}_3$};
\node[var,right=\davesep of PL4] (X4) {$\hat{x}_4$};
\node[var,left=\davesep of PL5] (X5) {$\hat{x}_5$};
\node[var,above=\davesep of EQ1] (Y1) {$\hat{y}_1$};
\node[var,above=\davesep of EQ2] (Y2) {$\hat{y}_2$};
\node[var,left=\davesep of EQ3] (Y3) {$\hat{y}_3$};
\node[var,left=\davesep of EQ4] (Y4) {$\hat{y}_4$};
\node[var,right=\davesep of EQ5] (Y5) {$\hat{y}_5$};
\node[var,below=\davesep of EQ6] (Y6) {$\hat{y}_6$};
\draw[<-] (X1.east) \updown -- (PL1);
\draw[<-] (X2.south) \leftright -- (PL2);
\draw[<-] (X3.west) \updown -- (PL3);
\draw[<-] (X4.west) \updown -- (PL4);
\draw[<-] (X5.east) \updown -- (PL5);
\draw[->] (Y1.south) \leftright -- (EQ1);
\draw[->] (Y2.south) \leftright -- (EQ2);
\draw[->] (Y3.east) \updown -- (EQ3);
\draw[->] (Y4.east) \updown -- (EQ4);
\draw[->] (Y5.west) \updown -- (EQ5);
\draw[->] (Y6.north) \leftright -- (EQ6);
\draw[<-] (PL1) -- (IEQ1);
\draw[<-] (PL1) -- (IEQ3);
\draw[<-] (PL2) -- (IEQ2);
\draw[<-] (PL2) -- (IEQ4);
\draw[<-] (PL3) -- (IEQ5);
\draw[<-] (PL4) -- (IEQ6);
\draw[<-] (PL2) -- (EQ1);
\draw[<-] (PL3) -- (EQ2);
\draw[<-] (PL4) -- (EQ4);
\draw[<-] (PL4) -- (EQ5);
\draw[<-] (PL5) -- (EQ3);
\draw[<-] (PL5) -- (EQ6);
\end{tikzpicture}
\caption{Normal realization of dual I/O behavior ${W}_{10} = \{(-\hat{\yb}M, \hat{\yb}) \mid \hat{\yb} \in \hat{\A}^E\}$.}
\label{NRIODD}
\end{figure}

 We may thus obtain an elementary normal realization of  ${W}_{10}$ by dualizing our earlier realization of $W_{01}$ in Figure \ref{NRIOD}, which results in the realization of Figure \ref{NRIODD}.   Again, we have included arrows on all edges to indicate cause-and-effect relationships;  note that all arrows are now reversed.

In Figure \ref{NRIODD}, a set $\hat{\yb} = \{\hat{y}_e, e \in E\}$ of $|E|$ external input variables  is associated with the edges $e \in E$ of $\G$.  Each edge variable $\hat{y}_e$ is replicated twice  via a repetition constraint of degree 3.  Each replica is passed on (with a sign inversion if $v = t(e)$) to the zero-sum constraint associated with one of the two vertices $h(e), t(e)$, thus making $\hat{y}_eM_{ev}$ the input to this zero-sum constraint.  At each vertex $v \in V$ of $\G$, the zero-sum constraint on all these incident  variables plus an output variable $\hat{x}_v$ enforces the constraint $\hat{x}_v = -(\hat{\yb}M)_v$;  thus $\hat{\xb} = -\hat{\yb}M$, as desired.  

\begin{figure}[h]
\setlength{\unitlength}{5pt}
\centering
%%%%%%%%%%%%%%
% Figure 12
%%%%%%%%%%%%%%
\begin{tikzpicture}
\node[box]                   (PL1) {$+$};
\node[box,right=\franksep of PL1 ] (PL2) {$+$};
\node[box,right=\franksep of PL2 ] (PL3) {$+$};
\node[box,below=\franksep of PL2 ] (PL4) {$+$};
\node[box,below=\franksep of PL1 ] (PL5) {$+$};
\node[var,left=\davesep of PL1] (X1) {$\hat{x}_1$};
\node[var,above=\davesep of PL2] (X2) {$\hat{x}_2$};
\node[var,right=\davesep of PL3] (X3) {$\hat{x}_3$};
\node[var,right=\davesep of PL4] (X4) {$\hat{x}_4$};
\node[var,left=\davesep of PL5] (X5) {$\hat{x}_5$};
\node[invert] (I12) at (barycentric cs:PL1=0.5,PL2=0.5) {};
\node[invert] (I15) at (barycentric cs:PL1=0.5,PL5=0.5) {};
\node[invert] (I23) at (barycentric cs:PL2=0.5,PL3=0.5) {};
\node[invert] (I24) at (barycentric cs:PL2=0.5,PL4=0.5) {};
\node[invert] (I34) at (barycentric cs:PL3=0.5,PL4=0.5) {};
\node[invert] (I45) at (barycentric cs:PL4=0.5,PL5=0.5) {};
\draw (X1.east) \updown -- (PL1);
\draw (X2.south) \leftright -- (PL2);
\draw (X3.west) \updown -- (PL3);
\draw (X4.west) \updown -- (PL4);
\draw (X5.east) \updown -- (PL5);
\draw (PL1) -- (I12);\draw (I12) -- (PL2);
\draw (PL1) -- (I15);\draw (I15) -- (PL5);
\draw (PL2) -- (I23);\draw (I23) -- (PL3);
\draw (PL2) -- (I24);\draw (I24) -- (PL4);
\draw (PL3) -- (I34);\draw (I34) -- (PL4);
\draw (PL4) -- (I45);\draw (I45) -- (PL5);
\end{tikzpicture}
\caption{Normal realization of boundary space $B_0 = \im \partial$ for graph $\G$ of Figure \ref{E1}.}
\label{NRB0}
\end{figure}
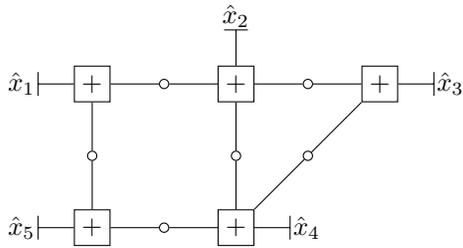

To get a realization of the boundary space $B_0 = \im \partial = \{\yb M \mid \yb \in \A^E\}$, we may simply remove the external variables $\hat{y}_e$ in Figure \ref{NRIODD}, while leaving the two internal replica variables representing $\pm \hat{y}_e$.  Since \begin{picture}(11,1)
\put(0, 0.5){\line(1,0){1}}
\put(1,-0.25){\framebox(1.5,1.5){$=$}}
\put(2.5, 0.15){$\circ$}
\put(3.25, 0.5){\line(1,0){1}}
\put(5.5, 0.2){$=$}
\put(7.5, 0.5){\line(1,0){1.2}}
\put(8.7, 0.15){$\circ$}
\put(9.25, 0.5){\line(1,0){1.25}}
\end{picture},
we may simply use the latter realization for each edge.  Thus we obtain the simple realization of $B_0$ shown in Figure \ref{NRB0}. (Strictly, this realization realizes $-B_0 = \{-\hat{\yb} M\}$, but since $B_0$ is an abelian group, we have $-B_0 = B_0$.)  Alternatively, since $B_0 = (Z^0)^\perp$, we may  obtain the realization of Figure \ref{NRB0} by dualizing the Figure \ref{NRZ0} realization of $Z^0$.

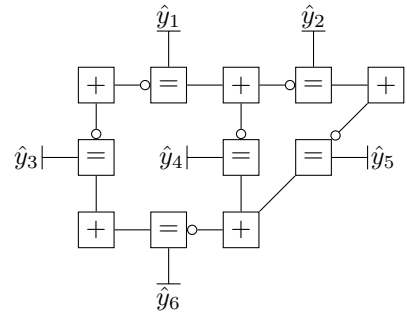
\begin{figure}[h]
\setlength{\unitlength}{5pt}
\centering
%%%%%%%%%%%%%%
% Figure 13
%%%%%%%%%%%%%%
\begin{tikzpicture}
\node[box]                  (PL1) {$+$};
\node[box,right=\davesep of PL1] (EQ1) {$=$};
\node[box,right=\davesep of EQ1] (PL2) {$+$};
\node[box,right=\davesep of PL2] (EQ2) {$=$};
\node[box,right=\davesep of EQ2] (PL3) {$+$};
\node[box,below=\davesep of PL1] (EQ3) {$=$};
\node[box,below=\davesep of PL2] (EQ4) {$=$};
\node[box,below=\davesep of EQ2] (EQ5) {$=$};
\node[box,below=\davesep of EQ3] (PL5) {$+$};
\node[box,right=\davesep of PL5] (EQ6) {$=$};
\node[box,right=\davesep of EQ6] (PL4) {$+$};
\node[invert,anchor=east]  (IEQ1) at (EQ1.west)  {};
\node[invert,anchor=east]  (IEQ2) at (EQ2.west)  {};
\node[invert,anchor=south] (IEQ3) at (EQ3.north) {};
\node[invert,anchor=south] (IEQ4) at (EQ4.north) {};
\node[invert,anchor=south west] (IEQ5) at (EQ5.north east) {};
\node[invert,anchor=west] (IEQ6) at (EQ6.east) {};
\node[var,above=\davesep of EQ1] (Y1) {$\hat{y}_1$};
\node[var,above=\davesep of EQ2] (Y2) {$\hat{y}_2$};
\node[var,left=\davesep of EQ3] (Y3) {$\hat{y}_3$};
\node[var,left=\davesep of EQ4] (Y4) {$\hat{y}_4$};
\node[var,right=\davesep of EQ5] (Y5) {$\hat{y}_5$};
\node[var,below=\davesep of EQ6] (Y6) {$\hat{y}_6$};
\draw (Y1.south) \leftright -- (EQ1);
\draw (Y2.south) \leftright -- (EQ2);
\draw (Y3.east) \updown -- (EQ3);
\draw (Y4.east) \updown -- (EQ4);
\draw (Y5.west) \updown -- (EQ5);
\draw (Y6.north) \leftright -- (EQ6);
\draw (PL1) -- (IEQ1);
\draw (PL1) -- (IEQ3);
\draw (PL2) -- (IEQ2);
\draw (PL2) -- (IEQ4);
\draw (PL3) -- (IEQ5);
\draw (PL4) -- (IEQ6);
\draw (PL2) -- (EQ1);
\draw (PL3) -- (EQ2);
\draw (PL4) -- (EQ4);
\draw (PL4) -- (EQ5);
\draw (PL5) -- (EQ3);
\draw (PL5) -- (EQ6);
\end{tikzpicture}
\caption{Normal realization of  $Z_1 = \ker \partial$  for graph $\G$ of Figure \ref{E1}.}
\label{NRZ1}
\end{figure}

Similarly, to get a realization of the zero-boundary space $Z_1 =  \ker \partial = (B^1)^\perp$, we may either zero the external vertex variables $\hat{x}_v$ in Figure  \ref{NRIODD}, or else dualize the realization of $B^1$ in Figure \ref{NRB1}.  By either method, we obtain the realization of $Z_1$ shown in Figure \ref{NRZ1}.  

\subsection{Cycles and bases for $Z_1$}\label{SPC}

By Theorem 4(b), $Z_1$ is a  $(|E|, \beta_1(\G))$ group code over $\hat{\A}$. We  now show that $Z_1$ is generated by a certain set of $\beta_1(\G)$ \emph{cycle vectors}. Thus $Z_1$ is  often  called the \emph{cycle space} of $\G$, and $\beta_1(\G)$ the \emph{nullity} of $\G$ \cite{Bollobas}.  Historically, this  seems to have been the  starting point of algebraic topology.

Given an edge $e \in E$ of a graph $\G$ with head and tail vertices $h(e)$ and $t(e)$, we define $-e$ as the \emph{reverse edge} with head and tail vertices $t(e)$ and $h(e)$, respectively.
A \emph{cycle} of a graph $\G$ is then a simple closed path  $\eb = (\pm e_1, \ldots, \pm e_n)$, consisting of a set of $n$ edges or reversed edges such that  $h(e_i) = t(e_{i+1})$ for all $i \in [1,n]$, where $e_{n+1} = e_1$, and no edge or vertex is repeated.

We then define a \emph{cycle vector} \cite{Bollobas} corresponding to a cycle $\eb = (\pm e_1, \ldots, \pm e_n)$ as any edge vector $\hat{\yb} \in \hat{\A}^E$ such that, for some $\hat{a} \in \hat{\A}$, $\hat{y}_e =  \hat{a}$ if $+e$ is an element of $\eb$, $\hat{y}_e =  -\hat{a}$ if $-e$ is an element of $\eb$, and $\hat{y}_e = 0$ otherwise.  The $|\hat{\A}|$ cycle vectors corresponding to a given cycle $\eb$ evidently form a one-dimensional \emph{cycle code} $\CC(\eb) \subseteq \hat{\A}^E$ that is effectively a repetition code with support $\eb$.

For example, our example graph $\G$ has a cycle $\eb = (e_1, e_4, e_6, -e_3)$ of length $n = 4$.  The corresponding cycle code is $\CC(\eb) = \{(\hat{a}, 0, -\hat{a}, \hat{a}, 0, \hat{a})  \mid \hat{a} \in \hat{\A}\}$.

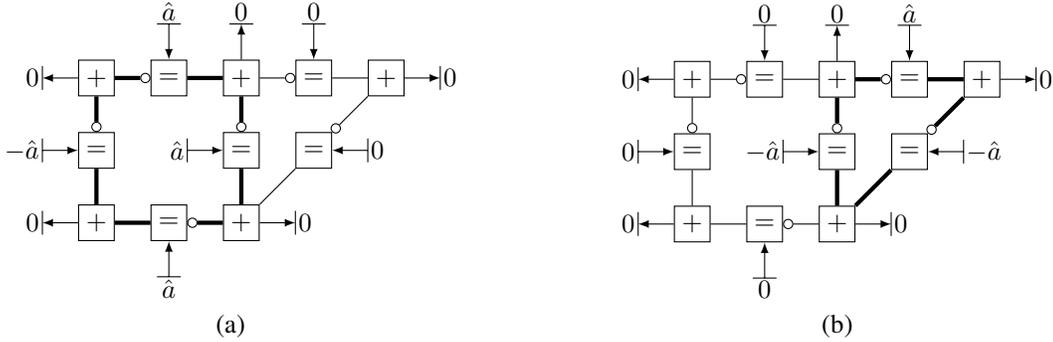
\begin{figure*}[h]
\setlength{\unitlength}{5pt}
\centering
\begin{tabular}{cc}
%%%%%%%%%%%%%%
% Figure 14(a)
%%%%%%%%%%%%%%
\begin{tikzpicture}
\node[box]                  (PL1) {$+$};
\node[box,right=\davesep of PL1] (EQ1) {$=$};
\node[box,right=\davesep of EQ1] (PL2) {$+$};
\node[box,right=\davesep of PL2] (EQ2) {$=$};
\node[box,right=\davesep of EQ2] (PL3) {$+$};
\node[box,below=\davesep of PL1] (EQ3) {$=$};
\node[box,below=\davesep of PL2] (EQ4) {$=$};
\node[box,below=\davesep of EQ2] (EQ5) {$=$};
\node[box,below=\davesep of EQ3] (PL5) {$+$};
\node[box,right=\davesep of PL5] (EQ6) {$=$};
\node[box,right=\davesep of EQ6] (PL4) {$+$};
\node[invert,anchor=east]  (IEQ1) at (EQ1.west)  {};
\node[invert,anchor=east]  (IEQ2) at (EQ2.west)  {};
\node[invert,anchor=south] (IEQ3) at (EQ3.north) {};
\node[invert,anchor=south] (IEQ4) at (EQ4.north) {};
\node[invert,anchor=south west] (IEQ5) at (EQ5.north east) {};
\node[invert,anchor=west] (IEQ6) at (EQ6.east) {};
\node[var,left=\davesep of PL1] (X1) {$0$};
\node[var,above=\davesep of PL2] (X2) {$0$};
\node[var,right=\davesep of PL3] (X3) {$0$};
\node[var,right=\davesep of PL4] (X4) {$0$};
\node[var,left=\davesep of PL5] (X5) {$0$};
\node[var,above=\davesep of EQ1] (Y1) {$\hat{a}$};
\node[var,above=\davesep of EQ2] (Y2) {$0$};
\node[var,left=\davesep of EQ3] (Y3) {$-\hat{a}$};
\node[var,left=\davesep of EQ4] (Y4) {$\hat{a}$};
\node[var,right=\davesep of EQ5] (Y5) {$0$};
\node[var,below=\davesep of EQ6] (Y6) {$\hat{a}$};
\draw[<-] (X1.east) \updown -- (PL1);
\draw[<-] (X2.south) \leftright -- (PL2);
\draw[<-] (X3.west) \updown -- (PL3);
\draw[<-] (X4.west) \updown -- (PL4);
\draw[<-] (X5.east) \updown -- (PL5);
\draw[->] (Y1.south) \leftright -- (EQ1);
\draw[->] (Y2.south) \leftright -- (EQ2);
\draw[->] (Y3.east) \updown -- (EQ3);
\draw[->] (Y4.east) \updown -- (EQ4);
\draw[->] (Y5.west) \updown -- (EQ5);
\draw[->] (Y6.north) \leftright -- (EQ6);
\draw[ultra thick] (PL1) -- (IEQ1);
\draw[ultra thick] (PL1) -- (IEQ3);
\draw (PL2) -- (IEQ2);
\draw[ultra thick] (PL2) -- (IEQ4);
\draw (PL3) -- (IEQ5);
\draw[ultra thick] (PL4) -- (IEQ6);
\draw[ultra thick] (PL2) -- (EQ1);
\draw (PL3) -- (EQ2);
\draw[ultra thick] (PL4) -- (EQ4);
\draw (PL4) -- (EQ5);
\draw[ultra thick] (PL5) -- (EQ3);
\draw[ultra thick] (PL5) -- (EQ6);
\end{tikzpicture}
& \hspace{10ex}
%%%%%%%%%%%%%%
% Figure 14(b)
%%%%%%%%%%%%%%
\begin{tikzpicture}
\node[box]                  (PL1) {$+$};
\node[box,right=\davesep of PL1] (EQ1) {$=$};
\node[box,right=\davesep of EQ1] (PL2) {$+$};
\node[box,right=\davesep of PL2] (EQ2) {$=$};
\node[box,right=\davesep of EQ2] (PL3) {$+$};
\node[box,below=\davesep of PL1] (EQ3) {$=$};
\node[box,below=\davesep of PL2] (EQ4) {$=$};
\node[box,below=\davesep of EQ2] (EQ5) {$=$};
\node[box,below=\davesep of EQ3] (PL5) {$+$};
\node[box,right=\davesep of PL5] (EQ6) {$=$};
\node[box,right=\davesep of EQ6] (PL4) {$+$};
\node[invert,anchor=east]  (IEQ1) at (EQ1.west)  {};
\node[invert,anchor=east]  (IEQ2) at (EQ2.west)  {};
\node[invert,anchor=south] (IEQ3) at (EQ3.north) {};
\node[invert,anchor=south] (IEQ4) at (EQ4.north) {};
\node[invert,anchor=south west] (IEQ5) at (EQ5.north east) {};
\node[invert,anchor=west] (IEQ6) at (EQ6.east) {};
\node[var,left=\davesep of PL1] (X1) {$0$};
\node[var,above=\davesep of PL2] (X2) {$0$};
\node[var,right=\davesep of PL3] (X3) {$0$};
\node[var,right=\davesep of PL4] (X4) {$0$};
\node[var,left=\davesep of PL5] (X5) {$0$};
\node[var,above=\davesep of EQ1] (Y1) {$0$};
\node[var,above=\davesep of EQ2] (Y2) {$\hat{a}$};
\node[var,left=\davesep of EQ3] (Y3) {$0$};
\node[var,left=\davesep of EQ4] (Y4) {$-\hat{a}$};
\node[var,right=\davesep of EQ5] (Y5) {$-\hat{a}$};
\node[var,below=\davesep of EQ6] (Y6) {$0$};
\draw[<-] (X1.east) \updown -- (PL1);
\draw[<-] (X2.south) \leftright -- (PL2);
\draw[<-] (X3.west) \updown -- (PL3);
\draw[<-] (X4.west) \updown -- (PL4);
\draw[<-] (X5.east) \updown -- (PL5);
\draw[->] (Y1.south) \leftright -- (EQ1);
\draw[->] (Y2.south) \leftright -- (EQ2);
\draw[->] (Y3.east) \updown -- (EQ3);
\draw[->] (Y4.east) \updown -- (EQ4);
\draw[->] (Y5.west) \updown -- (EQ5);
\draw[->] (Y6.north) \leftright -- (EQ6);
\draw (PL1) -- (IEQ1);
\draw (PL1) -- (IEQ3);
\draw[ultra thick] (PL2) -- (IEQ2);
\draw[ultra thick] (PL2) -- (IEQ4);
\draw[ultra thick] (PL3) -- (IEQ5);
\draw (PL4) -- (IEQ6);
\draw (PL2) -- (EQ1);
\draw[ultra thick] (PL3) -- (EQ2);
\draw[ultra thick] (PL4) -- (EQ4);
\draw[ultra thick] (PL4) -- (EQ5);
\draw (PL5) -- (EQ3);
\draw (PL5) -- (EQ6);
\end{tikzpicture} \\
(a) &  \hspace{10ex} (b) 
\end{tabular}
\caption{Cycle vectors corresponding to fundamental cycles including (a) $e_3$; (b) $e_5$.}
\label{Z1X}
\end{figure*}

It is easy to see that if $\yb \in \hat{\A}^E$ is any cycle vector, then $\partial(\yb) = \yb M = \zerob$;  \ie  every cycle vector is in $Z_1 = \ker \partial$.  For example, we show in Figure \ref{Z1X}(a) that if the cycle vector $(\hat{a}, 0, -\hat{a}, \hat{a}, 0, \hat{a})$ is the input to the I/O realization of Figure \ref{NRIODD}, then the  output vector is $\yb M = \zerob$.  Therefore:
 
\vspace{1ex}
\noindent
\textbf{Cycle Code Lemma}.  For every cycle $\eb$ of a connected graph $\G$, the cycle code $\CC(\eb)$ is a one-dimensional subcode of $Z_1$ with  support $\eb$. \qed \vspace{1ex}

We  will  now show that $Z_1$ has a ``basis" consisting of $\beta_1(\G)$ cycle codes $\CC(\eb)$.

We start with any spanning tree $\T = (V, E_\T)$ of $\G$.  We define $E_{\bar{\T}} = E \setminus E_\T$ as the complement of $E_\T$, so  $|E_{\bar{\T}}| = |E| - |V| +1 = \beta_1(\G)$. For  each  edge $e \in E_{\bar{\T}}$, there exists a \emph{fundamental cycle} $\eb_e$ of $\G$ comprising $e$ and the path connecting its final vertex  to its initial vertex  in $\T$.  Since the support of the cycle code $\CC(\eb_e)$ is $\eb_e \subseteq \{e\} \cup E_{\T}$, it follows that the $\beta_1(\G)$ cycle codes $\CC(\eb_e)$ are independent, since their supports are completely disjoint on $E_{\bar{\T}}$.  Thus we have proved:

\vspace{1ex}
\noindent
\textbf{Theorem 5} (\textit{``basis" for} $Z_1$).  For any graph $\G$ and any spanning tree $\T \subseteq \G$, $E_{\bar{\T}}$ is an information set for $Z_1$, and the $\beta_1(\G) = |E| - |V| + 1$  cycle codes $\CC(\eb_e), e \in E_{\bar{\T}},$ are a set of one-dimensional subspaces of $Z_1$ that form a systematic ``basis" for $Z_1$.  \qed \vspace{1ex}

For example, for our example graph $\G$, deleting the $\beta_1(\G) = 2$ edges $e_3$ and $e_5$ yields a spanning tree $\T$. The respective fundamental cycles in $\{e_3\} \cup E_\T$ and $\{e_5\} \cup E_\T$ are $\eb_3 = (e_1, e_4, e_6, -e_3))$ and $\eb_5 = (e_2, -e_5, -e_4)$.  Thus $\CC(\eb_3) = \{ (\hat{a}, 0, -\hat{a}, \hat{a}, 0, \hat{a}) \mid \hat{a} \in \hat{\A} \} $ and $\CC(\eb_5) = \{ (0, \hat{a}, 0, -\hat{a}, -\hat{a}, 0) \mid \hat{a} \in \hat{\A} \}$, as shown in Figure \ref{Z1X}.

We can now make precise our earlier statement that $\beta_1(\G)$ is the number of independent cycles in $\G$. Using a spanning tree $\T$ of $\G$, we have identified $\beta_1(\G)$ fundamental  cycles $\eb_e, e \in E_{\bar{\T}},$ such that $Z_1$ is generated by the cycle codes $\CC(\eb_e)$.  Thus for any  cycle $\eb$,  the cycle code $\CC(\eb) \subseteq Z_1$ is generated by the cycle codes  $\CC(\eb_e)$.  For example, our example graph $\G$ has  one other cycle, whose cycle vectors  $\{ (\hat{a}, \hat{a}, -\hat{a}, 0, \hat{a}, \hat{a}) \mid \hat{a} \in \hat{\A} \} $  are  sums of the corresponding vectors of $\CC(\eb_3)$ and $\CC(\eb_5)$.
 
 \subsection{Dual nonredundant input/output realizations}\label{DIOR}
 
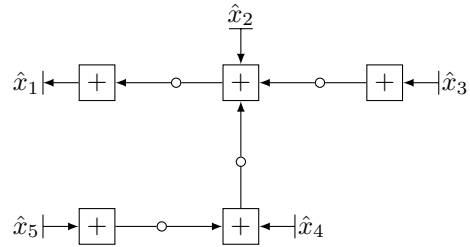
\begin{figure}[h]
\setlength{\unitlength}{5pt}
\centering
%%%%%%%%%%%%%%
% Figure 15
%%%%%%%%%%%%%%
\begin{tikzpicture}
\node[box]                   (PL1) {$+$};
\node[box,right=\franksep of PL1 ] (PL2) {$+$};
\node[box,right=\franksep of PL2 ] (PL3) {$+$};
\node[box,below=\franksep of PL2 ] (PL4) {$+$};
\node[box,below=\franksep of PL1 ] (PL5) {$+$};
\node[var,left=\davesep of PL1] (X1) {$\hat{x}_1$};
\node[var,above=\davesep of PL2] (X2) {$\hat{x}_2$};
\node[var,right=\davesep of PL3] (X3) {$\hat{x}_3$};
\node[var,right=\davesep of PL4] (X4) {$\hat{x}_4$};
\node[var,left=\davesep of PL5] (X5) {$\hat{x}_5$};
\node[invert] (I12) at (barycentric cs:PL1=0.45,PL2=0.55) {};
\node[invert] (I23) at (barycentric cs:PL2=0.45,PL3=0.55) {};
\node[invert] (I24) at (barycentric cs:PL2=0.45,PL4=0.55) {};
\node[invert] (I45) at (barycentric cs:PL4=0.45,PL5=0.55) {};
\draw[<-] (X1.east) \updown -- (PL1);
\draw[->] (X2.south) \leftright -- (PL2);
\draw[->] (X3.west) \updown -- (PL3);
\draw[->] (X4.west) \updown -- (PL4);
\draw[->] (X5.east) \updown -- (PL5);
\draw[<-] (PL1) -- (I12);\draw (I12) -- (PL2);
\draw[<-] (PL2) -- (I23);\draw (I23) -- (PL3);
\draw[<-] (PL2) -- (I24);\draw (I24) -- (PL4);
\draw[<-] (PL4) -- (I45);\draw (I45) -- (PL5);
\end{tikzpicture}
\caption{I/O realization of $B_0$, with information set $\{\hat{x}_2, \hat{x}_3, \hat{x}_4, \hat{x}_5\}$.}
\label{NRB0IO}
\end{figure}

As we have seen, the dual of an I/O realization is an I/O realization of the dual code with the complementary information set.  For example, the dual to the  I/O realization of $Z^0$ in Figure \ref{NRZ0IO} is the  I/O realization of the dual zero-sum code $B_0  = (Z^0)^\perp$ shown in Figure \ref{NRB0IO}, which is also controllable and observable.  

\begin{figure}[h]
\setlength{\unitlength}{5pt}
\centering
%%%%%%%%%%%%%%
% Figure 16
%%%%%%%%%%%%%%
\begin{tikzpicture}
\node[box]                   (PL1) {$+$};
\node[box,right=\franksep of PL1 ] (PL2) {$+$};
\node[box,right=\franksep of PL2 ] (PL3) {$+$};
\node[box,below=\franksep of PL2 ] (PL4) {$+$};
\node[box,below=\franksep of PL1 ] (PL5) {$+$};
\node[invert] (I12) at (barycentric cs:PL1=0.5,PL2=0.5) {};
\node[invert] (I15) at (barycentric cs:PL1=0.5,PL5=0.5) {};
\node[invert] (I23) at (barycentric cs:PL2=0.5,PL3=0.5) {};
\node[invert] (I24) at (barycentric cs:PL2=0.5,PL4=0.5) {};
\node[invert] (I34) at (barycentric cs:PL3=0.5,PL4=0.5) {};
\node[invert] (I45) at (barycentric cs:PL4=0.5,PL5=0.5) {};
\draw (PL1) -- (I12);\draw (I12) -- (PL2);
\draw (PL1) -- (I15);\draw (I15) -- (PL5);
\draw (PL2) -- (I23);\draw (I23) -- (PL3);
\draw (PL2) -- (I24);\draw (I24) -- (PL4);
\draw (PL3) -- (I34);\draw (I34) -- (PL4);
\draw (PL4) -- (I45);\draw (I45) -- (PL5);
\end{tikzpicture}
\caption{Realization of the unobservable behavior $\Bf^u$ of the Figure \ref{NRB0} realization of $B_0$.}
\label{UBFF}
\end{figure}
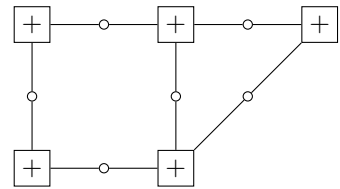

Alternatively, to obtain this I/O realization, we could have started with the realization of $B_0$ in Figure \ref{NRB0}.  The unobservable behavior $\Bf^u$ of this realization, shown in Figure \ref{UBFF}, has dimension $\dim \Bf^u = \beta_1(\G) = 2$;  \ie there are $\beta_1(\G) = 2$ internal degrees of freedom, corresponding to cycles in $Z_1 = \ker \partial$ that do not affect the vertex vector $\hat{\xb}$ (sometimes called  ``local symmetries").  Thus we  could  have   obtained Figure \ref{NRB0IO} by using two local reductions to break these  local  symmetries.

\begin{table*}[h]
\begin{center}
\begin{tabular}{|c|c|c|c|c|c|c|}
\hline
space & dimension  & realization  & observable? & $\Bf^u$ & controllable? & deg.\ unc.\ \\
\hline
        &&&&&&             \\[-0.20cm]
        $Z^0 = \ker d$ & $\beta_0(\G)$ & Fig.\ \ref{NRZ0} & yes & $\{0\}$ & no & $\beta_1(\G)$ \\
        $B^1 = \im d$ & $|E| - \beta_1(\G)$ & Fig.\ \ref{NRB1} & no & Fig.\ \ref{UBF} & yes & $0$ \\
        $Z_1 = \ker \partial$ & $\beta_1(\G)$ & Fig.\ \ref{NRZ1} & yes & $\{0\}$ & no &  $\beta_0(\G)$ \\
        $B_0 = \im \partial$ & $|V| - \beta_0(\G)$ & Fig.\ \ref{NRB0} & no & Fig.\ \ref{UBFF} & yes & $0$ \\
        \hline
\end{tabular}
\end{center}
\caption{Observability and controllability properties of  realizations  of $Z^0, B^1, Z_1$ and $B_0$.}
\label{T1}
\end{table*}

 It is shown in   \cite{F14, FGL12} that if a realization has  $\dim \Bf^u$ ``degrees of unobservability," then its dual realization has $\dim \Bf^u$ ``degrees of uncontrollability."  Thus the fact that the realization of $B_0$ in Figure \ref{NRB0} has $\beta_1(\G) = 2$ degrees of unobservability alternatively follows from the fact that the realization of $Z^0$ in Figure \ref{NRZ0} has $\beta_1(\G) = 2$ degrees of uncontrollability.  (We have found that  it is usually easiest to determine the controllability properties of a realization from the observability properties of its dual.)
 
 \newpage

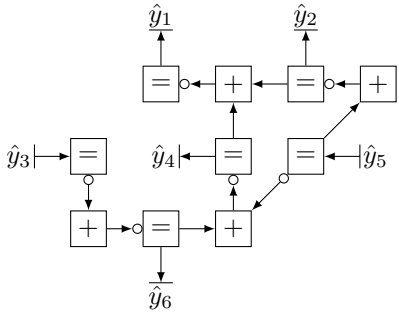
\begin{figure}[h]
\setlength{\unitlength}{5pt}
\centering
%%%%%%%%%%%%%%
% Figure 17
%%%%%%%%%%%%%%
\begin{tikzpicture}
\node[box] (EQ1) {$=$};
\node[box,right=\davesep of EQ1] (PL2) {$+$};
\node[box,right=\davesep of PL2] (EQ2) {$=$};
\node[box,right=\davesep of EQ2] (PL3) {$+$};
\node[box,below=\davesep of PL2] (EQ4) {$=$};
\node[box,below=\davesep of EQ2] (EQ5) {$=$};
\node[box,below=\davesep of EQ4] (PL4) {$+$};
\node[box,left=\davesep of PL4] (EQ6) {$=$};
\node[box,left=\davesep of EQ6] (PL5) {$+$};
\node[box,above=\davesep of PL5] (EQ3) {$=$};
\node[invert,anchor=west]  (IEQ1) at (EQ1.east)  {};
\node[invert,anchor=west]  (IEQ2) at (EQ2.east)  {};
\node[invert,anchor=north] (IEQ3) at (EQ3.south) {};
\node[invert,anchor=north] (IEQ4) at (EQ4.south) {};
\node[invert,anchor=north east] (IEQ5) at (EQ5.south west) {};
\node[invert,anchor=east] (IEQ6) at (EQ6.west) {};
\node[var,above=\davesep of EQ1] (Y1) {$\hat{y}_1$};
\node[var,above=\davesep of EQ2] (Y2) {$\hat{y}_2$};
\node[var,left=\davesep of EQ3] (Y3) {$\hat{y}_3$};
\node[var,left=\davesep of EQ4] (Y4) {$\hat{y}_4$};
\node[var,right=\davesep of EQ5] (Y5) {$\hat{y}_5$};
\node[var,below=\davesep of EQ6] (Y6) {$\hat{y}_6$};
\draw[<-] (Y1.south) \leftright -- (EQ1);
\draw[<-] (Y2.south) \leftright -- (EQ2);
\draw[->] (Y3.east) \updown -- (EQ3);
\draw[<-] (Y4.east) \updown -- (EQ4);
\draw[->] (Y5.west) \updown -- (EQ5);
\draw[<-] (Y6.north) \leftright -- (EQ6);
\draw[->] (PL2) -- (IEQ1);
\draw[<-] (PL2) -- (EQ2);
\draw[<-] (PL2) -- (EQ4);
\draw[->] (PL3) -- (IEQ2);
\draw[<-] (PL3) -- (EQ5);
\draw[<-] (PL4) -- (EQ6);
\draw[->] (PL4) -- (IEQ4);
\draw[<-] (PL4) -- (IEQ5);
\draw[<-] (PL5) -- (IEQ3);
\draw[->] (PL5) -- (IEQ6);
\end{tikzpicture}
\caption{I/O realization of    $Z_1 = \ker \partial$, with information set $\{\hat{y}_3, \hat{y}_5\}$.}
\label{NRZ1IO}
\end{figure}

Finally, the dual to the I/O realization of $B^1$ in Figure \ref{NRB1IO} is a  controllable and observable I/O realization of the dual code $Z_1 = \ker \partial$, which is a systematic $(|E|, \beta_1(\G))$ group code over $\hat{\A}$, using the complementary information set $E_{\bar{\T}}$, as shown in Figure \ref{NRZ1IO}.

Alternatively, we could have started with the realization of $Z_1$ in Figure \ref{NRZ1}.  As we have just discussed, since this realization is the dual to the unobservable realization of $B^1$ in Figure \ref{NRB1} with $\dim \Bf^u = 1$, it must have one ``degree of uncontrollability."
The reader may verify that the realization of Figure \ref{NRZ1} continues to function correctly if any single zero-sum constraint 
(\begin{picture}(2,1)
\put(0.25,-0.25){\framebox(1.5,1.5){$+$}}
\end{picture})
is removed.  For example, Figure \ref{NRZ1IO} is Figure \ref{NRZ1} with the top left  constraint removed.  

We   summarize the  observability and controllability properties of the realizations of $Z^0, B^1, Z_1$ and $B_0$ in Figures \ref{NRZ0},  \ref{NRB1}, \ref{NRZ1} and \ref{NRB0}  in Table \ref{T1} (where ``deg.\ unc." denotes ``degrees of uncontrollability").

\subsection{Homology spaces}

In algebraic topology, \emph{homology spaces} are quotient spaces (resp.\ quotient groups) whose dimensions (resp.\ ``free ranks") are topological invariants of complexes.  We will not delve deeply into this topic, but for future reference we give the homology spaces for the elementary one-dimensional complexes that we have studied in this section, and we exhibit  nice dual realizations for them.

For a  graph $\G = (V, E)$, the \emph{zeroth cohomology space} is defined as $H^0 = Z^0$.  We have seen that if $\G$ is connected, then $\dim Z^0 = 1$, so $\dim H^0 = 1$ for all connected graphs.

The \emph{zeroth homology space} of $\G$ is defined as  the quotient space $H_0 = C_0/B_0$.  Since $B_0 = (Z^0)^\perp = (H^0)^\perp$, $H_0$ is the dual space to $H^0$, and thus has the same dimension, namely $\dim H_0 = 1$, again for all connected graphs.

More generally, it is easy to see that if $\G$ is not connected, but rather consists of $\beta_0(\G) > 1$ connected components, then $H^0 = Z^0$ consists of the direct sum of $\beta_0(\G)$ independent repetition codes, one defined on each of the $\beta_0(\G)$ components of $\G$ (see the discussion in connection with Figure \ref{NRZ0}).  Thus in general $\dim H^0 = \beta_0(\G)$.  Dually, $B_0 = (Z^0)^\perp$ consists of the direct sum of $\beta_0(\G)$ independent zero-sum codes, one defined on each of the $\beta_0(\G)$ components of $\G$, so in general $\dim B_0 = |V| - \beta_0(\G)$, and $\dim H_0 = \beta_0(\G)$.  

For a  graph $\G$, the \emph{first  homology space} of  $\G$ is defined as  $H_1 = Z_1$.  We have seen that  if $\G$ is connected, then $\dim Z_1 = \beta_1(\G)$, so $\dim H_1 = \beta_1(\G)$ for all connected graphs.  Also, the \emph{first cohomology space} of  $\G$ is defined as $H^1 = C^1/B^1$.  Since $B^1 = (Z_1)^\perp = (H_1)^\perp$, $H^1$ is the dual space to $H_1$, and thus has the same dimension, namely $\dim H^1 = \beta_1(\G)$,  for all connected graphs.

More generally, it is easy to see that if $\G$ consists of $\beta_0(\G) > 1$ connected components, then $H_1 = Z_1$ consists of the direct sum of $\beta_0(\G)$ independent codes of dimensions $\beta_1(\G_i)$, one defined on each of the $\beta_0(\G)$ components $\G_i = (V_i, E_i)$ of $\G$.  Thus in general 
$\dim H_1 = \sum_i \beta_1(\G_i) = \sum_i (|E_i| - |V_i| + 1) = |E| - |V| + \beta_0(\G) =  \beta_1(\G).
$
Dually, $B^1 = (Z_1)^\perp$ consists of the direct sum of $\beta_0(\G)$ independent  codes of dimensions $|E_i| - \beta_1(\G_i)$, one defined on each component $\G_i$ of $\G$, so in general $\dim B^1 = |E| - \beta_1(\G)$, and $\dim H^1 = \beta_1(\G)$.

\begin{figure*}[h]
\setlength{\unitlength}{5pt}
\centering
\begin{tabular}{cc}
%%%%%%%%%%%%%%
% Figure 18(a)
%%%%%%%%%%%%%%
\begin{tikzpicture}
\node[box]                   (EQ1) {$=$};
\node[box,right=\franksep of EQ1 ] (EQ2) {$=$};
\node[box,right=\franksep of EQ2 ] (EQ3) {$=$};
\node[box,below=\franksep of EQ2 ] (EQ4) {$=$};
\node[box,below=\franksep of EQ1 ] (EQ5) {$=$};
\draw (EQ1) -- (EQ2);
\draw (EQ1) -- (EQ5);
\draw (EQ2) -- (EQ3);
\draw (EQ2) -- (EQ4);
\draw (EQ3) -- (EQ4);
\draw (EQ4) -- (EQ5);
\end{tikzpicture} &  \hspace{10ex}
%%%%%%%%%%%%%%
% Figure 18(b)
%%%%%%%%%%%%%%
\begin{tikzpicture}
\node[box]                   (PL1) {$+$};
\node[box,right=\franksep of PL1 ] (PL2) {$+$};
\node[box,right=\franksep of PL2 ] (PL3) {$+$};
\node[box,below=\franksep of PL2 ] (PL4) {$+$};
\node[box,below=\franksep of PL1 ] (PL5) {$+$};
\node[invert] (I12) at (barycentric cs:PL1=0.5,PL2=0.5) {};
\node[invert] (I15) at (barycentric cs:PL1=0.5,PL5=0.5) {};
\node[invert] (I23) at (barycentric cs:PL2=0.5,PL3=0.5) {};
\node[invert] (I24) at (barycentric cs:PL2=0.5,PL4=0.5) {};
\node[invert] (I34) at (barycentric cs:PL3=0.5,PL4=0.5) {};
\node[invert] (I45) at (barycentric cs:PL4=0.5,PL5=0.5) {};
\draw (PL1) -- (I12);\draw (I12) -- (PL2);
\draw (PL1) -- (I15);\draw (I15) -- (PL5);
\draw (PL2) -- (I23);\draw (I23) -- (PL3);
\draw (PL2) -- (I24);\draw (I24) -- (PL4);
\draw (PL3) -- (I34);\draw (I34) -- (PL4);
\draw (PL4) -- (I45);\draw (I45) -- (PL5);
\end{tikzpicture} \\
(a) &  \hspace{10ex} (b) \\
\end{tabular}
\caption{Dual realizations whose  behaviors represent (a) the zeroth cohomology space $H^0$ of $\G$; (b) the first homology space $H_1$ of $\G$.}
\label{IB}
\end{figure*}
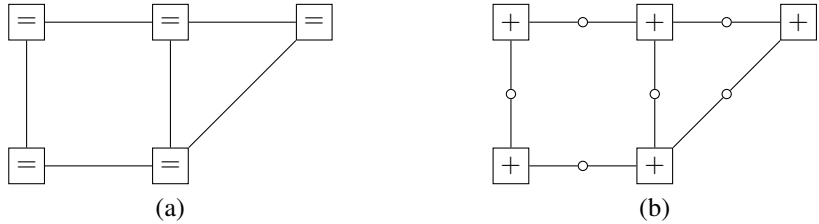

The properties of these homology spaces are elegantly captured by the unobservable behaviors $\Bf^u$ of Figures \ref{UBF} and \ref{UBFF}, respectively, which we  recapitulate in Figure \ref{IB}.  We note that these realizations are each others' duals. 

  It seems to us that the simple,  elegant and dual realizations  of Figure \ref{IB} represent the Platonic essences of  the zeroth cohomology space $H^0$ and the first homology space $H_1$ of a graph $\G$.

\section{Ising-type models}\label{Sec3}

We  now show how to extend these graphical models of  algebraic topology spaces to realize  partition functions of Ising and related models of statistical physics.

\subsection{Ising models and Ising-type models}

In statistical physics, an \emph{Ising model} is defined as follows:
\begin{itemize}
\item Particles are associated with the vertex set $V$ of a connected graph $\G = (V, E)$.
\item Each particle may be in one of two states (often called ``spins").  The state space is $\A = \Z_2$, and the state of the particle at the $v$th vertex is denoted by $x_v \in \A$.  The \emph{configuration space} is thus $\A^V$. 
\item Two particles $v, v' \in V$ interact directly only if they are joined by an edge  $e \in E$; \ie if $(v,v') = (h(e),t(e))$, or \emph{vice versa}, where $\{h(e), t(e)\}$ are the two vertices at the  ends of edge $e$.  Their \emph{interaction energy} $E_e(x_v, x_{v'})$ is $-J_e$ if $x_v = x_{v'}$ and $J_e$ if $x_v \neq x_{v'}$, where $J_e$ is the \emph{interaction strength} for edge $e$.  In other words, 
$E_e(x_v, x_{v'}) = -(-1)^{x_v - x_{v'}}J_e =  -(-1)^{x_{h(e)}- x_{t(e)}} J_e.$
\item The probability of a configuration $\xb \in \A^V$ is then given by the \emph{Boltzmann distribution}
$$p(\xb) = \frac{1}{Z} \prod_{e \in E} \exp \left(\beta J_e(-1)^{x_{h(e)}- x_{t(e)}}\right),$$
where $\beta$ is the \emph{inverse temperature}, and
 the  \emph{partition function} $Z$  is\footnote{This  $Z$ has nothing to do with the topological spaces $Z^0, Z_0, \ldots$ of the previous section.}
$$
Z = \sum_{\xb \in \A^V} \prod_{e \in E} \exp \left(\beta J_e(-1)^{x_{h(e)}- x_{t(e)}}\right).
$$
\end{itemize}

\vspace{-1ex}
Evidently the probability of a configuration $\xb \in \A^V$ depends only on the corresponding edge configuration $\yb(\xb) = d(\xb) = M\xb \in \A^E$.  Thus 
$$p(\xb) = \frac{1}{Z} \prod_{e \in E} \exp \left(\beta J_e(-1)^{y_e(\xb)}\right),$$
where $y_e(\xb) = M_e \xb = x_{h(e)}- x_{t(e)}$.  Moreover,
$$
Z  = \sum_{\xb \in \A^V} \prod_{e \in E} e^{\beta J_e(-1)^{y_e(\xb)}}= |\A| \sum_{\yb \in B^1} \prod_{e \in E} e^{\beta J_e(-1)^{y_e}},
$$
where $B^1 = \im d$, and $|\A| = |\ker d|$ is the number of configurations in $\A^V$ that map to each $\yb \in B^1$.

More generally, we  define an \emph{Ising-type model} as a statistical model based on a graph $\G = (V,E)$ in which 

\begin{itemize}
\item Particles are associated with the vertex set $V$ and have states $x_v \in \A$, where the state space $\A$ may be any finite abelian group;
\item The probability of a configuration $\xb \in \A^V$ depends only on the corresponding edge configuration $\yb(\xb) = d(\xb) = M\xb \in \A^E$, and is given by 
$$p(\xb) = \frac{1}{Z} \prod_{e \in E} f_e\left(y_e(\xb)\right)$$
for some set of \emph{edge-weighting functions} $\{f_e(y_e), e \in E\}$, where
$$Z = \sum_{\xb \in \A^V} \prod_{e \in E} f_e(y_e(\xb)) = |\A| \sum_{\yb \in B^1} \prod_{e \in E} f_e(y_e).$$
\end{itemize}

For example, a \emph{Potts model} is an Ising-type model in which $\A = \Z_q$ for $q > 2$, and $f_e(0) = \exp (\beta J_e)$, while $f_e(y_e) = 1$ for $y_e \neq 0$.  Much broader generalizations are evidently possible, but we do not know to what extent they may have been studied in statistical physics.

The usual object of study is the partition function $Z$ as a function of inverse temperature $\beta$.
In an Ising model, as $\beta \to 0$ (\ie in the high-temperature limit), the weight $w_e(\xb) = \exp (\beta J_e(-1)^{y_e(\xb)})$ tends to 1, independent of $y_e(\xb)$, so $Z \rightarrow |\A|^{|V|}$ and $p(\xb) \rightarrow |\A|^{-|V|}$ for all $\xb \in \A^V$;  \ie in physical terms, the interaction between neighboring vertices disappears.  On the other hand, as $\beta \to \infty$ (\ie in the low-temperature limit), and if $J_e > 0$ (the \emph{ferromagnetic} case), then the weight  $w_e(\xb)$ is large if $y_e = 0$ ($x_{h(e)} = x_{t(e)}$) and small if $y_e \neq 0$ ($x_{h(e)} \neq x_{t(e)}$), which tends to force these two values to agree.  If $\beta \to \infty$ and if $J_e < 0$ (the \emph{antiferromagnetic} case), then the weight  $w_e(\xb)$ becomes small  if $y_e = 0$ ($x_{h(e)} = x_{t(e)}$) and large if $y_e \neq 0$ ($x_{h(e)} \neq x_{t(e)}$), which tends to force these two values to disagree.

\subsection{From normal realizations to edge-weighted NFGs}\label{Sec3.2}

Following \cite{ML13, AV16}, we will now show how the partition function of an Ising-type model may be represented by a normal factor graph (NFG), and in particular by an \emph{edge-weighted NFG} (EWNFG).

{Normal factor graphs}  \cite{L04, AM11, FV11} build on the concepts of normal realizations \cite{F01} and factor graphs \cite{KFL01}.  NFGs are used to model functions that can be represented as real- or complex-valued sums of products, such as partition functions.    (For more on  partition functions of NFGs, see \cite{FV11}.)
  
  The semantics of NFGs are  similar to those of normal realizations.  Again, a \emph{normal factor graph}  is a graphical model based on a graph $G = (V, E, H)$, in which  the edges $e \in E$ represent internal variables $s_e \in \A_e$, and the \emph{half-edges} $h \in H$ represent external variables $a_h \in \A_h$.  However, the vertices $v \in V$ now represent complex-valued functions $f_v(\sb_v, \ab_v)$ of the values $(\sb_v, \ab_v)$ of all variables that correspond to the edges and half-edges that are incident on vertex $v$. 
  
  We will assume that all variable alphabets are finite abelian groups.  The \emph{internal} and \emph{external configuration spaces} are then $\A_E = \prod_E \A_e$ and $\A_H = \prod_H \A_h$, respectively.   
   The NFG then represents the sum of products
  $$
  Z(\ab) = \sum_{\sb \in \A_E} \prod_{v \in V} f_v(\sb_v, \ab_v),
  $$
called the \emph{partition function} (or ``exterior function" \cite{AM11, AV16}, or ``partition sum" \cite{AV16}) of the NFG.
  
  NFGs generalize normal realizations in the following sense.  Given a normal realization based on $G = (V, E, H)$, if each vertex constraint code $\CC_v$ is replaced by its indicator function $\delta_{\CC_v}$ (\ie $\delta_{\CC_v}(\sb_v, \ab_v) = 1$ if $(\sb_v, \ab_v) \in \CC_v$, else $\delta_{\CC_v}(\sb_v, \ab_v) = 0$), then
    $$ Z(\ab) = \sum_{\sb \in \A_E} \delta_\Bf(\sb, \ab), $$
    where $\delta_\Bf(\sb, \ab)$ is the indicator function of the behavior $\Bf = \{(\sb, \ab) \mid \mathrm{all~constraints~satisfied}\}$ of the normal realization.  Thus $Z(\ab) > 0$ if and only if $\ab$ is in the external behavior $\CC$, which is the projection $\CC = \Bf_{|\A_H}$ of $\Bf$ onto the {external configuration space} $\A_H$.  
    
    If all constraint codes $\CC_v$ are group codes, then the behavior and the external behavior are group codes $\Bf \subseteq \A_E \times \A_H$ and $\CC \subseteq \A_H$, respectively.  Moreover, if $\Bf$ is finite, then by the group property the number of elements of $\Bf$ that map to each element of $\CC$ is the same, namely $|\Bf^u| = |\Bf|/|\CC|$, where 
$$\Bf^u = \Bf_{:\A_E} = \left\{(\sb, \zerob) \in \Bf \mid \sb \in \A_E\right\}$$ is the \emph{unobservable behavior} of the realization.  
     Therefore $Z(\ab) = |\Bf^u| \delta_\CC(\ab)$.

      In summary:
   
   \vspace{1ex}
   \noindent
   \textbf{Theorem 6} (\textit{normal realization as an NFG}).  If all  alphabets are finite abelian groups, then a   normal realization with external behavior $\CC$ and unobservable behavior $\Bf^u$ may be interpreted as a  normal factor graph whose partition function is $Z(\ab) = |\Bf^u| \delta_\CC(\ab)$.  \qed \vspace{1ex}
   
   Next, we extend this definition as follows.
An \emph{edge-weighted NFG} consists of a normal realization of a  group code $\CC$ as above, in which all internal functions $\{f_v, v \in V\}$ are indicator functions $\delta_{\CC_v}$ of  group codes $\CC_v$, plus a set $\{f_h, h \in H\}$ of edge-weighting functions $f_h$ attached to each external half-edge $h \in H$ of the normal realization.  The resulting NFG has no external variables, and its partition function is evidently the complex number
$$Z = \sum_{\ab \in \A_H} Z(\ab) \fb(\ab) =  |\Bf^u| \sum_{\ab \in \CC} \fb(\ab),$$
where $\Bf^u$ is the unobservable behavior of the normal realization, and $\fb(\ab) = \prod_{h \in H} f_h(a_h)$. 

As \cite{AV16, ML13} have observed, the partition functions of Ising-type models are naturally represented by such edge-weighted NFGs.  
In particular, the partition function of an {Ising-type model} may be represented as an EWNFG based on a normal realization of  the coboundary space $B^1 = \im d$ of a graph $\G = (V, E)$ over the finite group alphabet $\A$, and an appropriate set of edge-weighting functions $\{f_e(y_e), e \in E\}$.  
The partition function of such a model is thus
$$Z = |\Bf^u| \sum_{\yb \in B^1} \fb(\yb) = |\Bf^u| \sum_{\yb \in B^1} \prod_{e \in E} f_e(y_e),$$
where $|\Bf^u|$ is the size of the unobservable behavior of the normal realization. 

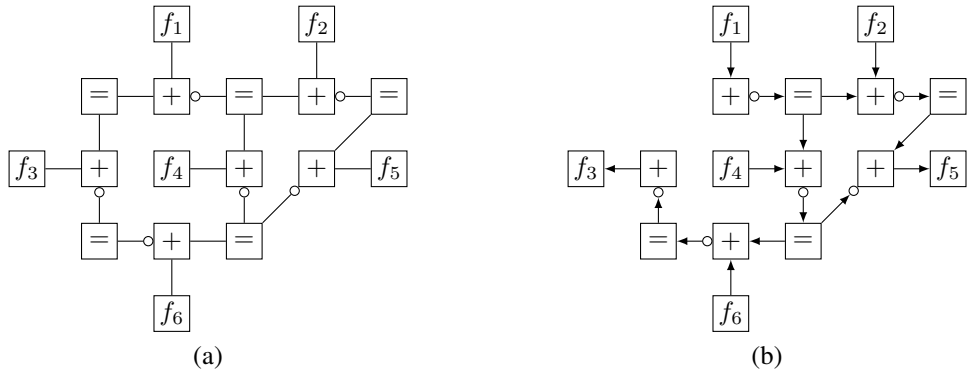
\begin{figure*}[h]
\setlength{\unitlength}{5pt}
\centering
\begin{tabular}{cc}
%%%%%%%%%%%%%%
% Figure 19(a)
%%%%%%%%%%%%%%
\begin{tikzpicture}
\node[box]                  (EQ1) {$=$};
\node[box,right=\davesep of EQ1] (PL1) {$+$};
\node[box,right=\davesep of PL1] (EQ2) {$=$};
\node[box,right=\davesep of EQ2] (PL2) {$+$};
\node[box,right=\davesep of PL2] (EQ3) {$=$};
\node[box,below=\davesep of EQ1] (PL3) {$+$};
\node[box,below=\davesep of EQ2] (PL4) {$+$};
\node[box,below=\davesep of PL2] (PL5) {$+$};
\node[box,below=\davesep of PL3] (EQ5) {$=$};
\node[box,right=\davesep of EQ5] (PL6) {$+$};
\node[box,right=\davesep of PL6] (EQ4) {$=$};
\node[invert,anchor=west]  (IPL1) at (PL1.east)  {};
\node[invert,anchor=west]  (IPL2) at (PL2.east)  {};
\node[invert,anchor=north] (IPL3) at (PL3.south) {};
\node[invert,anchor=north] (IPL4) at (PL4.south) {};
\node[invert,anchor=north east] (IPL5) at (PL5.south west) {};
\node[invert,anchor=east] (IPL6) at (PL6.west) {};
\node[box,above=\davesep of PL1] (Y1) {$f_1$};
\node[box,above=\davesep of PL2] (Y2) {$f_2$};
\node[box,left=\davesep of PL3] (Y3) {$f_3$};
\node[box,left=\davesep of PL4] (Y4) {$f_4$};
\node[box,right=\davesep of PL5] (Y5) {$f_5$};
\node[box,below=\davesep of PL6] (Y6) {$f_6$};
\draw (Y1.south)  -- (PL1);
\draw (Y2.south)  -- (PL2);
\draw (Y3.east)  -- (PL3);
\draw (Y4.east)  -- (PL4);
\draw (Y5.west)  -- (PL5);
\draw (Y6.north) -- (PL6);
\draw (EQ1) -- (PL1);
\draw (EQ1) -- (PL3);
\draw (EQ2) -- (IPL1);
\draw (EQ2) -- (PL2);
\draw (EQ2) -- (PL4);
\draw (EQ3) -- (IPL2);
\draw (EQ3) -- (PL5);
\draw (EQ4) -- (PL6);
\draw (EQ4) -- (IPL4);
\draw (EQ4) -- (IPL5);
\draw (EQ5) -- (IPL3);
\draw (EQ5) -- (IPL6);
\end{tikzpicture}
&  \hspace{10ex}
%%%%%%%%%%%%%%
% Figure 19(b)
%%%%%%%%%%%%%%
\begin{tikzpicture}
\node[box] (PL1) {$+$};
\node[box,right=\davesep of PL1] (EQ2) {$=$};
\node[box,right=\davesep of EQ2] (PL2) {$+$};
\node[box,right=\davesep of PL2] (EQ3) {$=$};
\node[box,below=\davesep of EQ2] (PL4) {$+$};
\node[box,below=\davesep of PL2] (PL5) {$+$};
\node[box,below=\davesep of PL4] (EQ4) {$=$};
\node[box,left=\davesep of EQ4] (PL6) {$+$};
\node[box,left=\davesep of PL6] (EQ5) {$=$};
\node[box,above=\davesep of EQ5] (PL3) {$+$};
\node[invert,anchor=west]  (IPL1) at (PL1.east)  {};
\node[invert,anchor=west]  (IPL2) at (PL2.east)  {};
\node[invert,anchor=north] (IPL3) at (PL3.south) {};
\node[invert,anchor=north] (IPL4) at (PL4.south) {};
\node[invert,anchor=north east] (IPL5) at (PL5.south west) {};
\node[invert,anchor=east] (IPL6) at (PL6.west) {};
\node[box,above=\davesep of PL1] (Y1) {$f_1$};
\node[box,above=\davesep of PL2] (Y2) {$f_2$};
\node[box,left=\davesep of PL3] (Y3) {$f_3$};
\node[box,left=\davesep of PL4] (Y4) {$f_4$};
\node[box,right=\davesep of PL5] (Y5) {$f_5$};
\node[box,below=\davesep of PL6] (Y6) {$f_6$};
\draw[->] (Y1.south)  -- (PL1);
\draw[->] (Y2.south) -- (PL2);
\draw[<-] (Y3.east) -- (PL3);
\draw[->] (Y4.east) -- (PL4);
\draw[<-] (Y5.west) -- (PL5);
\draw[->] (Y6.north) -- (PL6);
\draw[<-] (EQ2) -- (IPL1);
\draw[->] (EQ2) -- (PL2);
\draw[->] (EQ2) -- (PL4);
\draw[<-] (EQ3) -- (IPL2);
\draw[->] (EQ3) -- (PL5);
\draw[->] (EQ4) -- (PL6);
\draw[<-] (EQ4) -- (IPL4);
\draw[->] (EQ4) -- (IPL5);
\draw[->] (EQ5) -- (IPL3);
\draw[<-] (EQ5) -- (IPL6);
\end{tikzpicture}\\
(a) &  \hspace{10ex} (b) \\
\end{tabular}
\caption{EWNFGs for scaled partition function of Ising-type model based on normal realizations of $B^1 = \im d$ in (a) Figure \ref{NRB1};  (b)  Figure \ref{NRB1IO}.}
\label{NFGE}
\end{figure*}

For example, Figure \ref{NFGE} shows the EWNFGs derived from the normal realizations of $B^1 = \im d$ of Figures \ref{NRB1} and \ref{NRB1IO}, respectively (recall that Figure  \ref{NRB1IO} is an I/O realization using the information set $\yb_\T = \{y_1, y_2, y_4, y_6\}$).

We  recall that for the normal realization underlying Figure \ref{NFGE}(a), we have $|\Bf^u| = |\A|$, whereas for that of Figure \ref{NFGE}(b),  $|\Bf^u| = 1$.  Thus  the partition function of the EWNFG of Figure \ref{NFGE}(a) is
  $ |\A| \sum_{\yb \in B^1} \fb(\yb) = Z$, whereas the partition function of
Figure \ref{NFGE}(b) is $ \sum_{\yb \in B^1} \fb(\yb) = Z/|\A|$.  The two EWNFGs are thus equivalent up to a scale factor of $|\A|$.

Notice that such EWNFGs alternatively model the communications scenario in which the possible transmitted sequences are  codewords $\yb$ from the linear code $B^1$, and the relative (unscaled) likelihood of each possible symbol $y_e \in \A$ is $f_e(y_e)$.\footnote{Indeed, it was the recognition of this analogy by Sourlas \cite{S89}  in 1989 that led to the first connections between coding theory and the Ising models of statistical physics.

Hakimi and Bredeson  \cite{HB68} conducted an early investigation into  whether graphical binary linear codes such as $B^1$ could be useful for data communications.}

The dimension of $B^1$ is $\dim B^1 = |E| - \beta_1(\G) = |V| - 1$, the number of edges in a spanning tree $\T$ of $\G$.  Thus $Z$ may be computed as in \cite{Mex} by choosing a spanning tree $\T \subseteq \G$, letting  $\yb_\T$ run freely through $\A^{E_\T}$,  extending each $\yb_\T$ to the corresponding    $\yb \in B^1$ via a linear transformation,  computing $f_e(\yb) = \prod_{e \in E} f_e(y_e)$,  summing  these contributions, and finally multiplying by $|\A|$.
If we use an  I/O realization of $B_1$ as in Figure \ref{NRB1IO}, then the extension of
each $\yb_\T$ to the corresponding    $\yb \in B^1$ occurs ``automatically" by propagation through the I/O realization.

Molkaraie \cite{M15, M16} has proposed an importance-sampling algorithm to estimate the partition function $Z$, using the information set $\yb_\T$.  The idea is to choose a series of samples $\yb_\T \in \A^{E_\T}$   according to an auxiliary probability distribution $p(\yb_\T)$, extend each such  $\yb_\T$ to the corresponding $\yb \in B^1$, and  then compute $f_e(\yb)$ for each sample.
For the auxiliary probability distribution, he proposes
$p(\yb_\T) = f_e(\yb_\T)/Z_\T$, where
 $$f_e(\yb_\T) = \prod_{e \in E_\T} f_e(y_e);  \qquad
 Z_\T = \sum_{\yb_\T \in \A^{E_\T}} f_e(\yb_\T). $$ 
Again, an  I/O realization of $B_1$ could be used to compute this extension of $\yb_\T$ to $\yb$ ``automatically."
 
 \subsection{Dual realizations}
  
 Molkaraie and Loeliger \cite{ML13}  observed that it is sometimes easier to compute or estimate the partition function $Z$ of an EWNFG by using the dual EWNFG.  We now revisit this observation in the context of this paper.
 
The  \emph{normal factor graph duality theorem} (NFGDT) \cite{AM11, F11b, FV11} says that the dual of  an NFG whose partition function is  $Z$ is an NFG whose partition function is the Fourier transform $\hat{Z}$ of $Z$, up to a certain scale factor.  In the Appendix, we give the simplest proof we know of this very powerful and general result, using a simple Edge Replacement Lemma.

The dual of an EWNFG based on a normal realization of some group or linear code $\CC$ and edge-weighting functions $\{f_e, e \in E\}$ is evidently an EWNFG based on the dual normal realization  with the Fourier-transformed edge weighting functions $\{\hat{f}_e, e \in E\}$.  

Since the partition function of an EWNFG is  a complex number $Z$, its 
Fourier transform is simply $\hat{Z} = Z$; \ie the Fourier transform of a number (namely, a complex-valued function of no variables) is  that number.

In the Appendix, we consider interpreting a dual normal realization as an NFG.  Considering all relevant scale factors, we show that

   \vspace{1ex}
   \noindent
   \textbf{Theorem 7} (\textit{Dual normal realization as an NFG}).  The dual of a finite abelian group normal realization with  behavior $\Bf$ and external behavior $\CC$ may be interpreted as an NFG with partition function  $|\Bf||\A_E||\CC_V|^{-1} \delta_{\CC^\perp}(\hat{\ab}),$
     where $|\A_E| = \prod_E |\A_e|$ and $\CC_V = \prod_V |\CC_v|$.  \qed \vspace{1ex}

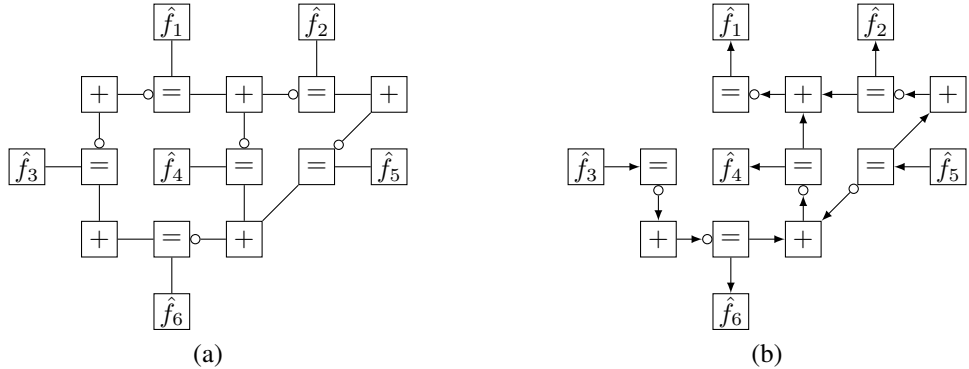
\begin{figure*}[h]
\setlength{\unitlength}{5pt}
\centering
\begin{tabular}{cc}
%%%%%%%%%%%%%%
% Figure 20(a)
%%%%%%%%%%%%%%
\begin{tikzpicture}
\node[box]                  (PL1) {$+$};
\node[box,right=\davesep of PL1] (EQ1) {$=$};
\node[box,right=\davesep of EQ1] (PL2) {$+$};
\node[box,right=\davesep of PL2] (EQ2) {$=$};
\node[box,right=\davesep of EQ2] (PL3) {$+$};
\node[box,below=\davesep of PL1] (EQ3) {$=$};
\node[box,below=\davesep of PL2] (EQ4) {$=$};
\node[box,below=\davesep of EQ2] (EQ5) {$=$};
\node[box,below=\davesep of EQ3] (PL5) {$+$};
\node[box,right=\davesep of PL5] (EQ6) {$=$};
\node[box,right=\davesep of EQ6] (PL4) {$+$};
\node[invert,anchor=east]  (IEQ1) at (EQ1.west)  {};
\node[invert,anchor=east]  (IEQ2) at (EQ2.west)  {};
\node[invert,anchor=south] (IEQ3) at (EQ3.north) {};
\node[invert,anchor=south] (IEQ4) at (EQ4.north) {};
\node[invert,anchor=south west] (IEQ5) at (EQ5.north east) {};
\node[invert,anchor=west] (IEQ6) at (EQ6.east) {};
\node[box,above=\davesep of EQ1] (Y1) {$\hat{f}_1$};
\node[box,above=\davesep of EQ2] (Y2) {$\hat{f}_2$};
\node[box,left=\davesep of EQ3] (Y3) {$\hat{f}_3$};
\node[box,left=\davesep of EQ4] (Y4) {$\hat{f}_4$};
\node[box,right=\davesep of EQ5] (Y5) {$\hat{f}_5$};
\node[box,below=\davesep of EQ6] (Y6) {$\hat{f}_6$};
\draw (Y1.south)  -- (EQ1);
\draw (Y2.south)  -- (EQ2);
\draw (Y3.east)  -- (EQ3);
\draw (Y4.east)  -- (EQ4);
\draw (Y5.west)  -- (EQ5);
\draw (Y6.north) -- (EQ6);
\draw (PL1) -- (IEQ1);
\draw (PL1) -- (IEQ3);
\draw (PL2) -- (IEQ2);
\draw (PL2) -- (IEQ4);
\draw (PL3) -- (IEQ5);
\draw (PL4) -- (IEQ6);
\draw (PL2) -- (EQ1);
\draw (PL3) -- (EQ2);
\draw (PL4) -- (EQ4);
\draw (PL4) -- (EQ5);
\draw (PL5) -- (EQ3);
\draw (PL5) -- (EQ6);
\end{tikzpicture}
&  \hspace{10ex}
%%%%%%%%%%%%%%
% Figure 20(b)
%%%%%%%%%%%%%%
\begin{tikzpicture}
\node[box] (EQ1) {$=$};
\node[box,right=\davesep of EQ1] (PL2) {$+$};
\node[box,right=\davesep of PL2] (EQ2) {$=$};
\node[box,right=\davesep of EQ2] (PL3) {$+$};
\node[box,below=\davesep of PL2] (EQ4) {$=$};
\node[box,below=\davesep of EQ2] (EQ5) {$=$};
\node[box,below=\davesep of EQ4] (PL4) {$+$};
\node[box,left=\davesep of PL4] (EQ6) {$=$};
\node[box,left=\davesep of EQ6] (PL5) {$+$};
\node[box,above=\davesep of PL5] (EQ3) {$=$};
\node[invert,anchor=west]  (IEQ1) at (EQ1.east)  {};
\node[invert,anchor=west]  (IEQ2) at (EQ2.east)  {};
\node[invert,anchor=north] (IEQ3) at (EQ3.south) {};
\node[invert,anchor=north] (IEQ4) at (EQ4.south) {};
\node[invert,anchor=north east] (IEQ5) at (EQ5.south west) {};
\node[invert,anchor=east] (IEQ6) at (EQ6.west) {};
\node[box,above=\davesep of EQ1] (Y1) {$\hat{f}_1$};
\node[box,above=\davesep of EQ2] (Y2) {$\hat{f}_2$};
\node[box,left=\davesep of EQ3] (Y3) {$\hat{f}_3$};
\node[box,left=\davesep of EQ4] (Y4) {$\hat{f}_4$};
\node[box,right=\davesep of EQ5] (Y5) {$\hat{f}_5$};
\node[box,below=\davesep of EQ6] (Y6) {$\hat{f}_6$};
\draw[<-] (Y1.south) -- (EQ1);
\draw[<-] (Y2.south) -- (EQ2);
\draw[->] (Y3.east) -- (EQ3);
\draw[<-] (Y4.east) -- (EQ4);
\draw[->] (Y5.west) -- (EQ5);
\draw[<-] (Y6.north) -- (EQ6);
\draw[->] (PL2) -- (IEQ1);
\draw[<-] (PL2) -- (EQ2);
\draw[<-] (PL2) -- (EQ4);
\draw[->] (PL3) -- (IEQ2);
\draw[<-] (PL3) -- (EQ5);
\draw[<-] (PL4) -- (EQ6);
\draw[->] (PL4) -- (IEQ4);
\draw[<-] (PL4) -- (IEQ5);
\draw[<-] (PL5) -- (IEQ3);
\draw[->] (PL5) -- (IEQ6);
\end{tikzpicture} \\
(a) &  \hspace{10ex} (b) \\
\end{tabular}
\caption{Dual EWNFGs for scaled Fourier transform of partition function of Ising-type model, based on normal realizations of $Z_1 = \ker \partial$ in (a) Figure \ref{NRZ1};  (b)  Figure \ref{NRZ1IO}.}
\label{DNFGE}
\end{figure*}

As discussed in the Appendix, Theorems 6 and 7 imply that:
\begin{itemize}
\item[(a)] The external behavior of the dual normal realization is $\CC^\perp$.  Thus, when all alphabets are finite abelian groups, we obtain the normal realization duality theorem (NRDT) as a corollary.
\item[(b)] The size of the unobservable behavior of the dual normal realization is $|\hat{\Bf}^u| = |\Bf||\A_E||\CC_V|^{-1}$.  Thus the dual NR is observable and the NR is controllable if and only if $|\Bf||\A_E||\CC_V|^{-1} = 1$.  Interestingly, this yields the \emph{controllability test} of \cite{FGL12} for the finite abelian group case.
\end{itemize}

For Ising-type models,  dualization of a normal realization of $B^1 = \im d$ as in Figure \ref{NRB1} or \ref{NRB1IO} gives a normal realization of $Z_1 = \ker \partial = (B^1)^\perp$ as in Figure \ref{NRZ1} or Figure \ref{NRZ1IO}.  Correspondingly, dualization of the corresponding edge-weighted NFGs in Figure \ref{NFGE} yields the dual EWNFGs shown in Figure \ref{DNFGE}.

  We recall that Figure  \ref{NRZ1IO} is an I/O realization using the information set $\yb_{\bar{\T}} = \{\hat{y}_3, \hat{y}_5\}$.
The dual partition function may be computed or sampled like the primal function, by letting $\yb_{\bar{\T}}$ range freely on the complement $E_{\bar{\T}}$ of the edge set $E_\T$ of any spanning tree $\T$ of $\G$.  Thus if we choose the same  tree $\T$ for the primal and dual realizations, then the two information sets are complements of each other.
 
 In the Appendix, we compute the  scale factor  for dual Ising-type models based on a graph $\G = (V,E)$.  If the partition function of the primal edge-weighted NFG as in Figure  \ref{NFGE}(a) is  $Z$, then $\hat{Z} = Z$, and the partition function of the dual edge-weighted NFG as in Figure  \ref{DNFGE}(a) is  
 $$|\A|^{|E| - |V|} \hat{Z} = |\A|^{\beta_1(\G) - 1} \hat{Z}.$$
 (This result was  derived previously by Molkaraie  \cite{Mex}.)\footnote{
Since the scale factor is independent of edge weights, one way of computing it is to calculate partition functions for the particular case where $f_e(y_e) = 1$ for all $y_e \in \A$ and for all $e \in E$;  then all configurations in Figure \ref{NFGE}(a) have weight 1, so $Z = |\A|^{|V|}$.  Now $\hat{f}_e(\hat{y}_e) = |\A|\delta(y_e)$, so only the all-zero configuration contributes to $\hat{Z}$, with weight $|\A|$ for each edge $e \in E$, so the partition function of Figure  \ref{DNFGE}(a) is  $|\A|^{|E|}$.  Since $|\A|^{|E|}$ is $|\A|^{|E| - |V|}$ times $Z = |\A|^{|V|}$,  the scale factor must be $|\A|^{|E| - |V|}$.}

Dualization turns hard constraints into soft constraints, and \emph{vice versa}.  For example, an interaction weight function $f_e(y_e)$ represents a \emph{strict (equality) constraint} if $f_e(y_e) \propto \delta_{\{0\}}(y_e)$, for then only configurations $\yb$ with $y_e = 0$ contribute to the partition function.  On the other hand, $f_e(y_e)$ represents \emph{no constraint} if $f_e(y_e) \propto 1$ for all $y_e$, for then it makes the same contribution for every configuration $\yb$.  Since the dual code to $\{0\}$ is the universe code $\A$,  the Fourier transform of a strict-constraint function is a no-constraint function, and \emph{vice versa}.
Similarly, in an Ising-type model, the Fourier transform of a low-temperature interaction weight function $f_e(y_e)$ is a high-temperature weight function, and \emph{vice versa}. 
 For this reason, an expression for $\hat{Z}$ is sometimes called a \emph{high-temperature expansion} of $Z$. High-temperature constraints are softer than low-temperature constraints, and have fewer long-range correlations, so convergence of Monte Carlo estimates is faster and less random \cite{ML13}.

\subsection{Example:  Single-cycle graph}

Following \cite{ML13}, we now give an example of  dual realizations of a partition function $Z$ and its Fourier transform $\hat{Z}$ on a  single-cycle graph.  In statistical physics, a single-cycle graph arises in a one-dimensional (1D) Ising-type model with periodic boundary conditions.

A single-cycle graph $\G$ of length $n$ has $n$ vertices $v_i, i \in \Z_n$, and $n$ edges $e_i, i \in \Z_n$, such that edge $e_i$ connects vertices $v_i$ and $v_{i+1}$ (with index arithmetic  in $\Z_n$, so  $e_{n-1}$ connects  $v_{n-1}$ and $v_0$).  

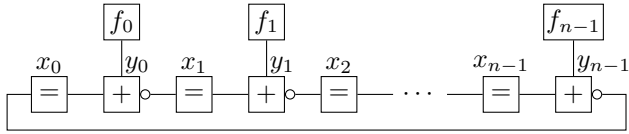
\begin{figure}[h]
\setlength{\unitlength}{5pt}
\centering
%%%%%%%%%%%%%%
% Figure 21
%%%%%%%%%%%%%%
\begin{tikzpicture}
\node[box] (EQ0) {$=$};
\node[box,right=\davesep of EQ0] (PL0) {$+$};
\node[box,right=\davesep of PL0] (EQ1) {$=$};
\node[box,right=\davesep of EQ1] (PL1) {$+$};
\node[box,right=\davesep of PL1] (EQ2) {$=$};
\node[right=\davesep of EQ2] (CDOTS) {$\cdots$};
\node[box,right=\davesep of CDOTS] (EQn) {$=$};
\node[box,right=\davesep of EQn] (PLn) {$+$};
\node[invert,anchor=west] (IPL0) at (PL0.east) {};
\node[invert,anchor=west] (IPL1) at (PL1.east) {};
\node[invert,anchor=west] (IPLn) at (PLn.east) {};
\node[box,above=\davesep of PL0] (F0) {$f_0$};
\node[box,above=\davesep of PL1] (F1) {$f_1$};
\node[box,above=\davesep of PLn] (Fn) {$f_{n-1}$};
\draw (EQ0) -- (PL0);
\draw (IPL0) -- (EQ1);
\draw (EQ1) -- (PL1);
\draw (IPL1) -- (EQ2);
\draw (EQ2) -- (CDOTS);
\draw (CDOTS) -- (EQn);
\draw (EQn) -- (PLn);
\draw (F0) -- (PL0);
\draw (F1) -- (PL1);
\draw (Fn) -- (PLn);
\coordinate[xshift=-2ex] (L) at (EQ0.west);
\coordinate[xshift=2ex] (R) at (IPLn.east);
\draw (IPLn.east) -- (R) -- ++(0,-\davesep) -| (L) -- (EQ0);
\node[inner sep=1pt,anchor=south] at (EQ0.north) {$x_0$};
\node[inner sep=1pt,anchor=south] at (EQ1.north) {$x_1$};
\node[inner sep=1pt,anchor=south] at (EQ2.north) {$x_2$};
\node[inner sep=1pt,anchor=south] at (EQn.north) {$x_{n-1}$};
\node[inner sep=1pt,anchor=south west] at (PL0.north) {$y_0$};
\node[inner sep=1pt,anchor=south west] at (PL1.north) {$y_1$};
\node[inner sep=1pt,anchor=south west] at (PLn.north) {$y_{n-1}$};
\end{tikzpicture}
\caption{NFG for partition function of 1D Ising-type model with periodic boundary conditions.}
\label{SCG}
\end{figure}

Figure \ref{SCG} shows an NFG with partition function $Z$ for a 1D Ising-type model of length $n$ over $\A = \Z_q$ with periodic boundary conditions.  The realization is based on an image realization of $B^1 = \im d$, which  is the $(n, n-1)$ zero-sum code over $\Z_q$.

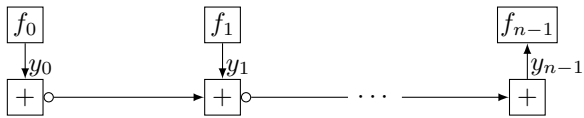
\begin{figure}[h]
\setlength{\unitlength}{5pt}
\centering
%%%%%%%%%%%%%%
% Figure 22
%%%%%%%%%%%%%%
\begin{tikzpicture}
\node[box] (PL0) {$+$};
\node[box,right=1.5\franksep of PL0] (PL1) {$+$};
\node[right=\franksep of PL1] (CDOTS) {$\cdots$};
\node[box,right=\franksep of CDOTS] (PLn) {$+$};
\node[invert,anchor=west] (IPL0) at (PL0.east) {};
\node[invert,anchor=west] (IPL1) at (PL1.east) {};
\node[box,above=\davesep of PL0] (F0) {$f_0$};
\node[box,above=\davesep of PL1] (F1) {$f_1$};
\node[box,above=\davesep of PLn] (Fn) {$f_{n-1}$};
\draw[->] (IPL0) -- (PL1);
\draw (IPL1) -- (CDOTS);
\draw[->] (CDOTS) -- (PLn);
\draw[->] (F0) -- (PL0);
\draw[->] (F1) -- (PL1);
\draw[<-] (Fn) -- (PLn);
\node[inner sep=1pt,anchor=south west] at (PL0.north) {$y_0$};
\node[inner sep=1pt,anchor=south west] at (PL1.north) {$y_1$};
\node[inner sep=1pt,anchor=south west] at (PLn.north) {$y_{n-1}$};
\end{tikzpicture}
\caption{NFG for 1D Ising-type model with fixed boundary conditions $x_0 = x_n = 0$.}
\label{SCGIO}
\end{figure}

Since $Z^0 = \ker d$ is the $(n, 1)$ repetition code over $\A$, every codeword of $B^1 = \im d$ is the image of  $|\A|$ different vertex vectors $\xb + a \oneb, a \in \A$.  Thus, if we like, we may fix any vertex variable to zero, say ${x}_0 = 0$, without affecting the partition function, up to a scale factor of $|\A|$.  We may also replace each repetition constraint of degree 2 plus its neighboring sign inverter by an inverting edge, since 
\begin{picture}(11,1)
\put(0.8, 0.5){\line(1,0){1}}
\put(1.8,-0.25){\framebox(1.5,1.5){$=$}}
\put(0, 0.15){$\circ$}
\put(3.3, 0.5){\line(1,0){1}}
\put(5, 0.2){$=$}
\put(8.05, 0.5){\line(1,0){1.2}}
\put(7.5, 0.15){$\circ$}
\put(9.25, 0.5){\line(1,0){1.25}}
\end{picture}.
This results in the observable I/O realization of Figure \ref{SCGIO}, whose underlying graph $\T$, a spanning tree of $\G$, is a cycle-free chain graph.
This shows that a 1D Ising-type model defined on $\T$ with fixed boundary conditions (\ie $x_0 = x_n = 0$) has the same partition function (up to a scale factor of $|\A|$) as the same model with periodic boundary conditions (\ie $x_0 = x_n$).

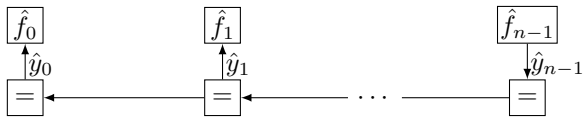
\begin{figure}[h]
\setlength{\unitlength}{5pt}
\centering
%%%%%%%%%%%%%%
% Figure 23
%%%%%%%%%%%%%%
\begin{tikzpicture}
\node[box] (EQ0) {$=$};
\node[box,right=1.5\franksep of EQ0] (EQ1) {$=$};
\node[right=\franksep of EQ1] (CDOTS) {$\cdots$};
\node[box,right=\franksep of CDOTS] (EQn) {$=$};
\node[box,above=\davesep of EQ0] (F0) {$\hat{f}_0$};
\node[box,above=\davesep of EQ1] (F1) {$\hat{f}_1$};
\node[box,above=\davesep of EQn] (Fn) {$\hat{f}_{n-1}$};
\draw[<-] (EQ0) -- (EQ1);
\draw[<-] (EQ1) -- (CDOTS);
\draw (CDOTS) -- (EQn);
\draw[<-] (F0) -- (EQ0);
\draw[<-] (F1) -- (EQ1);
\draw[->] (Fn) -- (EQn);
\node[inner sep=1pt,anchor=south west] at (EQ0.north) {$\hat{y}_0$};
\node[inner sep=1pt,anchor=south west] at (EQ1.north) {$\hat{y}_1$};
\node[inner sep=1pt,anchor=south west] at (EQn.north) {$\hat{y}_{n-1}$};
\end{tikzpicture} 
\caption{Dual NFG for 1D Ising-type model.}
\label{SCGD}
\end{figure}

Figure \ref{SCGD} shows the dual NFG, whose partition function is (up to scale)  the Fourier transform $\hat{Z}$ of the partition function $Z$ of Figure \ref{SCG}.  The underlying normal realization is a dual I/O realization of $Z_1 = \ker \partial$,  the $(n, 1)$ repetition code over $\hat{\A}$. Thus
$  \hat{Z} \propto \sum_{\hat{y} \in \hat{\A}} \prod_{i} \hat{f}_i(\hat{y}).  $
Computing the partition function of the dual NFG is clearly much easier than computing $Z$. 

\subsection{Ising-type models with an external field}

An Ising-type model may be generalized to include an external field  as follows \cite{M15}.
 In addition to the interaction energies $E_e(x_v, x_{v'})$,  there is an \emph{external field energy} $E_v(x_v)$ for  each particle $v \in V$.  
The \emph{external field weight function} is   defined as $g_v(x_v) = \exp\bigl( -\beta E_v(x_v)\bigr)$, where $\beta > 0$ is again the {inverse temperature}.
 The probability of a configuration $\xb \in \A^V$ is then given by the {Boltzmann distribution}
$$p(\xb) = \frac{1}{Z} \left[\prod_{e \in E} f_e(x_{h(e)} - x_{t(e)})\right]\left[ \prod_{v \in V} g_v(x_v) \right],$$
where the  {partition function} is now defined as
$$
Z = \sum_{\xb \in \A^V} \left[\prod_{e \in E} f_e(x_{h(e)} - x_{t(e)})\right]\left[ \prod_{v \in V} g_v(x_v) \right].
$$

\begin{figure}[h]
\setlength{\unitlength}{5pt}
\centering
%%%%%%%%%%%%%%
%% Figure 24
%%%%%%%%%%%%%%
\begin{tikzpicture}
\node[box]                  (EQ1) {$=$};
\node[box,right=\davesep of EQ1] (PL1) {$+$};
\node[box,right=\davesep of PL1] (EQ2) {$=$};
\node[box,right=\davesep of EQ2] (PL2) {$+$};
\node[box,right=\davesep of PL2] (EQ3) {$=$};
\node[box,below=\davesep of EQ1] (PL3) {$+$};
\node[box,below=\davesep of EQ2] (PL4) {$+$};
\node[box,below=\davesep of PL2] (PL5) {$+$};
\node[box,below=\davesep of PL3] (EQ5) {$=$};
\node[box,right=\davesep of EQ5] (PL6) {$+$};
\node[box,right=\davesep of PL6] (EQ4) {$=$};
\node[invert,anchor=west]  (IPL1) at (PL1.east)  {};
\node[invert,anchor=west]  (IPL2) at (PL2.east)  {};
\node[invert,anchor=north] (IPL3) at (PL3.south) {};
\node[invert,anchor=north] (IPL4) at (PL4.south) {};
\node[invert,anchor=north east] (IPL5) at (PL5.south west) {};
\node[invert,anchor=east] (IPL6) at (PL6.west) {};
\node[box,left=\davesep of EQ1] (X1) {$g_1$};
\node[box,above=\davesep of EQ2] (X2) {$g_2$};
\node[box,right=\davesep of EQ3] (X3) {$g_3$};
\node[box,right=\davesep of EQ4] (X4) {$g_4$};
\node[box,left=\davesep of EQ5] (X5) {$g_5$};
\node[box,above=\davesep of PL1] (Y1) {$f_1$};
\node[box,above=\davesep of PL2] (Y2) {$f_2$};
\node[box,left=\davesep of PL3] (Y3) {$f_3$};
\node[box,left=\davesep of PL4] (Y4) {$f_4$};
\node[box,right=\davesep of PL5] (Y5) {$f_5$};
\node[box,below=\davesep of PL6] (Y6) {$f_6$};
\draw[->] (X1.east) -- (EQ1);
\draw[->] (X2.south) -- (EQ2);
\draw[->] (X3.west) -- (EQ3);
\draw[->] (X4.west) -- (EQ4);
\draw[->] (X5.east) -- (EQ5);
\draw[<-] (Y1.south) -- (PL1);
\draw[<-] (Y2.south) -- (PL2);
\draw[<-] (Y3.east) -- (PL3);
\draw[<-] (Y4.east) -- (PL4);
\draw[<-] (Y5.west) -- (PL5);
\draw[<-] (Y6.north) -- (PL6);
\draw[->] (EQ1) -- (PL1);
\draw[->] (EQ1) -- (PL3);
\draw[->] (EQ2) -- (IPL1);
\draw[->] (EQ2) -- (PL2);
\draw[->] (EQ2) -- (PL4);
\draw[->] (EQ3) -- (IPL2);
\draw[->] (EQ3) -- (PL5);
\draw[->] (EQ4) -- (PL6);
\draw[->] (EQ4) -- (IPL4);
\draw[->] (EQ4) -- (IPL5);
\draw[->] (EQ5) -- (IPL3);
\draw[->] (EQ5) -- (IPL6);
\end{tikzpicture}
\caption{EWNFG for partition function of Ising-type model with an external field, based on normal realization of I/O behavior $W^{01} = \{({\xb}, d({\xb}) \mid {\xb} \in \A^V\}$ in Figure \ref{NRIOD}.}
\label{NFGEE}
\end{figure}
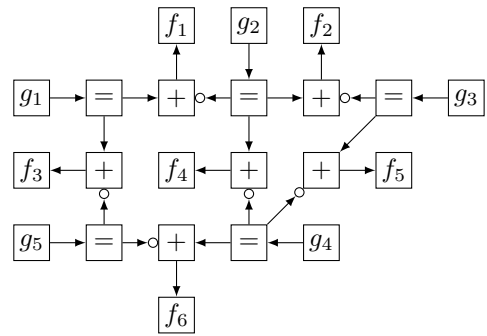

This partition function may be realized by starting with a realization of $Z$ as in Figure \ref{NFGE}(a), and attaching to each  vertex $v \in V$ a function vertex representing the external field weight function $g_v({x}_v)$, as shown in Figure \ref{NFGEE}.  The resulting EWNFG  has no external variables, and evidently realizes  the partition function of the model  including an external field. 

The normal realization underlying this EWNFG is  that of the I/O behavior $W^{01} = \{({\xb}, d({\xb}) \mid {\xb} \in \A^V\}$ of the coboundary operator $d$,  as  illustrated in Figure \ref{NRIOD}.  This behavior is a systematic $(|E| + |V|, |V|)$ group code over $\A$, for which the vertex set $V$ is an information set.

\begin{figure}[h]
\setlength{\unitlength}{5pt}
\centering
%%%%%%%%%%%%%%
% Figure 25
%%%%%%%%%%%%%%
\begin{tikzpicture}
\node[box]                  (PL1) {$+$};
\node[box,right=\davesep of PL1] (EQ1) {$=$};
\node[box,right=\davesep of EQ1] (PL2) {$+$};
\node[box,right=\davesep of PL2] (EQ2) {$=$};
\node[box,right=\davesep of EQ2] (PL3) {$+$};
\node[box,below=\davesep of PL1] (EQ3) {$=$};
\node[box,below=\davesep of PL2] (EQ4) {$=$};
\node[box,below=\davesep of EQ2] (EQ5) {$=$};
\node[box,below=\davesep of EQ3] (PL5) {$+$};
\node[box,right=\davesep of PL5] (EQ6) {$=$};
\node[box,right=\davesep of EQ6] (PL4) {$+$};
\node[invert,anchor=east]  (IEQ1) at (EQ1.west)  {};
\node[invert,anchor=east]  (IEQ2) at (EQ2.west)  {};
\node[invert,anchor=south] (IEQ3) at (EQ3.north) {};
\node[invert,anchor=south] (IEQ4) at (EQ4.north) {};
\node[invert,anchor=south west] (IEQ5) at (EQ5.north east) {};
\node[invert,anchor=west] (IEQ6) at (EQ6.east) {};
\node[box,left=\davesep of PL1] (X1) {$\hat{g}_1$};
\node[box,above=\davesep of PL2] (X2) {$\hat{g}_2$};
\node[box,right=\davesep of PL3] (X3) {$\hat{g}_3$};
\node[box,right=\davesep of PL4] (X4) {$\hat{g}_4$};
\node[box,left=\davesep of PL5] (X5) {$\hat{g}_5$};
\node[box,above=\davesep of EQ1] (Y1) {$\hat{f}_1$};
\node[box,above=\davesep of EQ2] (Y2) {$\hat{f}_2$};
\node[box,left=\davesep of EQ3] (Y3) {$\hat{f}_3$};
\node[box,left=\davesep of EQ4] (Y4) {$\hat{f}_4$};
\node[box,right=\davesep of EQ5] (Y5) {$\hat{f}_5$};
\node[box,below=\davesep of EQ6] (Y6) {$\hat{f}_6$};
\draw[<-] (X1.east)  -- (PL1);
\draw[<-] (X2.south)  -- (PL2);
\draw[<-] (X3.west)  -- (PL3);
\draw[<-] (X4.west)  -- (PL4);
\draw[<-] (X5.east)  -- (PL5);
\draw[->] (Y1.south)  -- (EQ1);
\draw[->] (Y2.south)  -- (EQ2);
\draw[->] (Y3.east)  -- (EQ3);
\draw[->] (Y4.east)  -- (EQ4);
\draw[->] (Y5.west)  -- (EQ5);
\draw[->] (Y6.north)  -- (EQ6);
\draw[<-] (PL1) -- (IEQ1);
\draw[<-] (PL1) -- (IEQ3);
\draw[<-] (PL2) -- (IEQ2);
\draw[<-] (PL2) -- (IEQ4);
\draw[<-] (PL3) -- (IEQ5);
\draw[<-] (PL4) -- (IEQ6);
\draw[<-] (PL2) -- (EQ1);
\draw[<-] (PL3) -- (EQ2);
\draw[<-] (PL4) -- (EQ4);
\draw[<-] (PL4) -- (EQ5);
\draw[<-] (PL5) -- (EQ3);
\draw[<-] (PL5) -- (EQ6);
\end{tikzpicture}
\caption{Dual EWNFG for scaled partition function of Ising-type model with an external field, based on normal realization of dual I/O behavior ${W}_{10} = \{(-\partial(\hat{\yb}), \hat{\yb}) \mid \hat{\yb} \in \hat{\A}^E\}$ in Figure \ref{NRIODD}.}
\label{DNFGEE}
\end{figure}
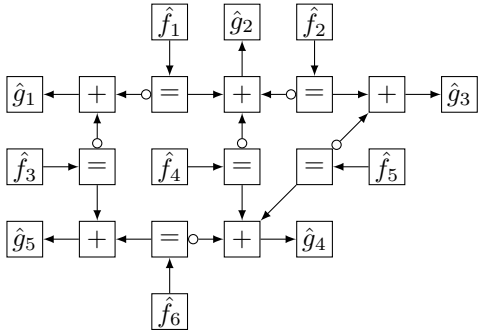

Consequently, the dual EWNFG is  based on the normal realization of the dual I/O behavior ${W}_{10} = \{(-\partial(\hat{\yb}), \hat{\yb}) \mid \hat{\yb} \in \hat{\A}^E\}$ of the boundary operator $\partial$,  as  illustrated in Figure \ref{NRIODD}.  This behavior is a systematic $(|E| + |V|, |E|)$ group code over $\A$, for which the edge set $E$ is an information set.
The resulting dual EWNFG shown in Figure \ref{DNFGEE}.  Explicitly, its partition function is
$$  \sum_{(\hat{\xb}, \hat{\yb}) \in W_{10}} \prod_{v \in V} \hat{g}_v(\hat{x}_v) \prod_{e \in E} \hat{f}_e(\hat{y}_e), $$
which is called a \emph{high-temperature expansion} of $Z$.

For the special case of the Ising model  ($\A = \Z_2$) with a constant  external field, there are well-known explicit formulas for this high-temperature expansion \cite{NM53, JS93}.  The above  expression generalizes these high-temperature expansions to general finite abelian group alphabets $\A$ and non-constant external fields $H_v$.

However, 
since $|E| = |V| - 1 + \beta_1(\G)$, the dual I/O behavior will be higher-dimensional than the primal I/O behavior if $\beta_1(\G) > 1$;  \ie if $\G$ has more than one cycle.  Thus with an external field it will generally be more complex to compute the partition function of the dual EWNFG than that of the primal EWNFG.  We will now explore methods of reducing this increased complexity.

\subsection{Alternative and hybrid I/O realizations}\label{Sec3.6}

We see that whereas in the primal domain the presence of an external field adds only one dimension to the realization of $Z$ (the difference between the dimensions of $B^1$ and $W^{01}$), in the dual domain it adds $|V| - 1$ dimensions (the difference between the dimensions of $Z_1$ and $W_{10}$).
We  now suggest alternative realizations of the partition function $Z$ and its Fourier transform $\hat{Z}$ for  Ising-type models with an external field, with the objective of simplifying their calculation.

As discussed above, $Z$ is the partition function of an EWNFG based on a normal realization of the I/O behavior $W^{01} = \{({\xb}, d({\xb}) \mid {\xb} \in \A^V\}$ of the coboundary operator $d$,  as  illustrated in Figure \ref{NRIOD}, where  $W^{01}$ is a linear $(|E| + |V|, |V|)$ code over $\A$.  

We have seen that the vertex vector $\xb \in \A^V$ is an obvious information set for $W^{01}$.  However, we will now show that for any $v \in V$ and any edge set $E_\T$ such that $\T$ is a spanning tree of $\G$,  $(x_v, \yb_\T) \in \A^{\{v\}} \times \A^{E_\T}$ is an information set for  $W^{01}$.  Moreover, we will give an I/O realization of $W^{01}$ based on this information set.

We recall that the projection of $W^{01}$ onto the edge configuration space $\A^E$ is $B^1 = \im d$, which is an $(|E|, |E| - \beta_1(\G)) = (|E|, |V| - 1)$ systematic group code with information set $E_\T$ for any spanning tree $\T \subseteq \G$.  Hence for any $(\xb, \yb) \in W^{01}$, the edge vector $\yb$ may be realized by taking $\yb_\T$ as the input to an I/O realization of $B^1$, such as the realization illustrated in Figure \ref{NRB1IO};  the output is then the unique $\yb_{\bar{\T}}$ such that $\yb = (\yb_\T, \yb_{\bar{\T}}) \in B^1$.

Now  we may extend  an I/O realization of $B^1$ such as Figure \ref{NRB1IO} to obtain a vertex vector $\xb(\yb_\T)$ as another output, thus obtaining a pair $(\xb(\yb_\T), \yb) \in W^{01}$. Note that we have shown that any single vertex variable $x_v$  in Figure \ref{NRB1IO} can be fixed to 0 (\ie $x_v(\yb_\T) = 0$, where $v \in V$ may be chosen arbitrarily), so  $(\xb_{\bar{v}}(\yb_\T), \yb) \in (W^{01})_{:\bar{v}}$.  Moreover, the set of all pairs $(\xb, \yb) \in W^{01}$ that have edge vector $\yb$ is the coset $(\xb + Z^0, \yb)$ of $Z^0 \times \{\zerob\}$ that contains $(\xb, \yb)$, where $Z^0 = \ker d$ is the $(|E|, 1)$ repetition code over $\A $.

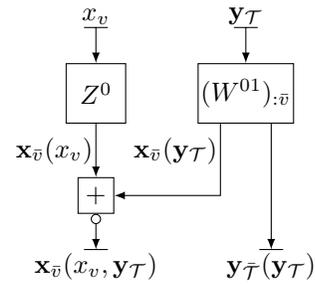
\begin{figure}[h]
\setlength{\unitlength}{5pt}
\centering
%%%%%%%%%%%%%%
% Figure 26
%%%%%%%%%%%%%%
\begin{tikzpicture}
\node[box,minimum size=5ex] (Z0) {$Z^0$};
\node[box,minimum height=5ex,right=2*\davesep of Z0] (W01) {$(W^{01})_{:\bar{v}}$};
\node[box,below=1.5\davesep of Z0] (PL) {$+$};
\node[var,above=\davesep of Z0] (xv) {$x_v$};
\node[var,above=\davesep of W01] (yt) {$\mathbf{y}_{\mathcal{T}}$};
\node[var,below=\davesep of PL] (xx) {$\mathbf{x}_{\bar{v}}(x_v,\mathbf{y}_{\mathcal{T}})$};
\node[var] at (W01.310 |- xx) (yy) {$\mathbf{y}_{\bar{\mathcal{T}}}(\mathbf{y}_{\mathcal{T}})$};
\node[invert,anchor=north]  (IPL) at (PL.south)  {};
\draw[->] (xv.south) \leftright -- (Z0);
\draw[->] (yt.south) \leftright -- (W01);
\draw[->] (W01.230) |- (PL);
\draw[<-] (yy.north) \leftright -- (W01.310);
\draw[<-] (xx.north) \leftright -- (IPL);
\draw[->] (Z0) -- node[var,left,anchor=east] (L) {$\mathbf{x}_{\bar{v}}(x_v)$} (PL);
\node[var,anchor=east] at (W01.230 |- L) {$\mathbf{x}_{\bar{v}}(\mathbf{y}_{\mathcal{T}})$};
\end{tikzpicture}
\caption{I/O realization of the I/O behavior $W^{01} = \{\xb, d(\xb)\}$ with information set $(x_v, \yb_\T)$.}
\label{NRW01IO}
\end{figure}

It follows that $W^{01}$ may be generated by the normal realization shown in Figure \ref{NRW01IO}.  On the left, a single input vertex variable $x_v \in \A^{\{v\}}$ generates the unique vertex vector $\xb_{\bar{v}}(x_v) = (x_v, \ldots, x_v) \in \A^{V \setminus \{v\}}$ such that $(x_v, \xb_{\bar{v}}(x_v)) \in Z_0$.  On the right, an input edge vector $\yb_\T \in \A^{E_\T}$ generates the unique edge vector $\yb_{\bar{\T}}(\yb_\T) \in \A^{E \setminus E_\T}$ such that $\yb(\yb_\T) = (\yb_\T, \yb_{\bar{\T}}(\yb_\T)) \in B^1$, and an associated vertex vector $\xb_{\bar{v}}(\yb_\T) \in \A^{V \setminus \{v\}}$ such that $(\xb(\yb_\T), \yb(\yb_\T)) \in W^{01}$, where $\xb(\yb_\T) = (0, \xb_{\bar{v}}(\yb_\T)) \in \A^V$.  The final output pair is $(\xb_{\bar{v}}(x_v, \yb_\T), \yb_{\bar{\T}}(\yb_\T))$, where $\xb_{\bar{v}}(x_v, \yb_\T) = \xb_{\bar{v}}(x_v) + \xb_{\bar{v}}(\yb_\T)$, which together with the input pair $(x_v, \yb_\T)$ give the unique pair  $(\xb, \yb) \in W^{01}$ that is consistent with the inputs $(x_v, \yb_\T) \in \A^{\{v\}} \times \A^{E_\T}$.

\begin{figure}[h]
\setlength{\unitlength}{5pt}
\centering
%%%%%%%%%%%%%%
% Figure 27
%%%%%%%%%%%%%%
\begin{tikzpicture}
\node[box,minimum size=5ex] (B0) {$B_0$};
\node[box,minimum height=5ex,right=2*\davesep of B0] (W10) {$(W_{10})_{|\bar{v}}$};
\node[box,below=1.5\davesep of B0] (EQ) {$=$};
\node[var,above=\davesep of B0] (xv) {$\hat{x}_v$};
\node[var,above=\davesep of W10] (yt) {$\hat{\mathbf{y}}_{\mathcal{T}}$};
\node[var,below=\davesep of EQ] (xx) {$\hat{\mathbf{x}}_{\bar{v}}$};
\node[var] at (W10.310 |- xx) (yy) {$\hat{\mathbf{y}}_{\bar{\mathcal{T}}}$};
\draw[<-] (xv.south) \leftright -- (B0);
\draw[<-] (yt.south) \leftright -- (W10);
\draw[<-] (W10.230) |- (EQ);
\draw[->] (yy.north) \leftright -- (W10.310);
\draw[->] (xx.north) \leftright -- (EQ);
\draw[<-] (B0) -- node[left,anchor=east] (L) {$\hat{\mathbf{x}}_{\bar{v}}$} (EQ);
\node[anchor=east] at (W10.230 |- L) {$\hat{\mathbf{x}}_{\bar{v}}$};
\end{tikzpicture}
\caption{I/O realization of the dual I/O behavior $W_{10} = \{-\partial(\hat{\yb}), \hat{\yb}\}$ with information set $(\hat{\xb}_{\bar{v}}, \hat{\yb}_{\bar{\T}})$.}
\label{NRW10IO}
\end{figure}
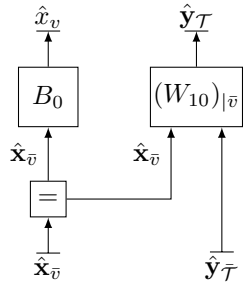

For a realization of the dual I/O map $W_{10} = (W^{01})^\perp$, we need merely dualize this representation, as shown in Figure \ref{NRW10IO}.  On the left, the dual of the repetition code $Z^0$ is the zero-sum code $B_0$.  On the right, the dual of the cross-section $(W^{01})_{:\bar{v}}$ is the projection $(W_{10})_{|\bar{v}}$, which may be realized by extending the I/O realization of $Z_1 = \ker \partial$ in Figure 17, which uses $\hat{\yb}_{\bar{\T}}$ as an information set, in a dual manner to the extension of Figure \ref{NRB1IO} discussed above, which involves adding $\hat{\xb}_{\bar{v}}$ as a second information set.  Finally, the remainder of the realization is dualized by replacing the zero-sum constraint by an equality constraint, and changing the directions of the arrows.\footnote{Superfluous sign inverters have been removed.}  Thus we obtain the realization of Figure  \ref{NRW10IO}.

\begin{figure}[h]
\setlength{\unitlength}{5pt}
\centering
%%%%%%%%%%%%%%
% Figure 28
%%%%%%%%%%%%%%
\begin{tikzpicture}
\node[box,minimum size=5ex] (B0) {$B_0$};
\node[box,minimum height=5ex,right=2*\davesep of B0] (W10) {$(W_{10})_{|\bar{v}}$};
\node[box,below=1.5\davesep of B0] (EQ) {$=$};
\node[box,above=\davesep of B0] (xv) {$\hat{g}_v(\hat{x}_v)$};
\node[box,above=\davesep of W10] (yt) {$\hat{f}_{\mathcal{T}}(\hat{\mathbf{y}}_{\mathcal{T}})$};
\node[box,below=\davesep of EQ] (xx) {$\hat{g}_{\bar{v}}(\hat{\mathbf{x}}_{\bar{v}})$};
\node[box] at (W10.310 |- xx) (yy) {$\hat{f}_{\bar{\mathcal{T}}}(\hat{\mathbf{y}}_{\bar{\mathcal{T}}})$};
\draw (xv.south) -- node[left,anchor=east] {$\hat{x}_v$} (B0);
\draw (yt.south) -- node[right,anchor=west] {$\hat{\mathbf{y}}_{\mathcal{T}}$} (W10);
\draw (W10.230) |- (EQ);
\draw (yy.north) -- (W10.310);
\draw (xx.north) -- node[left,anchor=east] {$\hat{\mathbf{x}}_{\bar{v}}$} (EQ);
\draw (B0) -- node[left,anchor=east] (L) {$\hat{\mathbf{x}}_{\bar{v}}$} (EQ);
\node[anchor=east] at (W10.230 |- L) {$\hat{\mathbf{x}}_{\bar{v}}$};
\node[anchor=west] at (W10.310 |- L) {$\hat{\mathbf{y}}_{\bar{\mathcal{T}}}$};
\end{tikzpicture}
\caption{Dual EWNFG for Ising-type model.}
\label{NFGDU}
\end{figure}
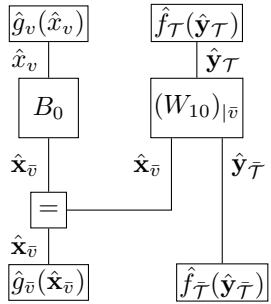

These normal realizations may be used to realize scaled versions of $Z$ and $\hat{Z}$ by attaching edge weights.  For example, the dual EWNFG of Figure \ref{NFGDU} realizes a scaled version of $\hat{Z}$ by attaching edge weights $\hat{g}(\hat{\xb})$ and $\hat{f}(\hat{\yb})$ to components of $\hat{\xb}$ and $\hat{\yb}$, respectively.

A hybrid alternative that may be attractive because of the simplicity of the repetition code $Z^0$ is shown in Figure \ref{NFGHY}.  Here, using the {Edge Replacement Lemma} (see Appendix), the left (vertex) side is realized in the primal domain, and the right (edge) side is realized in the dual domain, with a connection via a Fourier transform function (plus sign inverter) between the $(|V| - 1)$-dimensional primal and dual vertex vectors $\xb_{\bar{v}}$ and $\hat{\xb}_{\bar{v}}$.  (In this figure,  the scale factors of the Edge Replacement Lemma have been omitted.)

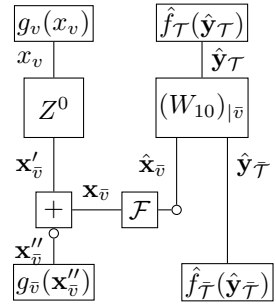
\begin{figure}[h]
\setlength{\unitlength}{5pt}
\centering
%%%%%%%%%%%%%%
% Figure 29
%%%%%%%%%%%%%%
\begin{tikzpicture}
\node[box,minimum size=5ex] (Z0) {$Z^0$};
\node[box,minimum height=5ex,right=2*\davesep of Z0] (W10) {$(W_{10})_{|\bar{v}}$};
\node[box,below=1.5\davesep of Z0] (PL) {$+$};
\node[box,above=\davesep of Z0] (xv) {$g_v(x_v)$};
\node[box,above=\davesep of W10] (yt) {$\hat{f}_{\mathcal{T}}(\hat{\mathbf{y}}_{\mathcal{T}})$};
\node[box,below=\davesep of PL] (xx) {$g_{\bar{v}}(\mathbf{x}_{\bar{v}}^{\prime\prime})$};
\node[box] at (W10.310 |- xx) (yy) {$\hat{f}_{\bar{\mathcal{T}}}(\hat{\mathbf{y}}_{\bar{\mathcal{T}}})$};
\node[invert,anchor=north]  (IPL) at (PL.south)  {};
\draw (xv.south) -- node[left,anchor=east] {$x_v$} (Z0);
\draw (yt.south) -- node[right,anchor=west] {$\hat{\mathbf{y}}_{\mathcal{T}}$} (W10);
\node[invert] (I) at (W10.230 |- PL) {};
\draw (W10.230) -- (I);
\node[box] (F) at (barycentric cs:I=0.7,PL=0.3) {$\mathcal{F}$};
\draw (F) -- (I);
\draw (PL) -- node[above] {$\mathbf{x}_{\bar{v}}$} (F);
\draw (yy.north) -- (W10.310);
\draw (xx.north) -- node[left,anchor=east] {$\mathbf{x}_{\bar{v}}^{\prime\prime}$} (IPL);
\draw (Z0) -- node[left,anchor=east] (L) {$\mathbf{x}_{\bar{v}}^{\prime}$} (PL);
\node[anchor=east] at (W10.230 |- L) {$\hat{\mathbf{x}}_{\bar{v}}$};
\node[anchor=west] at (W10.310 |- L) {$\hat{\mathbf{y}}_{\bar{\mathcal{T}}}$};
\end{tikzpicture}
\caption{Hybrid NFG for Ising-type model.}
\label{NFGHY}
\end{figure}

In this hybrid realization, the left side is 1-dimensional,  the right side is $\beta_1(\G)$-dimensional, and the two are linked by $|V| - 1$ edge variables, which require a $(|V| - 1)$-dimensional Fourier transform over $\A$.  For many alphabets $\A$--- \eg $\Z_2$--- ``fast Fourier transform" algorithms may be used.  Overall, such a hybrid realization may therefore yield a significant reduction in complexity.\footnote{Vontobel \cite{VPC} has pointed out that the ``Hamiltonian NFGs" in \cite[Fig.\ 4]{V11} are also hybrid realizations, in that they include both primal and dual parts, with Legendre transforms between them.}

\section{Two-dimensional algebraic topology}\label{Sec4}

We now consider graphs that have well-defined faces.  We will primarily consider connected planar graphs with no self-loops and no dangling edges, which is the simplest case.  

\vspace{-1ex}
\subsection{Two-dimensional complexes}

In elementary algebraic topology \cite{Bamberg}, a  graph $\G = (V,E)$ plus a set $F$ of \emph{faces} is called a \emph{two-dimensional complex} $\G^+ = (V, E, F)$.  A face $f \in F$ of $\G^+$ is called a \emph{two-dimensional object}.  

Faces are defined according to the topological space on which $\G$ is imagined to be drawn.  For instance, a \emph{planar graph} is one that can be drawn on a plane without any edges crossing.  The \emph{interior faces} of $\G$ are then the areas of the plane that are bounded by the edges of $\G$.
We will assume that  $\G$  has no ``dangling" degree-1  vertices;\footnote{The ``no dangling vertices" assumption is the dual  to the ``no self-loops" assumption.} then every edge of $\G$ bounds either  two interior faces of $\G$, if it is an \emph{interior edge}, or one interior face of $\G$, if it is an \emph{exterior edge}.

It is easy to see that in general the number of interior faces of a planar graph is  $|F| = \beta_1(\G)$.
For example, as a planar graph, our example graph $\G$ in Figure \ref{E1} has two interior faces, one interior edge, and five exterior edges.  
 
Alternatively,  a planar graph $\G$ may be considered to  be drawn on a sphere; then its ``exterior face" is also taken to be a face, and  every edge of $\G$ bounds precisely  two  faces of $\G$.   In this case, 
 $|F| = \beta_1(\G) + 1 = |E| - |V| +2$.  For example, as a planar graph on a sphere, our example graph $\G$ has $|V| = 5, |E| = 6$, and $|F| = 3$.

In algebraic topology, the vector space $C^2 = \A^F$ of column vectors over $\A$ indexed by $F$ is  called the space of \emph{2-cochains} of $\G^+$.  Again, we will call such vectors \emph{face vectors}, and write them as column vectors $\zb$ in matrix expressions.  The dual vector space $C_2 = \hat{\A}^F$ is called the space of \emph{2-chains}; again, we will call these \emph{dual face vectors}, write them as row vectors  $\hat{\zb} \in \hat{\A}^F$ in matrix expressions, and use the standard inner product. 

The \emph{boundary} of a face $f$ is a set $E(f) \subseteq E$ of directed edges.  Each face $f \in F$ is assumed to have an \emph{orientation}, perhaps arbitrary;  then each edge $e \in E(f)$ is given a sign $\alpha_{fe} =1$ if the face and  edge orientations are aligned, or $\alpha_{fe} =-1$ otherwise.  If $e \notin E(f)$, then $\alpha_{fe} = 0$.  The $\{0, \pm 1\}$-valued matrix $M_2 = \{\alpha_{fe} \mid f \in F, e \in E\}$ will be called the \emph{second incidence matrix} of $\G$.  (Hereafter, our original incidence matrix $M$ will be denoted  $M_1$ and called the \emph{first incidence matrix} of $\G$.)  Since an edge of $\G^+$ is incident on  two faces if it is an interior edge or on one face if it is an exterior edge, each of the $|E|$ columns of $M_2$ has one or two nonzero values.  The number of nonzero values in the $f$th row is the \emph{degree} $\delta_f$ of the face $f$, \ie the number of edges bounding $f$.

The \emph{second boundary operator} is defined as the homomorphism $\partial_2:  \hat{\A}^F \to \hat{\A}^E, \hat{\zb} \mapsto \hat{\zb} M_2$;  \ie the map whose matrix is the second incidence matrix $M_2$ of $\G$.  Its kernel  $Z_2 = \ker \partial_2$ is called the \emph{second zero-boundary space} of $\G^+$, and its image $B_1 = \im \partial_2$ is  the \emph{second boundary space} of $\G^+$. 

For example, our example graph $\G$ of Figure \ref{E1} may be taken as a planar graph $\G^+$ drawn on a plane, with two interior faces.  If each face is given a clockwise orientation, then its second incidence matrix  is 
$$
M_2 = \left[
\begin{array}{rrrrrr}
~1 & 0 & -1 & 1 & 0 & ~1 \\
0 & ~1 & 0 & -1 & 1 & 0
\end{array}
\right].  
$$ 
Note that only one edge is  interior, and that $Z_2 = \ker \partial_2$ is trivial, for this graph and  in general for planar graphs drawn on planes.  Thus $\dim B_1 = 2$.  In general, $\dim B_1 = |F| = \beta_1(\G)$.

However, if we consider our example graph to be drawn on a sphere, then $\G^+$ has three faces, and all edges become interior edges.  If the ``exterior face" is  given a counterclockwise orientation, then
$$
M_2 = \left[
\begin{array}{rrrrrr}
1 & 0 & -1 & 1 & 0 & 1 \\
0 & 1 & 0 & -1 & 1 & 0 \\
-1 & -1 & 1 & 0 & -1 & -1
\end{array}
\right].  
$$ 
Since all edges are now interior, each column of $M_2$ now has precisely 2 nonzero values $\pm 1$.
  Moreover, since the sum of the three rows is $\zerob \in \hat{\A}^E$, the second boundary operator $\partial_2$ now has a nontrivial kernel $Z_2$ of dimension 1.  Therefore $\dim Z_2 = 1$ in this case and, by similar arguments, for  planar graphs in general.  However, the image $B_1$ remains unchanged, regardless of whether we take $\G^+$ as a  graph on a plane or on a sphere.  Thus its dimension remains  $\dim B_1 = 2$ for this example, or $\dim B_1 = \beta_1(\G)$ for  planar graphs in general. 
  
Notice that the first two rows of $M_2$  are the edge vectors $\hat{\yb}(\pb_1), \hat{\yb}(\pb_2)$ corresponding to the two cycles $\pb_1, \pb_2$ that bound the two interior faces of $\G^+$, and therefore are elements of the zero-boundary space $Z_1$, the  kernel $Z_1$  of our original  \emph{first boundary operator} $\partial_1: \hat{\A}^E \to \hat{\A}^V, \hat{\yb} \mapsto \hat{\yb} M_1$.  It follows that $B_1 = Z_1$ in this example, and, by similar arguments, for planar graphs in general.
 
Dually, the image of our original \emph{first coboundary operator} $d_1: \A^V \to \A^E, \xb \mapsto M_1 \xb$ will continue to be denoted as $B^1$, and its kernel as $Z^0$.  Since $d_1$ is the adjoint homomorphism to $\partial_1$, $B^1 = (Z_1)^\perp$ and $Z^0 = (B_0)^\perp$ .  Similarly, we define the \emph{second coboundary operator} as the adjoint homomorphism to $\partial_2$--- \ie  the operator  $d_1: \A^E \to \A^F, {\yb} \mapsto M_2 {\yb}$.  By the Adjoint Homomorphism Lemma, the kernel $Z^1$ and image $B^2$ of $d_2$ are then  $Z^1 = (B_1)^\perp$ and $B^2 = (Z_2)^\perp$.

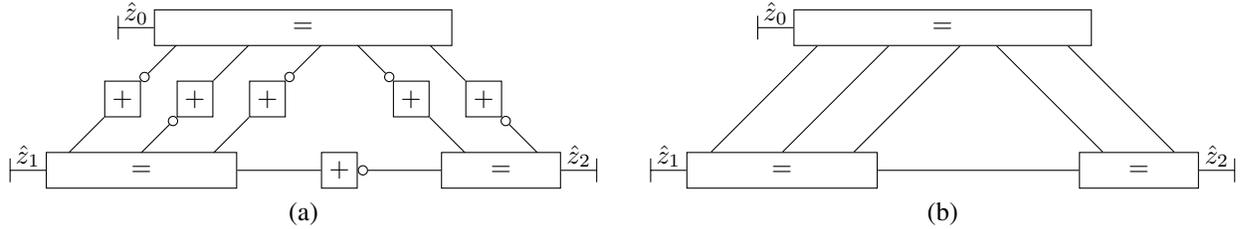
\begin{figure*}[h]
\setlength{\unitlength}{5pt}
\centering
\begin{tabular}{cc}
%%%%%%%%%%%%%%
% Figure 30 (a)
%%%%%%%%%%%%%%
\begin{tikzpicture}
\node[box] (PL1) {$+$};
\node[box,right=\davesep of PL1] (PL3) {$+$};
\node[box,right=\davesep of PL3] (PL6) {$+$};
\node[box,right=3*\davesep of PL6] (PL2) {$+$};
\node[box,right=\davesep of PL2] (PL5) {$+$};
\node[invert,anchor=south west] (IPL1) at (PL1.north east) {};
\node[invert,anchor=north east] (IPL3) at (PL3.south west) {};
\node[invert,anchor=south west] (IPL6) at (PL6.north east) {};
\node[invert,anchor=south east] (IPL2) at (PL2.north west) {};
\node[invert,anchor=north west] (IPL5) at (PL5.south east) {};
\draw (IPL1.north east) -- ++(45:1.4*\davesep);
\draw (PL3.north east) -- ++(45:1.4*\davesep);
\draw (IPL6.north east) -- ++(45:1.4*\davesep);
\draw (IPL2.north west) -- ++(135:1.4*\davesep);
\draw (PL5.north west) -- ++(135:1.4*\davesep);
\draw (PL1.south west) -- ++(225:1.4*\davesep);
\draw (IPL3.south west) -- ++(225:1.4*\davesep);
\draw (PL6.south west) -- ++(225:1.4*\davesep);
\draw (PL2.south east) -- ++(315:1.4*\davesep);
\draw (IPL5.south east) -- ++(315:1.4*\davesep);
\node[box,fill=white,minimum width=25ex,yshift=2*\davesep] (EQ1) at (barycentric cs:PL1=0.5,PL5=0.5) {$=$};
\node[box,fill=white,minimum width=16ex,yshift=-2*\davesep,xshift=-1.5*\davesep] (EQ2) at (barycentric cs:PL1=0.5,PL6=0.5) {$=$};
\node[box,fill=white,minimum width=10ex,yshift=-2*\davesep,xshift=1.5*\davesep] (EQ3) at (barycentric cs:PL2=0.5,PL5=0.5) {$=$};
\coordinate (MID) at (barycentric cs:PL6=0.5,PL2=0.5);
\node[box] (PL4) at (MID |- EQ2) {$+$};
\node[invert,anchor=west] (IPL4) at (PL4.east) {};
\draw (EQ2) -- (PL4);
\draw (IPL4) -- (EQ3);
\draw (EQ1.west) -- node[above,var] {$\hat{z}_0$} ++(-\davesep,0) \updown;
\draw (EQ2.west) -- node[above,var] {$\hat{z}_1$} ++(-\davesep,0) \updown;
\draw (EQ3.east) -- node[above,var] {$\hat{z}_2$} ++(\davesep,0) \updown;
%\node[var,anchor=east] at (PL1.west) {$\hat{y}_1$};
%\node[var,anchor=east] at (PL3.west) {$\hat{y}_3$};
%\node[var,anchor=east] at (PL6.west) {$\hat{y}_6$};
%\node[var,anchor=east] at (PL2.west) {$\hat{y}_2$};
%\node[var,anchor=east] at (PL5.west) {$\hat{y}_5$};
%\node[var,anchor=south] at (PL4.north) {$\hat{y}_4$};
\end{tikzpicture} 
&  \hspace{1ex}
%%%%%%%%%%%%%%
% Figure 30 (b)
%%%%%%%%%%%%%%
\begin{tikzpicture}
\coordinate (PL1);
\coordinate[right=\davesep+3ex of PL1] (PL3);
\coordinate[right=\davesep+3ex of PL3] (PL6);
\coordinate[right=3*\davesep+3ex of PL6] (PL2);
\coordinate[right=\davesep+3ex of PL2] (PL5);
\draw (PL1) -- ++(45:2.1*\davesep);
\draw (PL3) -- ++(45:2.1*\davesep);
\draw (PL6) -- ++(45:2.1*\davesep);
\draw (PL2) -- ++(135:2.1*\davesep);
\draw (PL5) -- ++(135:2.1*\davesep);
\draw (PL1) -- ++(225:2.1*\davesep);
\draw (PL3) -- ++(225:2.1*\davesep);
\draw (PL6) -- ++(225:2.1*\davesep);
\draw (PL2) -- ++(315:2.1*\davesep);
\draw (PL5) -- ++(315:2.1*\davesep);
\node[box,fill=white,minimum width=25ex,yshift=2*\davesep] (EQ1) at (barycentric cs:PL1=0.5,PL5=0.5) {$=$};
\node[box,fill=white,minimum width=16ex,yshift=-2*\davesep,xshift=-1.5*\davesep] (EQ2) at (barycentric cs:PL1=0.5,PL6=0.5) {$=$};
\node[box,fill=white,minimum width=10ex,yshift=-2*\davesep,xshift=1.5*\davesep] (EQ3) at (barycentric cs:PL2=0.5,PL5=0.5) {$=$};
\coordinate (MID) at (barycentric cs:PL6=0.5,PL2=0.5);
\draw (EQ2) -- (EQ3);
\draw (EQ1.west) -- node[above,var] {$\hat{z}_0$} ++(-\davesep,0) \updown;
\draw (EQ2.west) -- node[above,var] {$\hat{z}_1$} ++(-\davesep,0) \updown;
\draw (EQ3.east) -- node[above,var] {$\hat{z}_2$} ++(\davesep,0) \updown;
\end{tikzpicture} \\
(a) &  \hspace{1ex} (b) \\
\end{tabular}
\caption{$Z_2 = \ker \partial_2$ for $\G^+$: (a) normal realization;  (b) simplified normal realization.}
\label{H2}
\end{figure*}

\subsection{Homology spaces}
 
 In general, it is straightforward to show that in any two-dimensional complex, the second boundary space $B_1$ is a subspace of the first zero-boundary space $Z_1 = \ker \partial_1$, by showing that every row of $M_2$ is an edge vector $\hat{\yb}(\pb)$ corresponding to a cycle $\pb$.  Thus for a  general two-dimensional complex we have $B_1 \subseteq Z_1$.  The \emph{first homology space} is then defined as $H_1 = Z_1/B_1$.  
 
 For a two-dimensional complex $\G^+$ based on a planar graph $\G$, we have seen that $B_1 = Z_1$, so for our example planar graph and in general, we have $\dim H_1 = 0$, whether we take $\G^+$ as a  graph on a plane or on a sphere.

In a general  two-dimensional complex, the \emph{first cohomology space} is defined as $H^1 = Z^1/B^1 = (B_1)^\perp/(Z_1)^\perp$, which from linear algebra is the dual space to $H_1 = Z_1/B_1$.  Thus for a planar graph $\dim H^1 = \dim H_1 = 0$;  \ie the image $B^1$ of $d_1$ is equal to the kernel $Z^1$ of $d_2$.

 In this context, the \emph{second homology space} $H_2$ of $\G^+$ is defined as $Z_2$. We have seen that $\dim Z_2 = 0$ if $\G$ is a planar graph defined on a plane, but $\dim Z_2 = 1$ if $\G$ is defined on a sphere. The \emph{second cohomology space} is defined as $H^2 = C^2/B^2 = C^2/(Z_2)^\perp$, which is the dual space to $H_2 = Z_2$; thus $\dim H^2 = \dim H_2$.

The dimensions of the homology or cohomology spaces $H_0, H_1, H_2$ are thus $(1, 0, 0)$ if $\G$ is regarded as a planar graph on a plane, or $(1, 0, 1)$ if $\G$ is regarded as a planar graph on a sphere. We see that these dimensions are universal for all  graphs of these respective types \cite{Bamberg}.

Two-dimensional complexes become more interesting when $H_1 = Z_1/B_1$ is nontrivial;  \ie when there exist cycle vectors in $Z_1 = \ker \partial_1$ that are not second boundary vectors in $B_1 = \im \partial_2$.  For example, as shown in \cite{AV16}, if $\G$ is a 2-dimensional square lattice graph drawn on a torus, then $\dim H_1 =2$, since there are two independent cycle vectors that do not bound faces.

\subsection{Normal realizations}

Figure \ref{H2}(a) depicts a normal realization of  $Z_2 = \ker \partial_2$ when we view $\G^+$ as being drawn on a sphere, so there is an exterior face $f_0$ as well as the two interior faces $f_1, f_2$.  The three equality constraints correspond to the face variables $\hat{z}_0, \hat{z}_1, \hat{z}_2$, and the six zero-sum constraints correspond to the edge variables $\hat{y}_1, \ldots, \hat{y}_6$, which are all  set to 0 in this kernel realization.  One of the two incident face variables to each edge constraint is negated, according to the entries $\alpha_{fe}$ of  $M_2$.

Figure \ref{H2}(b) depicts a simplified realization of  $Z_2  = \ker \partial_2$.  Note the resemblance of this realization to that of $Z^0 = \ker d_1$ in Figure \ref{NRZ0} or \ref{NRZ0IO}.

\begin{figure}[h]
\setlength{\unitlength}{5pt}
\centering
%%%%%%%%%%%%%%
% Figure 31
%%%%%%%%%%%%%%
\begin{tikzpicture}
\node[invert] (PL1) {};
\node[invert,right=\davesep+3ex-3.5pt of PL1] (PL3) {};
\node[invert,right=\davesep+3ex-3.5pt of PL3] (PL6) {};
\node[invert,right=3*\davesep+3ex-3.5pt of PL6] (PL2) {};
\node[invert,right=\davesep+3ex-3.5pt of PL2] (PL5) {};
\draw (PL1) -- ++(45:2.1*\davesep);
\draw (PL3) -- ++(45:2.1*\davesep);
\draw (PL6) -- ++(45:2.1*\davesep);
\draw (PL2) -- ++(135:2.1*\davesep);
\draw (PL5) -- ++(135:2.1*\davesep);
\draw (PL1) -- ++(225:2.1*\davesep);
\draw (PL3) -- ++(225:2.1*\davesep);
\draw (PL6) -- ++(225:2.1*\davesep);
\draw (PL2) -- ++(315:2.1*\davesep);
\draw (PL5) -- ++(315:2.1*\davesep);
\node[box,fill=white,minimum width=25ex,yshift=2*\davesep] (EQ1) at (barycentric cs:PL1=0.5,PL5=0.5) {$+$};
\node[box,fill=white,minimum width=16ex,yshift=-2*\davesep,xshift=-1.5*\davesep] (EQ2) at (barycentric cs:PL1=0.5,PL6=0.5) {$+$};
\node[box,fill=white,minimum width=10ex,yshift=-2*\davesep,xshift=1.5*\davesep] (EQ3) at (barycentric cs:PL2=0.5,PL5=0.5) {$+$};
\coordinate (MID) at (barycentric cs:PL6=0.5,PL2=0.5);
\node[invert] (PL4) at (MID |- EQ2) {};
\draw (EQ2) -- (PL4);
\draw (PL4) -- (EQ3);
\draw (EQ1.west) -- node[above,var] {$z_0$} ++(-\davesep,0) \updown;
\draw (EQ2.west) -- node[above,var] {$z_1$} ++(-\davesep,0) \updown;
\draw (EQ3.east) -- node[above,var] {$z_2$} ++(\davesep,0) \updown;
\end{tikzpicture}
\caption{Dual normal realization of $B^2 = \im d_2$ for $\G^+$.}
\label{H2D}
\end{figure}
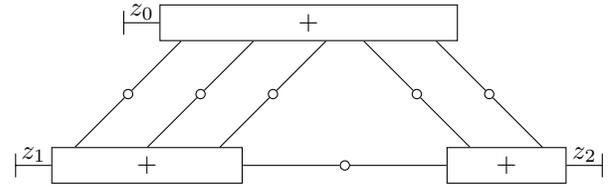

Similarly, Figure \ref{H2D} depicts the dual realization of  $B^2 = \im d_2$, which resembles that of $B_0 = \im \partial_1$ in Figure \ref{NRB0} or \ref{NRB0IO}.  We  now explain these resemblances.

\subsection{Dual graphs}

Let $\G = (V, E)$ be a  connected planar graph, let $\G^+ = (V, E, F)$  be the associated 2-dimensional complex when $\G$ is drawn on a sphere, and let $M_1$ and $M_2$ be the first and second incidence matrices of $\G^+$.  The number of faces is then $|F| = \beta_1(\G) + 1 = |E| - |V| + 2$, including the exterior face.  Assuming that $\G$ has no dangling  vertices,   every edge is incident on two distinct faces, a \emph{right face} $r(e) \in F$ and a \emph{left face} $\ell(e) \in F$.  (For our purposes, it does not matter which is  which.) 

Each row $(M_1)_e$ of $M_1$ then has two nonzero values,  $(M_1)_{eh(e)} = +1$ and  $(M_1)_{et(e)} = -1$.  Similarly,  each column $(M_2)_e$ of $M_2$ has two nonzero values,  $(M_2)_{er(e)} = +1$ and  $(M_2)_{e\ell(e)} = -1$.

It is thus natural to define the \emph{dual graph} $\hat{\G} = (F, E)$ as the  planar graph with the same edge set $E$, but with vertex and face sets interchanged, so $\hat{\G}^+ = (F, E, V)$ has incidence matrices $\hat{M}_1 = (M_2)^T$ and $\hat{M}_2 = (M_1)^T$ equal to the transposes of the original incidence matrices.  Thus $\beta_1(\hat{\G}) = |E| - |F| + 1 = |V| - 1 = |E| - \beta_1(\G).$

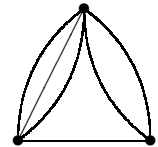
\begin{figure}[h]
\setlength{\unitlength}{5pt}
\centering
%%%%%%%%%%%%%%
% Figure 32
%%%%%%%%%%%%%%
\begin{picture}(20,9)(-1, 0)
\put(-0.5,-0.5){$\bullet$}
\put(9.5,-0.5){$\bullet$}
\put(4.5,9.5){$\bullet$}
\put(0,0){\line(1,0){10}}
\put(0,0){\line(1,2){5}}
\qbezier(0,0)(0,6.25)(5,10)
\qbezier(10,0)(10,6.25)(5,10)
\qbezier(0,0)(5,3.75)(5,10)
\qbezier(10,0)(5,3.75)(5,10)
\end{picture}
\caption{Dual graph $\hat{\G}$ to the graph $\G$ of Example 1.}
\label{DG}
\end{figure}

Geometrically, we may construct this dual graph by putting vertices of $\hat{\G}$  inside  each face of $\G$ (including the exterior face), and putting edges between two such vertices if and only if the corresponding faces share an edge, as illustrated for our example graph in Figure \ref{DG}.

\begin{figure*}[h]
\setlength{\unitlength}{5pt}
\centering
%%%%%%%%%%%%%%
% Figure 33(a)
%%%%%%%%%%%%%%
\begin{tabular}{cc}
\begin{tikzpicture}
\node[box] (EQ1) {$=$};
\node[box,right=\davesep of EQ1] (EQ3) {$=$};
\node[box,right=\davesep of EQ3] (EQ6) {$=$};
\node[box,right=3*\davesep of EQ6] (EQ2) {$=$};
\node[box,right=\davesep of EQ2] (EQ5) {$=$};
\node[invert,anchor=south west] (IEQ1) at (EQ1.north east) {};
\node[invert,anchor=north east] (IEQ3) at (EQ3.south west) {};
\node[invert,anchor=south west] (IEQ6) at (EQ6.north east) {};
\node[invert,anchor=south east] (IEQ2) at (EQ2.north west) {};
\node[invert,anchor=north west] (IEQ5) at (EQ5.south east) {};
\draw (IEQ1.north east) -- ++(45:1.4*\davesep);
\draw (EQ3.north east) -- ++(45:1.4*\davesep);
\draw (IEQ6.north east) -- ++(45:1.4*\davesep);
\draw (IEQ2.north west) -- ++(135:1.4*\davesep);
\draw (EQ5.north west) -- ++(135:1.4*\davesep);
\draw (EQ1.south west) -- ++(225:1.4*\davesep);
\draw (IEQ3.south west) -- ++(225:1.4*\davesep);
\draw (EQ6.south west) -- ++(225:1.4*\davesep);
\draw (EQ2.south east) -- ++(315:1.4*\davesep);
\draw (IEQ5.south east) -- ++(315:1.4*\davesep);
\node[box,fill=white,minimum width=25ex,yshift=2*\davesep] (PL1) at (barycentric cs:EQ1=0.5,EQ5=0.5) {$+$};
\node[box,fill=white,minimum width=16ex,yshift=-2*\davesep,xshift=-1.5*\davesep] (PL2) at (barycentric cs:EQ1=0.5,EQ6=0.5) {$+$};
\node[box,fill=white,minimum width=10ex,yshift=-2*\davesep,xshift=1.5*\davesep] (PL3) at (barycentric cs:EQ2=0.5,EQ5=0.5) {$+$};
\coordinate (MID) at (barycentric cs:EQ6=0.5,EQ2=0.5);
\node[box] (EQ4) at (MID |- PL2) {$=$};
\node[invert,anchor=west] (IEQ4) at (EQ4.east) {};
\draw (PL2) -- (EQ4);
\draw (IEQ4) -- (PL3);
\draw (EQ1.west) -- ++(-1ex,0) ++(0,-0.5ex) -- ++(0,1ex) node[var,anchor=south] {$y_1$};
\draw (EQ3.west) -- ++(-1ex,0) ++(0,-0.5ex) -- ++(0,1ex) node[var,anchor=south] {$y_3$};
\draw (EQ6.west) -- ++(-1ex,0) ++(0,-0.5ex) -- ++(0,1ex) node[var,anchor=south] {$y_6$};
\draw (EQ2.west) -- ++(-1ex,0) ++(0,-0.5ex) -- ++(0,1ex) node[var,anchor=310] {$y_2$};
\draw (EQ5.west) -- ++(-1ex,0) ++(0,-0.5ex) -- ++(0,1ex) node[var,anchor=300] {$y_5$};
\draw (EQ4.north) -- ++(0,1ex) ++(-0.5ex,0) -- node[var,above] {$y_4$} ++(1ex,0);
\end{tikzpicture}
 &  \hspace{1ex}
%%%%%%%%%%%%%%
% Figure 33(b)
%%%%%%%%%%%%%%
\begin{tikzpicture}
\node[box] (PL1) {$+$};
\node[box,right=\davesep of PL1] (PL3) {$+$};
\node[box,right=\davesep of PL3] (PL6) {$+$};
\node[box,right=3*\davesep of PL6] (PL2) {$+$};
\node[box,right=\davesep of PL2] (PL5) {$+$};
\node[invert,anchor=north east] (IPL1) at (PL1.south west) {};
\node[invert,anchor=south west] (IPL3) at (PL3.north east) {};
\node[invert,anchor=north east] (IPL6) at (PL6.south west) {};
\node[invert,anchor=north west] (IPL2) at (PL2.south east) {};
\node[invert,anchor=south east] (IPL5) at (PL5.north west) {};
\draw (PL1.north east) -- ++(45:1.4*\davesep);
\draw (IPL3.north east) -- ++(45:1.4*\davesep);
\draw (PL6.north east) -- ++(45:1.4*\davesep);
\draw (PL2.north west) -- ++(135:1.4*\davesep);
\draw (IPL5.north west) -- ++(135:1.4*\davesep);
\draw (IPL1.south west) -- ++(225:1.4*\davesep);
\draw (PL3.south west) -- ++(225:1.4*\davesep);
\draw (IPL6.south west) -- ++(225:1.4*\davesep);
\draw (IPL2.south east) -- ++(315:1.4*\davesep);
\draw (PL5.south east) -- ++(315:1.4*\davesep);
\node[box,fill=white,minimum width=25ex,yshift=2*\davesep] (EQ1) at (barycentric cs:PL1=0.5,PL5=0.5) {$=$};
\node[box,fill=white,minimum width=16ex,yshift=-2*\davesep,xshift=-1.5*\davesep] (EQ2) at (barycentric cs:PL1=0.5,PL6=0.5) {$=$};
\node[box,fill=white,minimum width=10ex,yshift=-2*\davesep,xshift=1.5*\davesep] (EQ3) at (barycentric cs:PL2=0.5,PL5=0.5) {$=$};
\coordinate (MID) at (barycentric cs:PL6=0.5,PL2=0.5);
\node[box] (PL4) at (MID |- EQ2) {$+$};
\node[invert,anchor=east] (IPL4) at (PL4.west) {};
\draw (EQ2) -- (IPL4);
\draw (PL4) -- (EQ3);
\draw (PL1.west) -- ++(-1ex,0) ++(0,-0.5ex) -- ++(0,1ex) node[var,anchor=south] {$\hat{y}_1$};
\draw (PL3.west) -- ++(-1ex,0) ++(0,-0.5ex) -- ++(0,1ex) node[var,anchor=south] {$\hat{y}_3$};
\draw (PL6.west) -- ++(-1ex,0) ++(0,-0.5ex) -- ++(0,1ex) node[var,anchor=south] {$\hat{y}_6$};
\draw (PL2.west) -- ++(-1ex,0) ++(0,-0.5ex) -- ++(0,1ex) node[var,anchor=310] {$\hat{y}_2$};
\draw (PL5.west) -- ++(-1ex,0) ++(0,-0.5ex) -- ++(0,1ex) node[var,anchor=300] {$\hat{y}_5$};
\draw (PL4.north) -- ++(0,1ex) ++(-0.5ex,0) -- node[var,above] {$\hat{y}_4$} ++(1ex,0);
\end{tikzpicture}  \\
(a) &  \hspace{1ex} (b) \\
\end{tabular}
\caption{(a) $Z^1 = \ker d_2$ for $\G^+$;  (b) $B_1 = \im \partial_2$ for $\G^+$.}
\label{Z1B1}
\end{figure*}
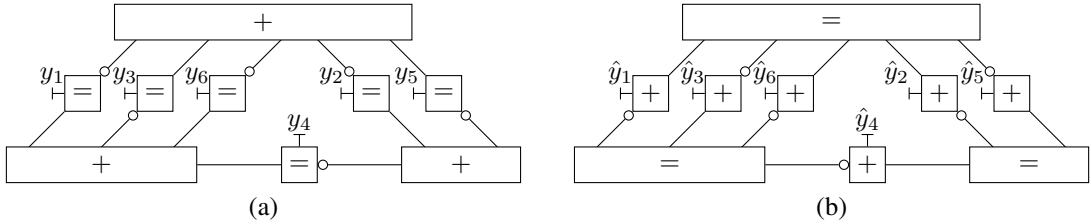

The similarity of this dual graph $\hat{\G}$ to the graphs of Figure \ref{H2} and \ref{H2D} is no accident.
For $\hat{\G}^+ = (F, E, V)$, the first boundary operator
$\hat{\partial}_1:  \hat{\A}^E \to \hat{\A}^F$ has matrix $\hat{M}_1 = (M_2)^T$, and thus maps the row vector $\hat{\yb} \in \hat{\A}^E$ to  $\hat{\zb} = \hat{\yb}(M_2)^T \in \hat{\A}^F$.  This map is evidently the same up to transposition as that of the second coboundary operator $d_2:  \A^E \to \A^F$ of $\G^+$, which maps the column vector ${\yb} \in \A^E$ to ${\zb} = M_2{\yb}$.  Similarly, $\hat{\partial}_2:  \hat{\A}^V \to \hat{\A}^E$ is the same map up to transposition as $d_1:  \A^V \to \A^E$, and the first and second coboundary operators, $\hat{d}_1:  \hat{\A}^F \to \hat{\A}^E$ and $\hat{d}_2:  \hat{\A}^E \to \hat{\A}^V$, are the same up to transposition as $\partial_2$ and $\partial_1$, respectively.

Thus the boundary and zero-boundary spaces for $\hat{\G}^+$ are simply the transposes of those spaces for $\G^+$:  
$\hat{B}_0 = (B^2)^T, \hat{Z}_1 = (Z^1)^T, \hat{B}_1 = (B^1)^T, \hat{Z}_2 = (Z^0)^T, \hat{B}^1 = (B_1)^T, \hat{Z}_0 = (Z_2)^T, \hat{B}^2 = (B_0)^T, \hat{Z}^1 = (Z_1)^T.$

Figure \ref{H2} may thus be seen as realizing either $Z_2 = \ker \partial_2$ for $\G^+$, or  $\hat{Z}^0 = \ker \hat{d}_1$ for $\hat{\G}^+$.  Similarly, Figure \ref{H2D} may be seen as realizing either $B^2 = \im d_2$ for $\G^+$, or  $\hat{B}_0 = \im \hat{\partial}_1$ for $\hat{\G}^+$. 

Using the dual graph $\hat{\G}$, we may also realize $Z^1 = \ker d_2$ for $\G^+$  as $\hat{Z}_1 = \ker \hat{\partial}_1$ for $\hat{\G}$, and $B_1 = \im \partial_2$ for $\G^+$  as $\hat{B}^1 = \im\hat{d}_1$ for $\hat{\G}$.
For example, Figures \ref{Z1B1}(a) and  \ref{Z1B1}(b) show realizations of $Z^1$ and $B_1$ for our example graph $\G^+$ as realizations of $\hat{Z}_1$ and $\hat{B}^1$ for its dual graph $\hat{\G}$ as in Figures \ref{NRZ1} or \ref{NRB1}.  Again, these are dual realizations;  \ie $Z^1 = (B_1)^\perp$.

However, for a two-dimensional complex $\G^+$ based on a planar graph $\G$, we also have $Z^1 = B^1 = \im d_1$ and $B_1 = Z_1 = \ker \partial_1$;  therefore we may alternatively realize $Z^1$ as an image realization based on $\G$ as in Figure \ref{NRB1}, or $B_1$ as a kernel realization based on $\G$ as in  Figure \ref{NRZ1}.  Since $\G$ has $|E|$ edges and $|V| = |E| - \beta_1(\G) + 1$ vertices, whereas $\hat{\G}$ has $|E|$ edges and $\beta_1(\G) + 1$ vertices, the representation based on $\G$ (resp.\ $\hat{\G}$) will in general be simpler if $\beta_1(\G) > |E|/2$ (resp.\ $\beta_1(\G) < |E|/2$).

\subsection{Realizations of partition functions for planar graphs}\label{Sec4.5}

Let $\G^+ = (V, E, F)$ be a two-dimensional complex based on a connected planar graph $\G$.  Then we have seen that the partition function $Z(\G)$ of an Ising-type model based on $\G$ with interaction weight functions $\{f_e(a) \mid e \in E, a \in \A\}$ may be represented  (up to scale) as the partition function of an EWNFG consisting of a realization of its first coboundary space $B^1 = \im d_1$ with edge weights $\{f_e(a)\}$.  We recall that $\dim B^1 = |V|-1 = |E| - \beta_1(\G)$.

As we have  seen, with a planar graph we have   
$B^1 = Z^1 = \ker d_2$.  Moreover, $Z^1$ may be realized as $\hat{Z}_1 = \ker \hat{\partial}_1$ for the dual graph $\hat{\G}$.  Hence $Z(\G)$ may also be represented (up to scale) by a
 realization of the first zero-boundary space $\hat{Z}_1 = \ker \hat{\partial}_1$ of $\hat{\G}$ with edge weights  $\{f_e(a)\}$.

Furthermore,  the Fourier transform $\hat{Z}(\G)$ of $Z(\G)$, which as we have seen is equal to $Z(\G)$ up to scale, may be represented (up to scale) by a kernel realization of its first zero-boundary space $Z_1 = \ker \partial_1$, with Fourier-transformed edge weights $\{\hat{f}_e(a)\}$.  We have $\dim Z_1 = \beta_1(\G)$, which can be less than $\dim B^1 = |V|-1 = |E| - \beta_1(\G)$, as we have seen in Figures \ref{SCG} and \ref{SCGD}.  The Fourier transform will in general convert a low-temperature  to a high-temperature model, and \emph{vice versa}.

If $\G$ is a planar graph, then 
$Z_1 = B_1$, and $B_1$ may be realized as $\hat{B}^1 = \im \hat{d}_1$ for the dual graph $\hat{\G}$.  Thus $\hat{Z}(\G)$ may also be represented (up to scale) by an image realization of the first coboundary space $\hat{B}^1 = \im \hat{d}_1$ of $\hat{\G}$, with Fourier-transformed edge weights $\{\hat{f}_e(a)\}$. 

Table \ref{T2} summarizes these four possible representations.

\begin{table*}[h]
\begin{center}
\begin{tabular}{|c|c|c|c|c|c|c|}
\hline
realizes & space   & dimension & realization type & graph & no.\ vertices & edge wts. \\
\hline
        &&&&&&             \\[-0.20cm]
$Z(\G)$ & $B^1 = \im d_1$ & $|E| - \beta_1(\G)$ &  Fig.\ \ref{NRB1} or \ref{NRB1IO} & $\G$ & $|E| - \beta_1(\G) + 1$ & $\{f_e(a)\}$ \\
$Z({\hat \G})$ & $\hat{Z}_1 = \ker \hat{\partial}_1$ & $|E| - \beta_1(\G)$  & Fig.\ \ref{NRZ1} or \ref{NRZ1IO} & $\hat{\G}$ & $\beta_1(\G) + 1$ & $\{f_e(a)\}$ \\

$\hat{Z}(\G)$ & $Z_1 = \ker \partial_1$ & $\beta_1(\G)$ &  Fig.\ \ref{NRZ1} or \ref{NRZ1IO} & $\G$ & $|E| - \beta_1(\G) + 1$ & $\{\hat{f}_e(a)\}$ \\
$\hat{Z}({\hat \G})$ & $\hat{B}^1 = \im \hat{d}_1$ &  $\beta_1(\G)$  & Fig.\ \ref{NRB1} or \ref{NRB1IO} & $\hat{\G}$ & $\beta_1(\G) + 1$ & $\{\hat{f}_e(a)\}$ \\
\hline
\end{tabular}
\end{center}
\caption{Representations of a partition function $Z$ (up to scale), using either a graph $\G$ or its dual graph $\hat{\G}$, and either edge weights $\{f_e(a)\}$ or dual weights $\{\hat{f}_e(a)\}$.}
\label{T2}
\end{table*}

Al-Bashabsheh and Vontobel \cite{AV16} show that the partition function $Z(\G)$ of an EWNFG based on a connected planar graph $\G$ and edge weights  $\{f_e(a)\}$ is equal up to scale to the partition function  of an EWNFG based on the dual graph $\hat{\G}$ with dual edge weights $\{\hat{f}_e(a)\}$.  This  follows from equating (up to scale) the first and fourth lines of  Table \ref{T2}.

For example, let $\G$ be a single-cycle graph of length $N$, which is a connected planar graph with $|V| = |E| = N$ and $\beta_1(\G) = 1$.  Its dual graph $\hat{\G}$ thus has only $|\hat{V}| = 2$ vertices, while  $\beta_1(\hat{\G}) = N - 1$.  Figure \ref{SCG} shows a representation of the partition function $Z(\G)$ using  an image realization of the $(N, N-1)$ linear code  $B^1 = \im d_1$ on the graph $\G$, and Figure \ref{SCGD} shows a representation of its Fourier transform $\hat{Z}(\G)$ (up to scale) using a kernel realization of the $(N, 1)$  code  $Z_1 = \ker \partial_1$ on $\G$.  We now see that $Z(\G)$ could  alternatively be represented (up to scale) by a kernel realization of $Z(\hat \G)$ using the $(N, N-1)$ code $\hat{Z}_1 = \ker \hat{\partial}_1$ with edge weights $\{f_e(a)\}$ on the dual graph $\hat{\G}$, or  alternatively by an image realization of $\hat Z(\hat \G)$ using the $(N, 1)$ code $\hat{B}^1 = \im \hat{d}_1$ with edge weights $\{\hat{f}_e(a)\}$ on $\hat{\G}$.

\section{Conclusion}

In this paper, we have presented an introduction to elementary algebraic topology using normal realizations, and, following \cite{AV16, ML13}, we have shown how such realizations may be used for calculating partition functions of Ising-type models.  Indeed, using dual realizations and dual graphs, we have given multiple alternative ways of representing such partition functions, summarized in Section \ref{Sec4.5}. 

While Molkaraie \emph{et al.} \cite{M15, M16, MG15, ML13}, have successfully exploited such alternatives in Monte Carlo simulations using importance sampling, much more could be done. In particular, in the presence of external fields, the hybrid models suggested in Section \ref{Sec3.6} should be explored further.

Our results are very general; in particular, they apply for any finite abelian group alphabet $\A$.  Thus one could explore Ising-type models with group alphabets more general  than $\Z_q$;  however, we have no idea whether such models would be of  interest to statistical physicists.

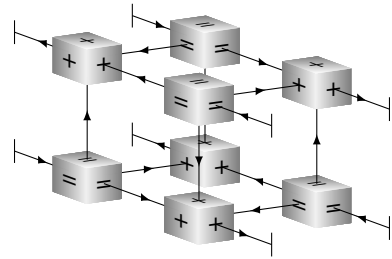
\begin{figure}[h]
\setlength{\unitlength}{5pt}
\centering
%%%%%%%%%%%%%%
% Figure 34
% this one is a bit fancy -- try playing with the sizes and angles....
%%%%%%%%%%%%%%
\begin{tikzpicture}
\newlength{\cubesize}\setlength{\cubesize}{3ex}  % cube size
\newlength{\cubesep}\setlength{\cubesep}{10ex}   % separation between cubes
\newcommand*\aangle{20}                          % left angle to horizontal
\newcommand*\bangle{8}                           % right angle to horizontal

\pgfmathsetmacro\sa{sin(\aangle)}
\pgfmathsetmacro\ca{cos(\aangle)}
\pgfmathsetmacro\sb{sin(\bangle)}
\pgfmathsetmacro\cb{cos(\bangle)}

%% 8 cubes each with 3 exposed faces: ABC, DEF, GHI, JKL, abc, def, ghi, jkl
%% top left = ABC, top back = DEF, top front = GHI, top right = JKL
%% bot left = abc, bot back = def, bot front = ghi, bot right = jkl
%% A is the left face, B is the right face, C is the top face, etc.

%\newcommand{\labelA}{A}
%\newcommand{\labelB}{B}
%\newcommand{\labelC}{C}
%\newcommand{\labelD}{D}
%\newcommand{\labelE}{E}
%\newcommand{\labelF}{F}
%\newcommand{\labelG}{G}
%\newcommand{\labelH}{H}
%\newcommand{\labelI}{I}
%\newcommand{\labelJ}{J}
%\newcommand{\labelK}{K}
%\newcommand{\labelL}{L}
%\newcommand{\labela}{a}
%\newcommand{\labelb}{b}
%\newcommand{\labelc}{c}
%\newcommand{\labeld}{d}
%\newcommand{\labele}{e}
%\newcommand{\labelf}{f}
%\newcommand{\labelg}{g}
%\newcommand{\labelh}{h}
%\newcommand{\labeli}{i}
%\newcommand{\labelj}{j}
%\newcommand{\labelk}{k}
%\newcommand{\labell}{l}

\newcommand{\labelA}{+}
\newcommand{\labelB}{+}
\newcommand{\labelC}{+}
\newcommand{\labelD}{\raisebox{-0.60ex}[0pt][0pt]{=}}
\newcommand{\labelE}{\raisebox{-0.60ex}[0pt][0pt]{=}}
\newcommand{\labelF}{\raisebox{-0.60ex}[0pt][0pt]{=}}
\newcommand{\labelG}{\raisebox{-0.60ex}[0pt][0pt]{=}}
\newcommand{\labelH}{\raisebox{-0.60ex}[0pt][0pt]{=}}
\newcommand{\labelI}{\raisebox{-0.60ex}[0pt][0pt]{=}}
\newcommand{\labelJ}{+}
\newcommand{\labelK}{+}
\newcommand{\labelL}{+}
\newcommand{\labela}{\raisebox{-0.60ex}[0pt][0pt]{=}}
\newcommand{\labelb}{\raisebox{-0.60ex}[0pt][0pt]{=}}
\newcommand{\labelc}{\raisebox{-0.60ex}[0pt][0pt]{=}}
\newcommand{\labeld}{+}
\newcommand{\labele}{+}
\newcommand{\labelf}{+}
\newcommand{\labelg}{+}
\newcommand{\labelh}{+}
\newcommand{\labeli}{+}
\newcommand{\labelj}{\raisebox{-0.60ex}[0pt][0pt]{=}}
\newcommand{\labelk}{\raisebox{-0.60ex}[0pt][0pt]{=}}
\newcommand{\labell}{\raisebox{-0.60ex}[0pt][0pt]{=}}

%% def
\draw (0,-\cubesep) ++(\bangle:\cubesep)
node[minimum height=\cubesize,minimum width=\cubesize,inner sep=0pt,outer sep=0pt,shade,shading angle=45,shading=axis,left color=black!50,right color=gray!10,text=black,cm={\ca,-\sa,0,1,(0,0)}] (d) {\labeld};
\node[minimum height=\cubesize,minimum width=\cubesize,inner sep=0pt,outer sep=0pt,shade,shading=axis,shading angle=45,right color=black!50,left color=gray!10,text=black,anchor=west,cm={\cb,\sb,0,1,(0,0)}] (e) at (d.east) {\labele};
\node[minimum size=\cubesize,minimum width=\cubesize,inner sep=0pt,outer sep=0pt,shade,shading=axis,shading angle=45,top color=black!50,bottom color=white,text=black,anchor=north west,cm={\cb,\sb,-\ca,\sa,(0,0)}] (f) at (d.north west) {\labelf};

% i/o line from face opposite to e
\path (e.center)  ++(-\aangle:-\cubesize) coordinate (eopp);
\draw[daveline,mid arrowa] (eopp) -- ++(-\aangle:-5ex) coordinate (eout);
\draw[daveline] (eout) \updown;

% redraw def to hide i/o line

\node[minimum height=\cubesize,minimum width=\cubesize,inner sep=0pt,outer sep=0pt,shade,shading angle=45,shading=axis,left color=black!50,right color=gray!10,text=black,cm={\ca,-\sa,0,1,(0,0)}] at (d) {\labeld};
\node[minimum height=\cubesize,minimum width=\cubesize,inner sep=0pt,outer sep=0pt,shade,shading=axis,shading angle=45,right color=black!50,left color=gray!10,text=black,anchor=west,cm={\cb,\sb,0,1,(0,0)}] at (d.east) {\labele};
\node[minimum size=\cubesize,minimum width=\cubesize,inner sep=0pt,outer sep=0pt,shade,shading=axis,shading angle=45,top color=black!50,bottom color=white,text=black,anchor=north west,cm={\cb,\sb,-\ca,\sa,(0,0)}] at (d.north west) {\labelf};

%% draw the neighbours of def, namely abc, jkl, DEF

%% abc
\node[minimum height=\cubesize,minimum width=\cubesize,inner sep=0pt,outer sep=0pt,shade,shading angle=45,shading=axis,left color=black!50,right color=gray!10,text=black,cm={\ca,-\sa,0,1,(0,0)}] (a) at (0,-\cubesep) {\labela};
\node[minimum height=\cubesize,minimum width=\cubesize,inner sep=0pt,outer sep=0pt,shade,shading=axis,shading angle=45,right color=black!50,left color=gray!10,text=black,anchor=west,cm={\cb,\sb,0,1,(0,0)}] (b) at (a.east) {\labelb};
\node[minimum size=\cubesize,minimum width=\cubesize,inner sep=0pt,outer sep=0pt,shade,shading=axis,shading angle=45,top color=black!50,bottom color=white,text=black,anchor=north west,cm={\cb,\sb,-\ca,\sa,(0,0)}] (c) at (a.north west) {\labelc};

%% jkl
\draw (0,-\cubesep) ++(-\aangle:\cubesep) ++(\bangle:\cubesep) node[minimum height=\cubesize,minimum width=\cubesize,inner sep=0pt,outer sep=0pt,shade,shading angle=45,shading=axis,left color=black!50,right color=gray!10,text=black,cm={\ca,-\sa,0,1,(0,0)}] (j) {\labelj};
\node[minimum height=\cubesize,minimum width=\cubesize,inner sep=0pt,outer sep=0pt,shade,shading=axis,shading angle=45,right color=black!50,left color=gray!10,text=black,anchor=west,cm={\cb,\sb,0,1,(0,0)}] (k) at (j.east) {\labelk};
\node[minimum size=\cubesize,minimum width=\cubesize,inner sep=0pt,outer sep=0pt,shade,shading=axis,shading angle=45,top color=black!50,bottom color=white,text=black,anchor=north west,cm={\cb,\sb,-\ca,\sa,(0,0)}] (l) at (j.north west) {\labell};

%% DEF
\node[minimum height=\cubesize,minimum width=\cubesize,inner sep=0pt,outer sep=0pt,shade,shading angle=45,shading=axis,left color=black!50,right color=gray!10,text=black,cm={\ca,-\sa,0,1,(0,0)}] (D) at (\bangle:\cubesep) {\labelD};
\node[minimum height=\cubesize,minimum width=\cubesize,inner sep=0pt,outer sep=0pt,shade,shading=axis,shading angle=45,right color=black!50,left color=gray!10,text=black,anchor=west,cm={\cb,\sb,0,1,(0,0)}] (E) at (D.east) {\labelE};
\node[minimum size=\cubesize,minimum width=\cubesize,inner sep=0pt,outer sep=0pt,shade,shading=axis,shading angle=45,top color=black!50,bottom color=white,text=black,anchor=north west,cm={\cb,\sb,-\ca,\sa,(0,0)}] (F) at (D.north west) {\labelF};

% i/o line to face opposite to b
\path (b.center) ++(-\aangle:-\cubesize) coordinate (bopp);
\draw[daveline,mid revarrowa] (bopp) -- ++(-\aangle:-5ex) coordinate (bout);
\draw[daveline] (bout) \updown;

% i/o line to face opposite E
\path (E.center) ++(-\aangle:-\cubesize) coordinate (Eopp);
\draw[daveline,mid revarrowa] (Eopp) --++(-\aangle:-5ex) coordinate (Eout);
\draw[daveline] (Eout) \updown;

%% interior lines from def to its neighbours

\path (a.center) ++(\bangle:\cubesize) coordinate (aopp);
\path (k.center) ++(-\aangle:-\cubesize) coordinate (kopp);
\path (F.center) ++(0,-\cubesize) coordinate (Fopp);
\draw[daveline,mid revarrowb] (d.center)--(aopp);
\draw[daveline,mid revarrowa] (e.center)--(kopp);
\draw[daveline,mid revarrowc] (f.center)--(Fopp);

% redraw abc
\node[minimum height=\cubesize,minimum width=\cubesize,inner sep=0pt,outer sep=0pt,shade,shading angle=45,shading=axis,left color=black!50,right color=gray!10,text=black,cm={\ca,-\sa,0,1,(0,0)}] at (a) {\labela};
\node[minimum height=\cubesize,minimum width=\cubesize,inner sep=0pt,outer sep=0pt,shade,shading=axis,shading angle=45,right color=black!50,left color=gray!10,text=black,anchor=west,cm={\cb,\sb,0,1,(0,0)}] at (a.east) {\labelb};
\node[minimum size=\cubesize,minimum width=\cubesize,inner sep=0pt,outer sep=0pt,shade,shading=axis,shading angle=45,top color=black!50,bottom color=white,text=black,anchor=north west,cm={\cb,\sb,-\ca,\sa,(0,0)}] at (a.north west) {\labelc};
%% redraw jkl
\draw (0,-\cubesep) ++(-\aangle:\cubesep) ++(\bangle:\cubesep) node[minimum height=\cubesize,minimum width=\cubesize,inner sep=0pt,outer sep=0pt,shade,shading angle=45,shading=axis,left color=black!50,right color=gray!10,text=black,cm={\ca,-\sa,0,1,(0,0)}] at (j) {\labelj};
\node[minimum height=\cubesize,minimum width=\cubesize,inner sep=0pt,outer sep=0pt,shade,shading=axis,shading angle=45,right color=black!50,left color=gray!10,text=black,anchor=west,cm={\cb,\sb,0,1,(0,0)}] at (j.east) {\labelk};
\node[minimum size=\cubesize,minimum width=\cubesize,inner sep=0pt,outer sep=0pt,shade,shading=axis,shading angle=45,top color=black!50,bottom color=white,text=black,anchor=north west,cm={\cb,\sb,-\ca,\sa,(0,0)}] at (j.north west) {\labell};
%% redraw DEF
\node[minimum height=\cubesize,minimum width=\cubesize,inner sep=0pt,outer sep=0pt,shade,shading angle=45,shading=axis,left color=black!50,right color=gray!10,text=black,cm={\ca,-\sa,0,1,(0,0)}] at (D)  {\labelD};
\node[minimum height=\cubesize,minimum width=\cubesize,inner sep=0pt,outer sep=0pt,shade,shading=axis,shading angle=45,right color=black!50,left color=gray!10,text=black,anchor=west,cm={\cb,\sb,0,1,(0,0)}] at (D.east) {\labelE};
\node[minimum size=\cubesize,minimum width=\cubesize,inner sep=0pt,outer sep=0pt,shade,shading=axis,shading angle=45,top color=black!50,bottom color=white,text=black,anchor=north west,cm={\cb,\sb,-\ca,\sa,(0,0)}] at (D.north west) {\labelF};

%% next draw ABC, ghi, JKL

%% ABC
\node[minimum height=\cubesize,minimum width=\cubesize,inner sep=0pt,outer sep=0pt,shade,shading angle=45,shading=axis,left color=black!50,right color=gray!10,text=black,cm={\ca,-\sa,0,1,(0,0)}] (A) at (0,0) {\labelA};
\node[minimum height=\cubesize,minimum width=\cubesize,inner sep=0pt,outer sep=0pt,shade,shading=axis,shading angle=45,right color=black!50,left color=gray!10,text=black,anchor=west,cm={\cb,\sb,0,1,(0,0)}] (B) at (A.east) {\labelB};
\node[minimum size=\cubesize,minimum width=\cubesize,inner sep=0pt,outer sep=0pt,shade,shading=axis,shading angle=45,top color=black!50,bottom color=white,text=black,anchor=north west,cm={\cb,\sb,-\ca,\sa,(0,0)}] (C) at (A.north west) {\labelC};
%% ghi
\draw (0,-\cubesep) ++(-\aangle:\cubesep)
node[minimum height=\cubesize,minimum width=\cubesize,inner sep=0pt,outer sep=0pt,shade,shading angle=45,shading=axis,left color=black!50,right color=gray!10,text=black,cm={\ca,-\sa,0,1,(0,0)}] (g) {\labelg};
\node[minimum height=\cubesize,minimum width=\cubesize,inner sep=0pt,outer sep=0pt,shade,shading=axis,shading angle=45,right color=black!50,left color=gray!10,text=black,anchor=west,cm={\cb,\sb,0,1,(0,0)}] (h) at (g.east) {\labelh};
\node[minimum size=\cubesize,minimum width=\cubesize,inner sep=0pt,outer sep=0pt,shade,shading=axis,shading angle=45,top color=black!50,bottom color=white,text=black,anchor=north west,cm={\cb,\sb,-\ca,\sa,(0,0)}] (i) at (g.north west) {\labeli};
%% JKL
\draw (0,0) ++(-\aangle:\cubesep) ++(\bangle:\cubesep) node[minimum height=\cubesize,minimum width=\cubesize,inner sep=0pt,outer sep=0pt,shade,shading angle=45,shading=axis,left color=black!50,right color=gray!10,text=black,cm={\ca,-\sa,0,1,(0,0)}] (J) {\labelJ};
\node[minimum height=\cubesize,minimum width=\cubesize,inner sep=0pt,outer sep=0pt,shade,shading=axis,shading angle=45,right color=black!50,left color=gray!10,text=black,anchor=west,cm={\cb,\sb,0,1,(0,0)}] (K) at (J.east) {\labelK};
\node[minimum size=\cubesize,minimum width=\cubesize,inner sep=0pt,outer sep=0pt,shade,shading=axis,shading angle=45,top color=black!50,bottom color=white,text=black,anchor=north west,cm={\cb,\sb,-\ca,\sa,(0,0)}] (L) at (J.north west) {\labelL};

% internal edges

\path (A.center) ++(\bangle:\cubesize) coordinate (Aopp);
\path (C.center) ++(0,-\cubesize) coordinate (Copp);
\path (g.center) ++(\bangle:\cubesize) coordinate (gopp);
\path (h.center) ++(-\aangle:-\cubesize) coordinate (hopp);
\path (K.center) ++(-\aangle:-\cubesize) coordinate (Kopp);
\path (L.center) ++(0,-\cubesize) coordinate (Lopp);
\draw[daveline,mid arrowb] (D.center) -- (Aopp);
\draw[daveline,mid arrowa] (E.center) -- (Kopp);
\draw[daveline,mid arrowb] (j.center) -- (gopp);
\draw[daveline,mid arrowc] (l.center) -- (Lopp);
\draw[daveline,mid arrowa] (b.center) -- (hopp);
\draw[daveline,mid arrowc] (c.center) -- (Copp);

% i/o line
\path (B.center) ++(-\aangle:-\cubesize) coordinate (Bopp);
\draw[daveline,mid arrowa] (Bopp) --++(-\aangle:-5ex) coordinate (Bout);
\draw[daveline] (Bout) \updown;

% redraw ABC, ghi, JKL
%% ABC
\node[minimum height=\cubesize,minimum width=\cubesize,inner sep=0pt,outer sep=0pt,shade,shading angle=45,shading=axis,left color=black!50,right color=gray!10,text=black,cm={\ca,-\sa,0,1,(0,0)}] at (A) {\labelA};
\node[minimum height=\cubesize,minimum width=\cubesize,inner sep=0pt,outer sep=0pt,shade,shading=axis,shading angle=45,right color=black!50,left color=gray!10,text=black,anchor=west,cm={\cb,\sb,0,1,(0,0)}] at (A.east) {\labelB};
\node[minimum size=\cubesize,minimum width=\cubesize,inner sep=0pt,outer sep=0pt,shade,shading=axis,shading angle=45,top color=black!50,bottom color=white,text=black,anchor=north west,cm={\cb,\sb,-\ca,\sa,(0,0)}] at (A.north west) {\labelC};
%% ghi
\node[minimum height=\cubesize,minimum width=\cubesize,inner sep=0pt,outer sep=0pt,shade,shading angle=45,shading=axis,left color=black!50,right color=gray!10,text=black,cm={\ca,-\sa,0,1,(0,0)}] at (g) {\labelg};
\node[minimum height=\cubesize,minimum width=\cubesize,inner sep=0pt,outer sep=0pt,shade,shading=axis,shading angle=45,right color=black!50,left color=gray!10,text=black,anchor=west,cm={\cb,\sb,0,1,(0,0)}] at (g.east) {\labelh};
\node[minimum size=\cubesize,minimum width=\cubesize,inner sep=0pt,outer sep=0pt,shade,shading=axis,shading angle=45,top color=black!50,bottom color=white,text=black,anchor=north west,cm={\cb,\sb,-\ca,\sa,(0,0)}] at (g.north west) {\labeli};
%% JKL
\node[minimum height=\cubesize,minimum width=\cubesize,inner sep=0pt,outer sep=0pt,shade,shading angle=45,shading=axis,left color=black!50,right color=gray!10,text=black,cm={\ca,-\sa,0,1,(0,0)}] at (J) {\labelJ};
\node[minimum height=\cubesize,minimum width=\cubesize,inner sep=0pt,outer sep=0pt,shade,shading=axis,shading angle=45,right color=black!50,left color=gray!10,text=black,anchor=west,cm={\cb,\sb,0,1,(0,0)}] at (J.east) {\labelK};
\node[minimum size=\cubesize,minimum width=\cubesize,inner sep=0pt,outer sep=0pt,shade,shading=axis,shading angle=45,top color=black!50,bottom color=white,text=black,anchor=north west,cm={\cb,\sb,-\ca,\sa,(0,0)}] at (J.north west) {\labelL};

%% final vertex GHI
\node[minimum height=\cubesize,minimum width=\cubesize,inner sep=0pt,outer sep=0pt,shade,shading angle=45,shading=axis,left color=black!50,right color=gray!10,text=black,cm={\ca,-\sa,0,1,(0,0)}] (G) at (-\aangle:\cubesep) {\labelG};
\node[minimum height=\cubesize,minimum width=\cubesize,inner sep=0pt,outer sep=0pt,shade,shading=axis,shading angle=45,right color=black!50,left color=gray!10,text=black,anchor=west,cm={\cb,\sb,0,1,(0,0)}] (H) at (G.east) {\labelH};
\node[minimum size=\cubesize,minimum width=\cubesize,inner sep=0pt,outer sep=0pt,shade,shading=axis,shading angle=45,top color=black!50,bottom color=white,text=black,anchor=north west,cm={\cb,\sb,-\ca,\sa,(0,0)}] (I) at (G.north west) {\labelI};

% internal edges

\path (G.center) ++(\bangle:\cubesize) coordinate (Gopp);
\path (H.center) ++(-\aangle:-\cubesize) coordinate (Hopp);
\path (I.center) ++(0,-\cubesize) coordinate (Iopp);
\draw[daveline,mid revarrowa] (B.center) -- (Hopp);
\draw[daveline,mid revarrowb] (J.center) -- (Gopp);
\draw[daveline,mid revarrowc] (i.center) -- (Iopp);

%% redraw GHI
\node[minimum height=\cubesize,minimum width=\cubesize,inner sep=0pt,outer sep=0pt,shade,shading angle=45,shading=axis,left color=black!50,right color=gray!10,text=black,cm={\ca,-\sa,0,1,(0,0)}] at (G) {\labelG};
\node[minimum height=\cubesize,minimum width=\cubesize,inner sep=0pt,outer sep=0pt,shade,shading=axis,shading angle=45,right color=black!50,left color=gray!10,text=black,anchor=west,cm={\cb,\sb,0,1,(0,0)}] at (G.east) {\labelH};
\node[minimum size=\cubesize,minimum width=\cubesize,inner sep=0pt,outer sep=0pt,shade,shading=axis,shading angle=45,top color=black!50,bottom color=white,text=black,anchor=north west,cm={\cb,\sb,-\ca,\sa,(0,0)}] at (G.north west) {\labelI};

% i/o lines

\draw[daveline,mid arrowa] (K.center) -- ++(-\aangle:5ex) coordinate (KOUT);
\draw[daveline] (KOUT) \updown;
\draw[daveline,mid revarrowa] (H.center) -- ++(-\aangle:5ex) coordinate (HOUT);
\draw[daveline] (HOUT) \updown;
\draw[daveline,mid revarrowa] (k.center) -- ++(-\aangle:5ex) coordinate (kOUT);
\draw[daveline] (kOUT) \updown;
\draw[daveline,mid arrowa] (h.center) -- ++(-\aangle:5ex) coordinate (hOUT);
\draw[daveline] (hOUT) \updown;

\end{tikzpicture}
\caption{``2-state" 3-cube realization of the binary (8,4,4) first-order Reed-Muller code.  (Figure courtesy of F. R. Kschischang.)}
\label{3CR}
\end{figure}

\newpage
For the field of codes on graphs, this development suggests exploring graphical models that are inspired more by algebraic topology than by traditional system-theory models (\eg trellises, tail-biting trellises, kernel and image representations).  For example, we have recently found a simple and elegant ``2-state" elementary normal realization of the $(8, 4, 4)$ first-order Reed-Muller code on a 3-cube graph, shown in Figure \ref{3CR}.  Are there similarly ``nice" realizations of more complex codes?

\section*{Acknowledgments}
I am very grateful to Ali  Al-Bashabsheh, Frank Kschischang, Mehdi Molkaraie, and Pascal Vontobel
for helpful comments on earlier drafts of this paper.

\clearpage
\section*{Appendix:  The NFG duality theorem}

The \emph{normal factor graph duality theorem} (NFGDT) \cite{AM11,F11b, FV11} is the key duality result for normal factor graphs (NFGs) over finite abelian groups.  It shows that the partition function of a dual NFG is equal to the Fourier transform of the partition function of the primal NFG, up to a scale factor which was shown  in \cite{AM11} to be $|\A_E|$ (see below for terminology and notation).

In this appendix we give the simplest proof we know of this result, and extend it to situations in which the NFG is based on a normal realization \cite{F01};  specifically, to NFGs that are based purely on normal realizations, and to  {edge-weighted NFGs} (EWNFGs), as discussed in the main text. We compute the appropriate scale factors for these cases.

\subsection*{A.1 \hspace{1ex} Fourier transforms over finite abelian groups}

Given an additive finite abelian group $\A$, its \emph{dual group} (or \emph{character group}) $\hat{\A}$ may be defined as the set of all homomorphisms $\hat{a}:  \A \to \R/\Z$, where $\R/\Z$ is the additive group of  real numbers modulo 1.  It is well known that $\hat{\A}$ is a finite abelian group that is isomorphic to $\A$.  Also, the dual group to $\hat{\A}$ is $\A$, where $a:  \hat{\A} \to \R/\Z$  is defined by $a(\hat{a}) = \hat{a}(a)$.

For $\hat{a} \in \hat{\A}, a \in \A$, we may define the ``inner product" (\emph{pairing}) $\inner{\hat{a}}{a} = \hat{a}(a) = a(\hat{a})$.  The  usual inner product properties--- \eg $\inner{\hat{a}}{0} = 0$, $\inner{\hat{a}}{-a} = -\inner{\hat{a}}{a}$, $\inner{\hat{a}}{a \pm b} = \inner{\hat{a}}{a} \pm \inner{\hat{a}}{b}$--- then follow from the properties of homomorphisms.  

If $f(a)$ is any complex-valued function $f: \A \to \C$, then its \emph{Fourier transform} $\hat{f}(\hat{a})$ is  the complex-valued function $\hat{f}: \hat{\A} \to \C$ defined by
$$\hat{f}(\hat{a}) = \sum_{a \in {\A}} f(a) e^{2\pi i{\inner{\hat{a}}{a}}}.$$
The matrix $\FF = \{e^{2\pi i{\inner{\hat{a}}{a}}} \mid a \in \A, \hat{a} \in \hat{\A}\}$ is called the \emph{Fourier transform matrix} over $\A$.  We will regard $\FF$ as a function of two variables with alphabets $\A$ and $\hat{\A}$.

If $\hat{g}(\hat{a})$ is any complex-valued function $\hat{g}: \hat{\A} \to \C$, then its \emph{inverse Fourier transform} $g(a)$ is  the complex-valued function  $g: \A \to \C$ defined by
$$g(a) = |\A|^{-1} \sum_{\hat{a} \in \hat{\A}} \hat{g}(\hat{a}) e^{-2\pi i \inner{\hat{a}}{a}}.$$
The matrix $\FF^{-1} = \{|\A|^{-1}e^{-2\pi i \inner{\hat{a}}{a}} \mid \hat{a} \in \hat{\A}, a \in \A\}$ is called the \emph{inverse Fourier transform matrix} over $\A$, and will also be regarded as a function of two variables with alphabets $\hat{\A}$ and ${\A}$.
 
 We may verify that $\FF^{-1}$ is in fact the inverse of $\FF$ by using the basic \emph{orthogonality relation}
$ |\A|^{-1} \sum_{\hat{a} \in \hat{\A}} e^{2 \pi i \inner{\hat{a}}{a}} = \delta_{{a}}.$

Given a subgroup $\CC \subseteq \A$, the \emph{orthogonal subgroup} $\CC^\perp \subseteq \hat{\A}$ is the set $\{\hat{a} \in \hat{\A} \mid \inner{\hat{a}}{a} = 0, \forall a \in \CC\}$.  The orthogonal subgroup to $\CC^\perp$ is $\CC$, and $|\CC||\CC^\perp| = |\A|$.  It is well known, and easy to prove, that  the Fourier transform of the indicator function $\delta_{\CC}$ of $\CC$ is the scaled indicator function $|\CC|\delta_{\CC^\perp}$ of $\CC^\perp$.\footnote{The  simple and lovely proof goes as follows:  (a) obvious for $\hat{a} \in \CC^\perp$, since $\inner{\hat{a}}{a} = 0$ for all $a \in \CC$;   (b) if $\hat{a} \notin \CC^\perp$, then $\inner{\hat{a}}{b} \neq 0$ for some ${b} \in \CC$; for this ${b}$, we have 
$e^{2 \pi i \inner{\hat{a}}{b}} (\sum_{a \in \CC} e^{2 \pi i \inner{\hat{a}}{a}}) = \sum_{a \in \CC} e^{2 \pi i \inner{\hat{a}}{a + b}} = \sum_{a \in \CC} e^{2 \pi i \inner{\hat{a}}{a}},$
since $\CC + b = \CC$; but since $e^{ 2 \pi i \inner{\hat{a}}{b}} \neq 1$, this equation can hold only if  $ \sum_{a \in \CC} e^{2 \pi i \inner{\hat{a}}{a}} = 0$.}  The above orthogonality relation  follows from the special case in which $\CC =  \{0\}$ and $\CC^\perp = \hat{\A}$. 

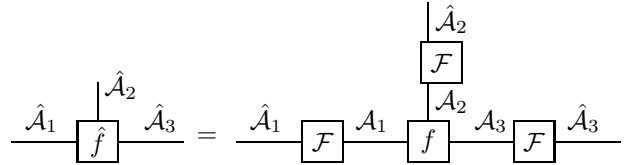
\begin{figure}[h]
\setlength{\unitlength}{5pt}
\centering
\begin{picture}(40,10)(3, 0)

\put(17,0){\line(1,0){5}}
\put(18,1){$\hat{\A}_1$}
\put(22,-1.5){\framebox(3,3){$\FF$}}
\put(25,0){\line(1,0){5}}
\put(26,1){${\A_1}$}
\put(30,-1.5){\framebox(3,3){$f$}}
\put(35,1){${\A_3}$}
\put(33,0){\line(1,0){5}}
\put(38,-1.5){\framebox(3,3){$\FF$}}
\put(41,0){\line(1,0){5}}
\put(42,1){$\hat{\A}_3$}
\put(31.5,1.5){\line(0,1){3}}
\put(32,2.5){${\A_2}$}
\put(31,4.5){\framebox(3,3){$\FF$}}
\put(31.5,7.5){\line(0,1){3}}
\put(32,8.5){$\hat{\A}_2$}

\put(14,0){$=$}

\put(0,0){\line(1,0){5}}
\put(1,1){$\hat{\A}_1$}
\put(5,-1.5){\framebox(3,3){$\hat{f}$}}
\put(6.5,1.5){\line(0,1){3}}
\put(7,3.5){$\hat{\A}_2$}
\put(10,1){$\hat{\A}_3$}
\put(8,0){\line(1,0){5}}
\end{picture}
\caption{The Fourier transform is separable.}
\label{Sep}
\end{figure}

Finally,  if $f(\ab)$ is a function of multiple variables $\ab = \{a_i, i \in \I\}$, then its Fourier transform $\hat{f}(\hat{\ab})$ is obtained by taking the Fourier transform of each variable separately, since $\inner{\hat{\ab}}{\ab} = \sum_\I \inner{\hat{a}_i}{a_i}$;  \ie the Fourier transform is \emph{separable}, as illustrated in Figure \ref{Sep}.

\subsection*{A.2 \hspace{1ex} Normal factor graphs}

A \emph{normal factor graph} (NFG) is based on a graph $\G = (V, E, H)$ consisting of a set of vertices indexed by a vertex index set $V$, a set of edges indexed by an edge index set $E$, and a set of \emph{half-edges}  indexed by a half-edge index set $H$.  With each edge $e \in E$ we associate an internal (state) variable $s_e$ whose alphabet is a finite abelian group denoted by $\A_e$, and with each half-edge we associate an external variable $a_h$ whose alphabet is a finite abelian group denoted by $\A_h$.  The \emph{internal variable configuration space} is defined as the Cartesian product $\A_E = \prod_E \A_e$.
 
 With each vertex $v \in V$ we associate a complex-valued function $f_v(\sb_v, \ab_v)$ of the variables corresponding to the edges and half-edges that are incident on vertex $v$.
The \emph{partition function} (or ``exterior function" \cite{AM11}) of the NFG is then defined as the following function of its external variables:
$$ Z(\ab) = \sum_{\sb \in \A_E} \prod_{v \in V} f_v(\sb_v, \ab_v). $$
 
The \emph{dual normal factor graph} to an NFG as defined above is based on the same graph $\G = (V, E, H)$, but with the following replacements:

\begin{itemize}
\item For each edge $e \in E$, the variable $s_e \in \A_e$ is replaced by a dual variable $\hat{s}_e \in \hat{\A}_e$, where $\hat{\A}_e$ denotes the dual group to $\A_e$;
\item For each half-edge $h \in H$, the variable $a_h \in \A_h$ is replaced by a dual variable $\hat{a}_h \in \hat{\A}_h$;
\item For each vertex $v \in V$, the function $f_v(\sb_v, \ab_v)$ is replaced by its Fourier transform $\hat{f}_v(\hat{\sb}_v, \hat{\ab}_v)$;
\item Lastly, each edge is replaced by a sign-inverting edge.
\end{itemize}

The \emph{normal factor graph duality theorem} (NFGDT) says that the partition function of the dual NFG is the Fourier transform of the partition function of the primal NFG, up to a scale factor that  will be  determined shortly.
The key to the proof of the NFGDT is the following {Edge Replacement Lemma} (an example of what is called a ``holographic transformation" in \cite{AM11, F11b, FV11}): 

\vspace{1ex}
\noindent
\textbf{Edge Replacement Lemma}.
In any NFG, any edge representing a variable whose alphabet is a finite abelian group $\A$ may be replaced by
\begin{picture}(8,1)
\put(0, 0.5){\line(1,0){1}}
\put(1,-0.25){\framebox(1.5,1.5){$\FF$}}
\put(2.5, 0.5){\line(1,0){1}}
\put(3.5, 0.15){$\circ$}
\put(4.3, 0.5){\line(1,0){1.2}}
\put(5.5,-0.25){\framebox(1.5,1.5){$\FF$}}
\put(7, 0.5){\line(1,0){1}}
\end{picture}, namely
a cascade of $\FF$, a sign inverter, and $\FF$,
plus a disconnected node \begin{picture}(4,1)
\put(0,-0.25){\framebox(4,1.75){$|\A|^{-1}$}}
\end{picture} that contributes a scale factor of  $|\A|^{-1}$ to the partition function, without changing the partition function.  

\vspace{1ex}
\noindent
\textit{Proof}:  By the basic orthogonality relation given above, we have $|\A|^{-1} \sum_{\hat{\A}}e^{2 \pi i \inner{\hat{a}}{a}}e^{-2 \pi i \inner{\hat{a}}{a'}} = \delta_{aa'}$.
\qed \vspace{1ex}

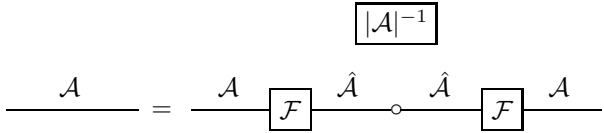
\begin{figure}[h]
\setlength{\unitlength}{5pt}
\centering
\begin{picture}(40,6)(2, 0)
\put(0,0){\line(1,0){10}}
\put(4,1){$\A$}
\put(11,-0.5){$=$}
\put(14,0){\line(1,0){6}}
\put(16,1){$\A$}
\put(20,-1.5){\framebox(3,3){$\FF$}}
\put(23,0){\line(1,0){6}}
\put(25,1){$\hat{\A}$}
\put(29, -0.5){$\circ$}
\put(32,1){$\hat{\A}$}
\put(29.8,0){\line(1,0){6.2}}
\put(36,-1.5){\framebox(3,3){$\FF$}}
\put(39,0){\line(1,0){6}}
\put(41,1){${\A}$}
\put(26.5,5){\framebox(6,3){$|\A|^{-1}$}}
\end{picture}
\caption{{Edge Replacement Lemma}:  an edge with  alphabet $\A$ may be replaced by a cascade of $\FF$, a sign inverter, and $\FF$, plus a disconnected node representing a scale factor $|\A|^{-1}$.}
\label{Eq3}
\end{figure}

This lemma is illustrated by Figure \ref{Eq3}.

The NFGDT then follows:

\vspace{1ex}
\noindent
\textbf{Theorem} (\emph{NFG duality}).  If $\G = (V, E, H)$ is a normal factor graph  with functions $\{f_v \mid v \in V\}$, internal variable alphabets $\{\A_e \mid e \in E\}$, and external variable alphabets $\{\A_h \mid h \in H\}$ whose partition function is $Z(\ab)$, then the dual normal factor graph with functions $\{\hat{f}_v \mid v \in V\}$, internal variable alphabets $\{\hat{\A}_e \mid e \in E\}$,  external variable alphabets $\{\hat{\A}_h \mid h \in H\}$, and with sign inverters inserted in each edge has partition function 
$|\A_E| \hat{Z}(\hat{\ab}),$ where $\hat{Z}(\hat{\ab})$ is the Fourier transform of $Z(\ab)$, and
 the scale factor is $|\A_E| = \prod_{E} |\A_e|$. 
 
 \vspace{1ex}
 \noindent
 \textit{Proof}:   If we have an NFG whose partition function is $Z(\ab)$, then by definition and by separability the Fourier transform $\hat{Z}(\hat{\ab})$ is the partition function of  the NFG that results when each external half-edge, representing a variable with alphabet $\A_h, h \in H$, is replaced by \begin{picture}(3.5,1)
\put(0, 0.5){\line(1,0){1}}
\put(1,-0.25){\framebox(1.5,1.5){$\FF$}}
\put(2.5, 0.5){\line(1,0){1}}
\end{picture}, namely a Fourier transform from $\A_h$ to the dual external variable alphabet $\hat{\A}_h$.  By the Edge Replacement Lemma, if we then replace each  edge $e \in E$, representing an internal variable with alphabet $\A_e$, by  \begin{picture}(8,1)
\put(0, 0.5){\line(1,0){1}}
\put(1,-0.25){\framebox(1.5,1.5){$\FF$}}
\put(2.5, 0.5){\line(1,0){1}}
\put(3.5, 0.15){$\circ$}
\put(4.3, 0.5){\line(1,0){1.2}}
\put(5.5,-0.25){\framebox(1.5,1.5){$\FF$}}
\put(7, 0.5){\line(1,0){1}}
\end{picture},
omitting the scale factor $|\A_e|^{-1}$,
then the resulting NFG has Fourier transform $|\A_E|\hat{Z}(\hat{\ab})$, where $|\A_E| = \prod_{E} |\A_e|$.
Now each function node $f_v, v \in V,$ in the NFG is surrounded by edge and half-edge segments of the form  \begin{picture}(3.5,1)
\put(0, 0.5){\line(1,0){1}}
\put(1,-0.25){\framebox(1.5,1.5){$\FF$}}
\put(2.5, 0.5){\line(1,0){1}}
\end{picture}.  By the definition of the Fourier transform and by separability, each such node and its surrounding Fourier transform functions realizes the Fourier-transformed function $\hat{f}_v$. 
\qed \vspace{1ex}
 
The scale factor $|\A_E|$ is thus the product of all internal variable alphabet (state space) sizes $|\A_e|$, as in the version of the NFGDT derived in  \cite{AM11}. 

\vspace{-1ex}
\subsection*{A.3 \hspace{1ex} Scale factors of code indicator functions}

In Section \ref{Sec3.2}, we consider interpreting a  normal realization as an NFG.  Edges and half-edges  represent the same internal and external variables, but a  constraint code $\CC_v$ is now interpreted as its indicator function $\delta_{\CC_v}$.  If the normal realization has external behavior $\CC$ and  unobservable behavior $\Bf^u$, then we conclude that the partition function  of the resulting NFG is as follows:
   
   \vspace{1ex}
   \noindent
   \textbf{Theorem 6} (\textit{Normal realization as an NFG}).  If all alphabets are finite abelian groups, then a normal realization with external behavior $\CC$ and unobservable behavior     $\Bf^u$ may be interpreted as a  normal factor graph whose partition function is $Z(\ab) = |\Bf^u| \delta_\CC(\ab)$.  \qed \vspace{1ex}
   
   By the NFG duality theorem, the dual NFG to the NFG of Theorem 6 has partition function 
   $$|\A_E|\hat Z(\hat \ab) = |\A_E| |\Bf^u| |\CC| \delta_{\CC^\perp}(\hat \ab). $$
The vertex functions of this  dual NFG  are the Fourier transforms of the primal vertex functions $f_v = \delta_{\CC_v}$, and therefore are of the form $\hat{f}_v = |\CC_v| \delta_{\CC_v^\perp}$.  The NFG of the dual NR of the NR of Theorem 6 is the same as this dual NFG, except that its function nodes are of the form $\delta_{\CC_v^\perp}$;  therefore its partition function is 
   $|\A_E||\CC_V|^{-1}\hat Z(\hat \ab) = |\A_E| |\Bf^u| |\CC| |\CC_V|^{-1}\delta_{\CC^\perp}(\hat \ab) = |\A_E| |\Bf|  |\CC_V|^{-1}\delta_{\CC^\perp}(\hat \ab), $
   where $|\CC_V| = \prod_V |\CC_v|$.  In summary:

   \vspace{1ex}
   \noindent
   \textbf{Theorem 7} (\textit{Dual normal realization as an NFG}).  The dual of a finite abelian group normal realization with  behavior $\Bf$ and external behavior $\CC$ may be interpreted as an NFG with partition function  $|\Bf||\A_E||\CC_V|^{-1} \delta_{\CC^\perp}(\hat{\ab}),$
     where $|\A_E| = \prod_E |\A_e|$ and $\CC_V = \prod_V |\CC_v|$.  \qed \vspace{1ex}

Now if we denote the external behavior of the dual normal realization as $\hat{\CC}$ and its unobservable behavior as $\hat{\Bf}^u$, then, from Theorem 6, its partition function as an NFG is   $|\hat{\Bf}^u| \delta_{\hat{\CC}}(\hat \ab)$.  We thus conclude that:
\begin{itemize}
\item[(a)] The external behavior $\hat{\CC}$ of the dual normal realization is $\CC^\perp$.  Thus,  when all alphabets are finite abelian groups, we obtain the normal realization duality theorem as a corollary.
\item[(b)] The size $|\hat{\Bf}^u|$ of the unobservable behavior of the dual normal realization is $|\Bf||\A_E||\CC_V|^{-1}$.
\end{itemize}

There is an interesting connection between  result (b) and the notions of observability and controllability of normal realizations defined in \cite{FGL12}.  A finite abelian group normal realization with unobservable behavior $\Bf^u$ is \emph{observable} if and only if $|\Bf^u| = 1$, and $\dim \Bf^u$ measures its ``degree of unobservability."  A finite abelian group normal realization whose dual has unobservable behavior $\hat{\Bf}^u$ is \emph{controllable} if and only if $|\hat{\Bf}^u| = 1$, and $\dim \hat{\Bf}^u$ measures its ``degree of uncontrollability."  Thus as a further corollary we have the \emph{controllability test} of \cite{FGL12}:  

\vspace{1ex}
\noindent
\textbf{Corollary} (\emph{Controllability test} \cite{FGL12}).  Given a finite abelian group normal realization with behavior $\Bf$, total constraint size $|\CC_V| = \prod_V |\CC_v|$, and total state space size $|\A_E| = \prod_{E} |\A_e|$, the unobservable behavior $\hat{\Bf}^u$ of the dual normal realization has size $|\hat{\Bf}^u| = |\Bf||\A_E||\CC_V|^{-1} \ge 1$.  Thus the realization is controllable if and only if $|\Bf| = |\CC_V|/|\A_E|$. \qed \vspace{1ex}

This controllability test may be understood as follows.  If all edges are removed from the realization, then its behavior is simply  $\CC_V = \prod_V \CC_v$, the Cartesian product of the behaviors $\CC_v$ of each of its disconnected nodes.  If we reinsert  the edge constraints, each of which is a degree-2 equality constraint between two variables with a common alphabet $\A_e$, then each such constraint will reduce the size of the behavior by a factor of $|\A_e|$, provided that it is independent of all previous constraints.  Thus $|\Bf| \ge |\CC_V|/|\A_E|$, with equality if and only if all constraints are independent.

\subsection*{A.4 \hspace{1ex} Scale factors for edge-weighted NFGs}

An \emph{edge-weighted NFG} consists of an NFG based on a normal realization of a linear or group code $\CC$ as above, in which all internal functions $f_v$ are indicator functions $\delta_{\CC_v}$ of linear or group codes $\CC_v$, plus edge-weighting functions $f_h$ attached to each external half-edge $h \in H$ of the normal realization.  The resulting NFG has no external variables, and its partition function is the constant
$Z = |\Bf^u| \sum_{\ab \in \CC} \fb(\ab),$
where $\Bf^u$ is the unobservable behavior of the normal realization as above, and $\fb(\ab) = \prod_H f_h(a_h)$.  

As noted above,  if $\CC$ is a linear or group code, 
then the Fourier transform of $\delta_{\CC}(\sb)$ is the scaled indicator function $|\CC|\delta_{\CC^\perp}(\hat{\sb})$.  Thus if we construct the ``dual" edge-weighted NFG  by replacing every node function $\delta_{\CC_v}(\sb)$  simply by $\delta_{\CC_v^\perp}(\hat{\sb})$ rather than by $|\CC_v|\delta_{\CC_v^\perp}(\hat{\sb})$, then the partition function will be reduced by a scale factor of $|\CC_V|^{-1} = \prod_{V} |\CC_v|^{-1}$.  

Thus we obtain the following corollary of the NFG duality theorem:

\vspace{1ex}
\noindent
\textbf{Theorem 8} (\emph{Edge-weighted NFG duality}).  Given an edge-weighted NFG based on a graph $G = (V,E,H)$ with internal functions $\{\delta_{\CC_v} \mid v \in V\}$, internal variable alphabets $\{\A_e \mid e \in E\}$,  external variable alphabets $\{\A_h \mid h \in H\}$, and edge-weighting functions $\{f_h \mid h \in H\}$ that realizes a partition function  $Z = |\Bf^u| \sum_{\ab \in \CC} \fb(\ab),$ then the ``dual edge-weighted NFG" with internal functions $\{\delta_{\CC_v^\perp} \mid v \in V\}$, internal variable alphabets $\{\hat{\A}_e \mid e \in E\}$,  external variable alphabets $\{\hat{\A}_h \mid h \in H\}$, external weighting functions $\{\hat{f}_h \mid h \in H\}$, and  sign inverters inserted in each edge realizes the partition function 
$$\frac{|\A_E|}{|\CC_V|} Z = |\hat{\Bf}^u| \sum_{\hat{\ab} \in \CC^\perp} \hat{\fb}(\hat{\ab}),$$
 where 
 $|\A_E| = \prod_{E} |\A_e|$, $|\CC_V| = \prod_v |\CC_v|$, and $|\hat{\Bf}^u| = |\Bf||\A_E||\CC_V|^{-1}.$  \qed \vspace{1ex}
 
For example, let us consider an Ising-type model based on a graph $\G = (V,E)$, with variable alphabet $\A$ and  edge-weighting functions $\{f_e(y_e), e \in E\}$, such as the edge-weighted NFG shown in Figure \ref{NFGE}(a).  The resulting edge-weighted NFG $G =  (V_G, E_G)$ actually has $|V_G| = |V| + 2|E|$ vertices, consisting of $|V|$ equality  functions, $|E|$ zero-sum functions, and $|E|$ edge-weighting functions;
 $|E_G| = 3|E|$ edges, each representing an internal variable with alphabet $\A$;
and no half-edges, so its partition function is a constant $Z$.

If we wish to compute the partition function of the dual edge-weighted NFG as shown in Figure \ref{DNFGE}(a) from the partition function $Z$ of the primal edge-weighted NFG, then we need to adjust $Z$ as follows.
 Since the  NFG has $|E_G| = 3|E|$ edges, we must multiply by the scale factor $|\A_E| = |\A|^{3|E|}$.
 Since the primal NFG has $|V|$ equality functions and $|E|$ zero-sum weight functions of degree 3 with total dimension $2|E|$, we need to divide by $|\CC_V| = |\A|^{|V| + 2 |E|}.$
Therefore the partition function of the dual EWNFG is $|\A|^{|E| - |V|}Z$.\footnote{Again, this result was  derived previously by Molkaraie  \cite{Mex}, and Footnote 8 gives an alternative derivation.}

\newpage

	\textbf{G. David Forney, Jr.} (M'61--F'73--LF'06) received the B.S.E. degree from Princeton University in 1961, and the Sc.D. degree from M.I.T. in 1965.

	From 1965-99 he was with the Codex Corporation and its successor, the Motorola Information Systems Group.  He is currently an Adjunct Professor Emeritus at M.I.T.

	Dr. Forney   received the 1992 IEEE Edison Medal, the 1995 IEEE Information Theory Society Claude E. Shannon Award, and the 2016 IEEE Medal of Honor.  He was elected a Fellow of the IEEE in 1973, a member of the National Academy of Engineering (U.S.A.) in 1983, a Fellow of the American Academy of Arts and Sciences in 1998, and a member of the National Academy of Sciences (U.S.A.) in 2003.

\end{document}